\RequirePackage{fix-cm}
\documentclass[smallextended]{svjour3}       
\smartqed  
\usepackage{graphicx}
\usepackage{longtable}
\usepackage{xcolor}
\usepackage[colorlinks]{hyperref}
\usepackage{natbib}
\usepackage{etoolbox}
\usepackage[figuresright]{rotating}

\hypersetup{
	colorlinks,
	linkcolor= blue,
	citecolor=blue,
	urlcolor=blue
}

\makeatletter

\patchcmd{\NAT@citex}
{\@citea\NAT@hyper@{%
		\NAT@nmfmt{\NAT@nm}%
		\hyper@natlinkbreak{\NAT@aysep\NAT@spacechar}{\@citeb\@extra@b@citeb}%
		\NAT@date}}
{\@citea\NAT@nmfmt{\NAT@nm}%
	\NAT@aysep\NAT@spacechar\NAT@hyper@{\NAT@date}}{}{}

\patchcmd{\NAT@citex}
{\@citea\NAT@hyper@{%
		\NAT@nmfmt{\NAT@nm}%
		\hyper@natlinkbreak{\NAT@spacechar\NAT@@open\if*#1*\else#1\NAT@spacechar\fi}%
		{\@citeb\@extra@b@citeb}%
		\NAT@date}}
{\@citea\NAT@nmfmt{\NAT@nm}%
	\NAT@spacechar\NAT@@open\if*#1*\else#1\NAT@spacechar\fi\NAT@hyper@{\NAT@date}}
{}{}

\makeatother
\setcitestyle{authoryear,open={(},close={)}} 
   \setcitestyle{aysep={}} 
\usepackage[utf8]{inputenc}
\usepackage[american]{babel}
\usepackage{amsmath}
\usepackage{amssymb}
\usepackage{array}
\usepackage{csquotes}

\usepackage{amsthm}
\usepackage{multirow}
\usepackage{algorithm}
\usepackage{fixltx2e}
\usepackage{amsthm}
\usepackage[inline]{enumitem}
\usepackage{caption}

\usepackage{todonotes}

\usepackage{tikz}

\usepackage{url}

\usepackage{xspace}

\newcommand{\cf}[0]{\textit{cf.}\@\xspace}
\newcommand{\ie}[0]{\textit{i.\,e.},\@\xspace}

\newcommand{\eg}[0]{\textit{e.\,g.},\@\xspace}

\newcommand{\etc}[0]{etc.\@\xspace}

\newcommand{\sect}[1]{Sec\-tion\xspace\ref{sec:#1}}

\newcommand{\fig}[1]{\figurename\xspace\ref{fig:#1}}

\newcommand{\tab}[1]{Table~\ref{tab:#1}}

\newcommand{\goal}[1]{\textbf{G#1}}
\newcommand{\rqs}[1]{\textbf{RQ#1}}

\newcommand{\qsone}[0]{\textit{Q1}\xspace}
\newcommand{\qstwo}[0]{\textit{Q2}\xspace}
\newcommand{\qsthree}[0]{\textit{Q3}\xspace}
\newcommand{\qsfour}[0]{\textit{Q4}\xspace}

\newcommand{\qssix}[0]{\textit{Q6}\xspace}
\newcommand{\qsseven}[0]{\textit{Q7}\xspace}
\newcommand{\qseight}[0]{\textit{Q8}\xspace}
\newcommand{\qsnine}[0]{\textit{Q9}\xspace}
\newcommand{\qsten}[0]{\textit{Q10}\xspace}
\newcommand{\qseleven}[0]{\textit{Q11}\xspace}
\newcommand{\qstwelve}[0]{\textit{Q12}\xspace}
\newcommand{\qsthirteen}[0]{\textit{Q13}\xspace}
\newcommand{\qsfourteen}[0]{\textit{Q14}\xspace}
\newcommand{\qsfifeteen}[0]{\textit{Q15}\xspace}
\newcommand{\qssixteen}[0]{\textit{Q16}\xspace}
\newcommand{\qsseventeen}[0]{\textit{Q17}\xspace}
\newcommand{\qseighteen}[0]{\textit{Q18}\xspace}
\newcommand{\qsnineteen}[0]{\textit{Q19}\xspace}
\newcommand{\qstwenty}[0]{\textit{Q20}\xspace}

\newcommand{\lsone}[0]{\textit{LS1}\xspace}
\newcommand{\lstwo}[0]{\textit{LS2}\xspace}
\newcommand{\lsthree}[0]{\textit{LS3}\xspace}
\newcommand{\lsfour}[0]{\textit{LS4}\xspace}
\newcommand{\lsfive}[0]{\textit{LS5}\xspace}
\newcommand{\lssix}[0]{\textit{LS6}\xspace}
\newcommand{\lsseven}[0]{\textit{LS7}\xspace}

\newcommand{\dqone}[0]{\textit{DQ1}\xspace}
\newcommand{\dqtwo}[0]{\textit{DQ2}\xspace}
\newcommand{\tqone}[0]{\textit{TQ1}\xspace}
\newcommand{\tqtwo}[0]{\textit{TQ2}\xspace}
\newcommand{\tqthree}[0]{\textit{TQ3}\xspace}
\newcommand{\tqfour}[0]{\textit{TQ4}\xspace}
\newcommand{\tqfive}[0]{\textit{TQ5}\xspace}
\newcommand{\tqsix}[0]{\textit{TQ6}\xspace}
\newcommand{\tqseven}[0]{\textit{TQ7}\xspace}
\newcommand{\tqeight}[0]{\textit{TQ8}\xspace}
\newcommand{\tqnine}[0]{\textit{TQ9}\xspace}
\newcommand{\fqone}[0]{\textit{FQ1}\xspace}

\newcommand{\fqsix}[0]{\textit{FQ6}\xspace}
\newcommand{\feone}[0]{\textit{FE1}\xspace}
\newcommand{\fetwo}[0]{\textit{FE2}\xspace}
\newcommand{\fethree}[0]{\textit{FE3}\xspace}
\newcommand{\fefour}[0]{\textit{FE4}\xspace}
\newcommand{\fefive}[0]{\textit{FE5}\xspace}
\newcommand{\fesix}[0]{\textit{FE6}\xspace}
\newcommand{\feseven}[0]{\textit{FE7}\xspace}
\newcommand{\feeight}[0]{\textit{FE8}\xspace}
\newcommand{\prqone}[0]{\textit{PRQ1}\xspace}

\newcommand{\prqfour}[0]{\textit{PRQ4}\xspace}
\newcommand{\poqone}[0]{\textit{POQ1}\xspace}

\newcommand{\poqfour}[0]{\textit{POQ4}\xspace}
\usepackage{subcaption}
\newcommand{\change}[1]{\textcolor{black}{#1}}
\begin{document}

\title{A User Study for Evaluation of Formal Verification Results and their Explanation at Bosch}

\titlerunning{A User Study for Evaluation of Formal Verification Results}        

\author{Arut Prakash Kaleeswaran     \and
        Arne Nordmann \and
        Thomas Vogel \and
        Lars Grunske
}

\authorrunning{Kaleeswaran et al.} 

\institute{Arut Prakash Kaleeswaran \at
              Cross-Domain Computing Solutions, Robert Bosch GmbH, Stuttgart, Germany, at the time of the study: Corporate Sector Research, Robert Bosch GmbH, Renningen, Germany \\
              \email{arutprakash.kaleeswaran@de.bosch.com}           
           \and
           Arne Nordmann \at
           Corporate Sector Research, Robert Bosch GmbH, Renningen, Germany\\
           	\email{arne.nordmann@neura-robotics.com}        
           	\and
           Thomas Vogel \& Lars Grunske \at
		Software Engineering Group, Humboldt-Universit \"{a}t zu Berlin, Berlin, Germany\\
			\email{\{thomas.vogel, grunske\}@informatik.hu-berlin.de}
}

\date{Received: date / Accepted: date}

\maketitle
\begin{abstract}
\textit{Context:} Ensuring safety for any sophisticated system is getting more complex due to the rising number of features and functionalities.
This calls for formal methods to entrust confidence in such systems. Nevertheless, using formal methods in industry is demanding because of their lack of usability and the difficulty of understanding verification results.
\textit{Objective:} We evaluate the acceptance of formal methods by Bosch automotive engineers, particularly whether the difficulty of understanding verification results can be reduced.
\textit{Method:} We perform two different exploratory studies. First, we conduct a user survey to explore challenges in identifying inconsistent specifications and using formal methods by Bosch automotive engineers. Second, we perform a one-group pretest-posttest experiment to collect impressions from Bosch engineers familiar with formal methods to evaluate whether understanding verification results is simplified by our counterexample explanation approach.
\textit{Results:} \change{The results from the user survey indicate that identifying refinement inconsistencies, understanding formal notations, and interpreting verification results are challenging. Nevertheless, engineers are still interested in using formal methods in real-world development processes because it could reduce the manual effort for verification. Additionally, they also believe formal methods could make the system safer. Furthermore, the one-group pretest-posttest experiment results indicate that engineers are more comfortable understanding the counterexample explanation than the raw model checker output.}
\textit{Limitations:} The main limitation of this study is the generalizability beyond the target group of Bosch automotive engineers.	

\end{abstract}

\section{Introduction}
\label{sec:introduction}

\begin{center}
\textit{The overarching goal of formal methods is to help engineers construct more reliable systems.} -- \cite{ClarkeW96}
\end{center}

\paragraph{\change{Motivation.}}

\change{Research on formal methods has been continuing for more than three decades~\citep{BowenB21}. Initial applications of formal methods are introduced by \cite{Wing90} and \cite{rushby1993formal}. In recent years, formal methods have been developed to analyze and verify complex safety-critical systems~\citep{FerrariB23,KALEESWARAN2022106800}. Especially automated verification techniques based on formal methods are promising candidates~\citep{BaierK08,GrumbergV08,ClarkeGKPV18,ClarkeHVB18} to ensure the functional safety, for instance, of automotive systems. Even though the use of formal methods is considered to be a promising solution, industries are still hesitant to use it for real-world projects due to its complexity \citep{Heitmeyer98,Abrial06, BicarreguiFLW09,KossakMGI14,JonesT22,FerrariB23}. To adopt formal methods in industry, usability and learnability are key factors \citep{BeekBFFGLM19,ReidCFHJL20}.}  There exist several approaches that ease the use of formal methods. For example, property specification patterns \citep{DwyerAC99,KonradC05,Grunske06,PostH12,AutiliGLPT15} and structured natural language \citep{GiannakopoulouP20} are convenient means to specify requirements to be translated into a temporal logic. Furthermore, the tools by \cite{Ratiu2021}, \cite{GerkingSDH15}, and \cite{BarbonLS19} support performing verification in integration with a model checker. In a recent survey, we provide an overview of the state of the art in research on explaining counterexamples and how engineers are supported in interpreting counterexamples \citep{KALEESWARAN2022106800}. 

With this work, we want to identify challenges and opinions on using formal methods in a concrete industrial setting, which is the identification of inconsistent specifications of safety-critical \change{automotive} systems at Bosch. 
\change{With the rising number of features and functionalities, ensuring safety of such systems is a complex task that can be supported by formal methods such as model checking. However, adopting formal methods in industry is demanding because of lack of usability and the difficulty of understanding verification results obtained by model checkers. Therefore,} we want to investigate whether a concrete approach to counterexample explanation eases the understanding of verification results, thus helping Bosch engineers in their daily work to specify and verify their specifications of safety-critical automotive systems. 

\paragraph{Contributions.}

In this work, we present the results obtained from two different user studies with Bosch automotive engineers: (\textit{Part\,1})~\textit{user survey} and (\textit{Part\,2})~\textit{one-group pretest-posttest experiment}. 41\,participants had taken part in the \textit{user survey} and 13\,participants in the \textit{one-group pretest-posttest experiment}. 
From the \textit{user survey}, we first collect challenges on identifying inconsistent specifications and concrete inconsistencies in a specification, and on maintaining refinement consistency between components. We further collect feedback and opinions in using formal methods. The results of the \textit{user survey} are analyzed quantitatively and qualitatively.

Formal methods are not completely new to Bosch. They are used to define requirements as pattern-based specifications to support verification during product development~\citep{PostMHP12,PostH12}. Additionally, we have presented a \emph{counterexample explanation approach} that attempts to ease the use of formal methods by reducing the manual work and difficulty of interpreting the verification results generated by model checkers \citep{KaleeswaranNVG20}. Thus, the motive of the \textit{one-group pretest-posttest experiment} is to evaluate our \emph{counterexample explanation approach} in an industrial setting at Bosch. Particularly, we investigate whether explaining a counterexample generated by a model checker in an understandable format would increase the use of formal methods among Bosch engineers. 

\change{We pre-registered our studies (Part 1 and 2) by a report submitted to and accepted by the Registered Reports Track at the International Conference on Software Maintenance and Evolution (ICSME)~2021\footnote{\url{https://icsme2021.github.io/cfp/RegisteredReportsTrack.html}} \citep{KaleeswaranUserStudy21}. Both studies have been conducted following the same research protocol as described in the registered report~\citep{KaleeswaranUserStudy21} without any modification to the design of the studies.}

\paragraph{Outline.}

We introduce the background and terminology in \sect{background}, and discuss related  user studies concerning formal methods in \sect{related_work}. 
We provide an overview of the counterexample explanation approach
in \sect{cex_exp}. 
In \sect{design}, we outline the research questions, the design, execution plan, target participants, and analysis plan of the studies. 
We present the results of the \textit{user survey} and \textit{one-group pretest-posttest experiment} in Sections~\ref{sec:phase1} and~\ref{sec:phase2}, and discuss and interpret them in \sect{discussion}. 
Threats to validity are discussed in \sect{ttv}. Finally, we conclude and outline the future work in \sect{conclusion}.

\change{\paragraph{Summary of results.} 
From the results of the user survey, we found out that understanding formal notations~(\sect{phase1:notation}), identifying inconsistent specifications and understanding inconsistencies (\sect{phas1:inconsistent_spec}), as well as maintaining the consistency of refined specifications and verifying the refinement consistency (\sect{phase1:refinement}) to be difficult for engineers. Further, the majority of participants answered positively that formal verification could support safety analysis and make a system safer (\sect{phase1:fm_safety}), formal methods are potential candidates to use in the real-world development process, and usage of formal methods could be increased by improving understanding of formal notations  (\sect{phase1:opinion:using}).}

\change{From the results of the one-group pretest-posttest experiment, we found that participants obtain a good understanding of the inconsistencies with our proposed counterexample explanation (Sections~\ref{sec:phase2:postest:result} and \ref{sec:phase2:posttest:tq}) Further, analyzing the feedback from the participants, the majority of them find that the counterexample explanation provides a better and quicker understanding than the counterexample generated by the model checker (Sections~\ref{sec:phase2:posttest:opinion}--\ref{sec:phase2:posttest:feedback}).} 
\section{Background and Terminology}
\label{sec:background}

In this section, we introduce the background and terminology of formal methods, model checking, contract-based design, which are relevant for our studies.

\paragraph{Formal methods.} By formal methods we refer to models, \eg SysML \citep{friedenthal2014practical}, and Kripke structures used as input to verification tools \citep{mcmillan1999smv}, formal specifications of requirements, \eg expressed in natural language-like statements using property specification patterns \citep{DwyerAC99}, or directly in a temporal logic such as Linear Temporal Logic (LTL) \citep{Pnueli77}, and to automated tools to perform the verification. Examples of such tools are model checkers, \eg NuSMV \citep{CimattiCGR00}, theorem provers, \eg Isabelle \citep{Paulson94}, and solvers, \eg Z3 \citep{MouraB08}. 

\paragraph{Model checking.} Considering the verification as a model checking problem, a specification is expressed in a temporal logic ($\phi$) and a system is modeled as a Kripke structure (\textit{K}). Both are the input for a model checker. The model checker verifies whether the given system model (\textit{K}) satisfies the given \textit{property or specification}~($\phi$), that is, $K\vDash$~$\phi$ \citep{BaierK08,ClarkeGKPV18}.
If $\phi$ is not satisfied by \textit{K}, the model checker generates a \textit{counterexample} describing an execution path in \textit{K} that leads from the initial system state to a state that violates~$\phi$, where each state consists of system variables with their values. Based on the counterexample, a user can manually localize the fault in $K$ that causes the violation of $\phi$.

\paragraph{Contract-based design.} Contract-based design (CBD)~\citep{CimattiT12, KaiserWOBNZ15} supports the automated  verification of refinement consistency and correctness. In CBD, model checking is used to identify whether the top-level requirements of a system are consistently refined along the refinement of the system to components.
Each component of a system is associated with a contract that precisely specifies the expected behavior of the component by assumptions, and the provided behavior by guarantees. 
If a component is refined to sub-components, its contract is also refined and assigned to its sub-components.
Thus, all of the sub-components should satisfy the expected behavior of the parent component. This corresponds to the correctness of the refined contracts and can be verified by model checking, which is known as the \textit{refinement check}~\citep{CimattiT12}.
\section{Related Work}
\label{sec:related_work}

In the following, we discuss existing user surveys focusing on formal methods. 
Interviews with users of formal methods are conducted by \cite{SnookH01} to collect the impact on using formal methods on the company, products, and the development process. Furthermore, the interviews focus on various software engineering aspects such as scalability, understandability, and tool support. \cite{RodriguesEMSGR18} perform a survey with 20\,participants who use formal specifications to solve limitations of informal specifications. \change{\cite{KhazeevACMBB19} conduct a survey with the students of the Software Engineering program to examine the AutoProof tool and highlight the challenges associated with formal approaches.}

There exists several surveys focusing mainly on particular application domains.
\cite{DavisCCFHHHMW13} conduct a survey with 31\,participants in the aerospace domain. The survey collects barriers in using formal methods and also propose mitigation measures for the identified barriers.
The studies by \cite{FerrariBMBFGPT19} and \cite{BeekBFFGLM19} focus on the railway domain. \cite{FerrariBMBFGPT19} summarizes results that are suitable for system modelling and verification from surveying 114\,primary studies, 44\,participants from academia and industry, and eight\,projects. Similarly, \cite{BeekBFFGLM19} collect the opinion of users on adopting formal methods in the same domain.

According to the recent user study by \cite{GleirscherM20} performed with 216\,par\-ticipants, the participants are inclined to use formal methods when sufficient training and tool support is available. 
They also present a systematic map of 35 existing studies that summarizes the opinions of the studies' authors on formal methods, as well as the motivation, research method, and results of the studies. 
Similarly, the study performed by \cite{GaravelBP20} with 130\,participants focus mainly on the use of formal methods in research, industry, and education in the past and present. Furthermore, the study summarizes the future direction of formal methods in industry, future target audience using formal methods, domains in which formal methods may have an impact, and competitive or alternative methods to formal methods. 

In contrast to these studies, 
(1)~we particularly focus on identifying challenges that engineers face in identifying inconsistent specifications rather than general challenges of using formal methods, and (2)~we conduct the study with engineers who work on real-world projects in the automotive domain.

\section{Counterexample Explanation in a Nutshell}
\label{sec:cex_exp}

Contract-based design (CBD)~\citep{CimattiT12} is a scalable solution to overcome a manual analysis by automatically and compositionally verifying the consistency of system and requirement refinements using model checking. Thus, CBD provides assurances for consistent refinements early in the development process, which promises to ease corresponding testing activities at later stages.
However, using such a formal method in industry is challenging due to usability issues, \eg the difficulty of understanding model checking results. Thus, we have proposed a counterexample explanation approach that eases the error comprehension of engineers---especially of non-experts in formal methods---if the refinement check fails. The approach generates a user-friendly explanation 
that localizes the fault at the levels of requirements and components. Examples of a CBD and explanations of counterexamples will be given in the context of the study in \sect{phase2}.

\begin{sidewaysfigure}
	\centering
	\vspace{11.5cm}
	\includegraphics[width=\textwidth]{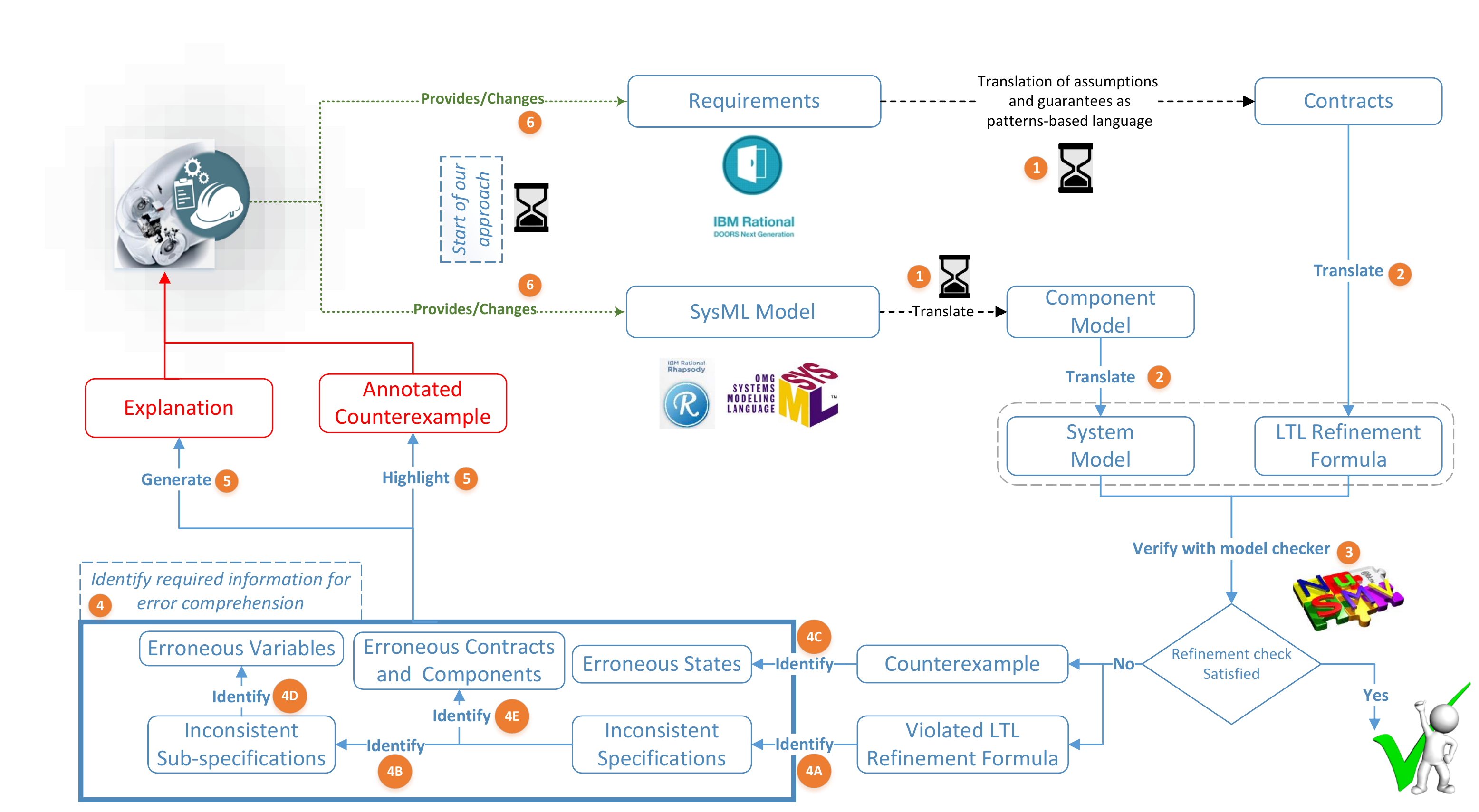}
	\caption{Overview of the counterexample explanation approach.}
	\label{fig:overview_method}
\end{sidewaysfigure}

The counterexample explanation approach comprises of six steps illustrated in \fig{overview_method}. Steps~\textcircled{\small 1} and~\textcircled{\small 6} are performed manually while the other steps are completely automated. Step~\textcircled{\small 1} is the translation of CBD by importing SysML models from Rhapsody and DNG requirements into FASTEN \citep{Ratiu2021}. FASTEN is an open-source platform to experiment with rigorous modeling of safety-critical systems. Translating the (largely informal) requirements from DNG into contracts is a manual effort, further detailed in the context of Bosch by \cite{PostMHP12,PostH12}.
CBD languages provided by FASTEN allow us to model component-based architectures, requirements as contracts (assumptions and guarantees), and refinements, hence creating a CBD. 
After the design, in Step~\textcircled{\small 2}, FASTEN automatically translates a CBD to a formal system model $K$ and refinement specification $\phi$ that allows model checking by NuSMV/nuXMV in Step~\textcircled{\small 3}. This refinement specification follows the scheme by \cite{CimattiT12} who define a refinement check by a set of LTL formulae \citep[Section~2]{KaleeswaranNVG20}. If the model checker identifies any refinement inconsistency during the verification, it returns the violated LTL refinement specification and the counterexample.

Taking the violated LTL refinement specification and counterexample as input, the counterexample explanation approach extracts erroneous parts in Step~\textcircled{\small 4}. To extract such parts, we identify (i)~the \emph{inconsistent specifications} in the violated LTL refinement specification, (ii)~\emph{inconsistent sub-specifications} in the inconsistent specifications, (iii)~\textit{erroneous contracts and components} by using the inconsistent specifications and by referring to the refinement formula, (iv)~\emph{erroneous states} in the counterexample, and (v)~\emph{erroneous variables} by using the inconsistent sub-specifications (cf. {\Large \textcircled{\small 4A}} -- {\Large \textcircled{\small 4E}} in \fig{overview_method}).

Finally in Step~\textcircled{\small 5}, a statement is generated explaining inconsistency along with the counterexample highlighting \emph{erroneous states} and \emph{erroneous variables}. The generated statement consists of the \textit{erroneous components}, violated contract information, and \emph{inconsistent specifications} expressed in a pattern-based language referring to requirements and SysML model being the initial user input of the approach (cf. \textcircled{\small 1} in \fig{overview_method}). Finally, the \emph{inconsistent sub-specifications} are highlighted in the pattern-based expression.
With the statement, the engineer gets a high-level understanding of the refinement inconsistency. Further, the counterexample with highlighted erroneous parts supports the engineer in understanding the erroneous behavior of the system.
Finally, the engineer can correct the refinement inconsistencies by remodeling the component model and changing the requirements (cf. \textcircled{\small 6} in \fig{overview_method}). To ensure the correctness of the changes, the engineer re-verifies the changed refinement.
\section{Research Method}
\label{sec:design}

In this section, we discuss the research questions, design, execution plan, target participants, and analysis plan of the studies. 

\subsection{Research Questions}
\label{sec:rq}

Our study aims to explore and understand the challenges in identifying inconsistent specifications, and the acceptance of formal methods by Bosch automotive engineers. Therefore, this user study has two significant goals: \textbf{(G1)}~to understand challenges faced by Bosch engineers when identifying inconsistent specifications, and challenges along with their opinions to use formal verification or formal methods in real-world development processes, and \textbf{(G2)}~to explore whether Bosch engineers are interested in using formal methods, particularly model checking, in real-world development processes if the difficulty of understanding model checking results is reduced by our counterexample explanation approach. Considering these two goals, we formulate the following three research questions:\\[-0.5em]

\noindent\emph{\rqs{1}: To what extent do engineers face challenges in identifying inconsistent specifications in formal models that are introduced during the refinement of a system?}\\[-0.5em]

\noindent With this RQ we want to investigate whether:
\begin{enumerate}[label={\textbf{(I\arabic*)}},leftmargin=*]
	\item Understanding formal notations is difficult for engineers.
	\item Identifying inconsistent specifications that are introduced during a refinement of a top-level specification is difficult.
\end{enumerate}

\noindent\emph{\rqs{2}: To what extent the identification of inconsistent specifications and usage of formal methods prove beneficial to a real-world development process?}\\[-0.5em]

\noindent With this RQ we want to investigate whether:
\begin{enumerate}[label={\textbf{(I\arabic*)}},leftmargin=*]
	  \setcounter{enumi}{3}
	\item Usage of formal verification or formal methods is beneficial in a real-world development process.
	\item Identifying inconsistent specifications is beneficial in a real-world development process.
\end{enumerate}

\noindent\emph{\rqs{3}: To what extent do engineers prefer to use formal methods (model checkers particularly) if the difficulty is reduced for understanding verification results to identify inconsistent specifications?}\\[-0.5em]

\noindent With this RQ we want to investigate whether:
\begin{enumerate}[label={\textbf{(I\arabic*)}},leftmargin=*]
		  \setcounter{enumi}{4}
	\item The counterexample explanation approach eases the comprehension when  compared to interpretation of the raw model checker output for engineers with a formal methods background.
	%
	%
	\item It is possible for engineers with a background in formal methods to identify and fix inconsistent specifications based on the counterexample explanation approach.
	\item The counterexample explanation approach can promote formal verification and usage of model checking in real-world development processes.	
\end{enumerate}

\subsection{Variables}
\label{sec:variables}

To attain the goals \textbf{G1} and \textbf{G2}, we perform two different types of exploratory user studies as shown in \fig{steps}. The first study is the \emph{user survey (Part\,1)}, and the second study is a \emph{one-group pretest-posttest experiment (Part\,2)}. 

\subsubsection{Variables of Part\,1: User Survey}
\label{sec:var_part1}

Our user survey evaluates the research questions \rqs{1} and \rqs{2}. The independent variables of \emph{Part\,1} are \emph{participants' professional background and experience}. The dependent variables are different for each research questions. For \rqs{1}, the dependent variable is the \emph{difficulty in understanding} that infers understanding formal notations and identifying inconsistent specifications by engineers are difficult. Similarly, the dependent variable for \rqs{2} is the \emph{increase in confidence in system safety}, that is, the identification of inconsistent specifications and use of formal methods in real-world development processes can make systems safer.

\subsubsection{Variables of Part\,2: One-Group Pretest-Posttest Design}
\label{sec:var_part2}

According to \cite{Babbie16}, an experimental stimulus (also called an intervention) is the independent variable. In the one-group pretest-posttest design, we use our counterexample explanation approach as an intervention. Therefore, it serves as the independent variable of \emph{Part\,2}. Further, we evaluate \rqs{3} based on the following four attributes that serve as dependent variables for \emph{Part\,2}:

\begin{enumerate}[leftmargin=*]
	\item \emph{Better understanding:} Does the counterexample explanation approach allow engineers to understand model checking results and identify inconsistencies more effectively?
	\item \emph{Quicker understanding:} Does the counterexample explanation approach allow engineers to understand model checking results and identify inconsistencies more efficiently?
	\item \emph{Confidence:} Does the counterexample explanation approach make engineers more confident in their understanding of the system and its inconsistency respective to safety?
	\item \emph{No value:} This attribute is inversely related to the above attributes. Will the counterexample explanation approach provide no or only minimal value to real-world projects?
\end{enumerate}

\begin{table}[b!]
	\centering
	\caption{Questionnaire of our user survey (Part\,1). A scale is either nominal (N) or ordinal (O). Labels are either not applicable (NA) or they refer to one of the scales (LS) defined in \tab{likert_scale}.}
	\label{tab:survey_questions}
	\resizebox{\textwidth}{!}{
	\begin{tabular}{|llll|}
		\hline
		\multicolumn{1}{|l|}{\textbf{\#}} &
		\multicolumn{1}{l|}{\textbf{Questions}} &
		\multicolumn{1}{l|}{\textbf{Scale}} &
		\textbf{Label} \\ \hline
		\multicolumn{4}{|l|}{\textbf{Demographic Questions}} \\ \hline
		\multicolumn{1}{|l|}{Q1} &
		\multicolumn{1}{l|}{Designation/Role/Assignment} &
		\multicolumn{1}{l|}{N} &
		NA \\ \hline
		\multicolumn{1}{|l|}{Q2} &
		\multicolumn{1}{l|}{Rate your knowledge of formal methods} &
		\multicolumn{1}{l|}{O} &
		LS1 \\ \hline
		\multicolumn{1}{|l|}{Q3} &
		\multicolumn{1}{l|}{\begin{tabular}[c]{@{}l@{}}How many years have you used formal methods in your daily work?\end{tabular}} &
		\multicolumn{1}{l|}{O} &
		LS2 \\ \hline
		\multicolumn{1}{|l|}{Q4} &
		\multicolumn{1}{l|}{\begin{tabular}[c]{@{}l@{}}Indicate the applications/products/projects you worked \\on using formal methods\end{tabular}} &
		\multicolumn{1}{l|}{N} &
		NA \\ \hline
		\multicolumn{1}{|l|}{Q5} &
		\multicolumn{1}{l|}{\begin{tabular}[c]{@{}l@{}}How many years have you worked in the safety domain?\end{tabular}} &
		\multicolumn{1}{l|}{N} &
		LS2 \\ \hline
		\multicolumn{1}{|l|}{Q6} &
		\multicolumn{1}{l|}{\begin{tabular}[c]{@{}l@{}}Indicate the applications/products/projects you worked on focusing \\on safety aspects\end{tabular}} &
		\multicolumn{1}{l|}{O} &
		NA \\ \hline
		\multicolumn{4}{|l|}{\textbf{Main Survey Questions}} \\ \hline
			\multicolumn{1}{|l|}{Q7} &
		\multicolumn{1}{l|}{\begin{tabular}[c]{@{}l@{}}How easy is it for you to understand formal notations?\end{tabular}} &
		\multicolumn{1}{l|}{N \& O} &
		LS3 \\ \hline
		\multicolumn{1}{|l|}{Q8} &
		\multicolumn{1}{l|}{\begin{tabular}[c]{@{}l@{}}What is your opinion on using formal verification?\end{tabular}} &
		\multicolumn{1}{l|}{N} &
		NA \\ \hline
		\multicolumn{1}{|l|}{Q9} &
		\multicolumn{1}{l|}{\begin{tabular}[c]{@{}l@{}}How easy is it for you to identify  inconsistent formal specifications \\(where the consistency is)?\end{tabular}} &
		\multicolumn{1}{l|}{N \& O} &
		LS3 \\ \hline
		\multicolumn{1}{|l|}{Q10} &
		\multicolumn{1}{l|}{\begin{tabular}[c]{@{}l@{}}How easy is it for you to identify the inconsistencies in formal\\ specifications (the cause for the inconsistency)?\end{tabular}} &
		\multicolumn{1}{l|}{N \& O} &
		LS3 \\ \hline
		\multicolumn{1}{|l|}{Q11} &
		\multicolumn{1}{l|}{\begin{tabular}[c]{@{}l@{}}How fast could you identify inconsistent formal specifications?\end{tabular}} &
		\multicolumn{1}{l|}{N \& O} &
		LS4 \\ \hline
		\multicolumn{1}{|l|}{Q12} &
		\multicolumn{1}{l|}{\begin{tabular}[c]{@{}l@{}}What are the challenges that you face in order to identify\\ inconsistent formal specifications?\end{tabular}} &
		\multicolumn{1}{l|}{N} &
		NA \\ \hline
		\multicolumn{1}{|l|}{Q13} &
		\multicolumn{1}{l|}{\begin{tabular}[c]{@{}l@{}}What sort of methods have you used to identify inconsistent\\ formal specifications?\end{tabular}} &
		\multicolumn{1}{l|}{N} &
		NA \\ \hline
		\multicolumn{1}{|l|}{Q14} &
		\multicolumn{1}{l|}{\begin{tabular}[c]{@{}l@{}}How easy is it for you to maintain consistency when refining\\ formal requirements for sub-components of a system architecture?\end{tabular}} &
		\multicolumn{1}{l|}{N \& O} &
		LS3 \\ \hline
		\multicolumn{1}{|l|}{Q15} &
		\multicolumn{1}{l|}{\begin{tabular}[c]{@{}l@{}}How hard is it for you to check the consistency  of formal requirements\\ that are associated with  components of a system architecture?\end{tabular}} &
		\multicolumn{1}{l|}{N \& O} &
		LS3 \\ \hline
		\multicolumn{1}{|l|}{Q16} &
		\multicolumn{1}{l|}{\begin{tabular}[c]{@{}l@{}}In your opinion, will the usage of formal verification make systems safer?\end{tabular}} &
		\multicolumn{1}{l|}{N \& O} &
		LS5 \\ \hline
		\multicolumn{1}{|l|}{Q17} &
		\multicolumn{1}{l|}{\begin{tabular}[c]{@{}l@{}}In your opinion, can formal verification be an add-on to the functional\\ safety methods to ensure safety?\end{tabular}} &
		\multicolumn{1}{l|}{N \& O} &
		LS5 \\ \hline
		\multicolumn{1}{|l|}{Q18} &
		\multicolumn{1}{l|}{\begin{tabular}[c]{@{}l@{}}In your opinion, how beneficial is the identification of inconsistent formal  \\specifications for a safety analysis?\end{tabular}} &
		\multicolumn{1}{l|}{N \& O} &
		LS5 \\ \hline
		\multicolumn{1}{|l|}{Q19} &
		\multicolumn{1}{l|}{\begin{tabular}[c]{@{}l@{}}Can you imagine using formal methods if understanding of formal \\notations is made easier?\end{tabular}} &
		\multicolumn{1}{l|}{N \& O} &
		LS5 \\ \hline
		\multicolumn{1}{|l|}{Q20} &
		\multicolumn{1}{l|}{\begin{tabular}[c]{@{}l@{}}Do you think formal methods are usable in real-world \\development processes?\end{tabular}} &
		\multicolumn{1}{l|}{N \& O} &
		LS5 \\ \hline
	\end{tabular}}
\end{table}

\change{For each of the four attributes, the participants are asked one question in the pretest and posttest in order to collect their opinions about the attributes. Table~\ref{tab:onegroup} lists the questions asked to the participants and the scale of possible answers in both tests. Particularly, the questions PRQ1 in the pretest and POQ1 in posttest target the attribute \textit{Better understanding}, PRQ2 and POQ2 target \textit{Quicker understanding}, PRQ3 and POQ3 target \textit{Confidence}, and PRQ4 and POQ4 target \textit{No value}. 
	Finally, comparing the results gathered from both the pretest and posttest, we can investigate the participants' opinions on understanding inconsistencies using the proposed counterexample explanation and the raw counterexample generated by the model checker.}

\subsection{Design of the User Study}
\label{sec:design_study}

In this section, we describe the design, questionnaires, and tools used for the \emph{user survey (Part\,1)} and the \emph{one-group pretest-posttest experiment (Part\,2)}.

\subsubsection{Part\,1: User Survey}
\label{sec:part1}

For \emph{Part\,1}, we use a cross sectional survey~\citep{KitchenhamP08} to collect data from engineers to achieve goal \goal{1}. For planning and conducting this user survey, we follow the guidelines by \cite{Neuman14} (majorly Chapter~7), \cite{KitchenhamP08}, and \cite{Fink03}. In addition, we follow the guidelines by \cite{RobsonM16} (Chapter~11), and \cite{Babbie16} (Chapter~9) for the questionnaire construction. Furthermore, we also refer to and adapt some of the questionnaires from existing user surveys \citep{GleirscherM20,GaravelBP20}. The questionnaire of our user survey is shown in \tab{survey_questions}. Responses to each question are either collected as qualitative statements, and/or they follow predefined eight-point Likert scales shown in \tab{likert_scale}.

\begin{table}
	\centering
	\caption{Scales used for our user study.}
	\label{tab:likert_scale}
	\resizebox{\textwidth}{!}{
		\begin{tabular}{|l|l|l|}
			\hline
			\textbf{Label} & \textbf{Type} & \textbf{Answer Scales}                                                                                                                                                                 \\ \hline
			LS1                                   & Expertise                          & \begin{tabular}[c]{@{}l@{}}Novice,  Advanced Beginner, Competent, Proficient, Expert, \\Mastery, Practical Wisdom, No Opinion\end{tabular}                                                                                                   \\ \hline
			LS2                                   &  Experience                 & \begin{tabular}[c]{@{}l@{}}$<$\,1, 1 to $<$\,2, 2 to $<$\,4, 4 to $<$\,6, 6 to $<$\,8, 8 to $<$\,10, $<$\,10, No Experience\end{tabular}                                                                                                                      \\ \hline
			LS3                                   & Agreement                          & \begin{tabular}[c]{@{}l@{}}Extremely Hard, Hard, Slightly   Hard, Neither Hard nor\\ Easy, Slightly Easy, Easy, Extremely Easy, No Opinion \end{tabular}                                                                                                               \\ \hline
			LS4                                   & Agreement                          & \begin{tabular}[c]{@{}l@{}}Extremely Fast, Fast, Slightly   Fast, Neither Fast nor Slow,\\ Slightly Slow, Slow, Extremely Slow, No Opinion     \end{tabular}                                                                                                           \\ \hline	
			LS5                                   & Likelihood                         & \begin{tabular}[c]{@{}l@{}}Definitely,  Very Probably,  Probably, Neither Probably nor \\Possibly,   Possibly,  Probably Not,  Definitely Not, No Opinion                 \end{tabular}                                                                                      \\ \hline
			LS6                                   & Agreement                         & \begin{tabular}[c]{@{}l@{}}Strongly Agree,  Agree,  Somewhat Agree,   Neither Agree nor-\\Disagree, Somewhat Disagree,  Disagree,Strongly Disagree, No Opinion       \end{tabular}                                                                                                \\ \hline
			LS7                                   & Usefulness                         & \begin{tabular}[c]{@{}l@{}}Exceptional, Excellent, Very Good, Good, Fair, Poor,Very Poor, No Opinion   \end{tabular}                                                                                                    \\ \hline
	\end{tabular}}
\end{table}

\begin{table}[!tbph]
	\centering
		\caption{Questionnaire of the one-group pretest-posttest study (Part\,2). A scale is either nominal (N) or ordinal (O). Labels are either not applicable (NA) or they refer to one of the scales (LS) defined in \tab{likert_scale}.}
	\label{tab:onegroup}
	\resizebox{\textwidth}{!}{
	\begin{tabular}{|l|l|l|l|}
		\hline
		\textbf{\#} & \textbf{Questions}                                                                                                            & \textbf{Scale}        & \textbf{Label} \\ \hline
		\multicolumn{4}{|l|}{\textbf{Demographic Questions}} \\ \hline
		\multicolumn{1}{|l|}{DQ1} &
		\multicolumn{1}{l|}{Rate your knowledge of formal methods} &
		\multicolumn{1}{l|}{O} &
		LS1 \\ \hline
		\multicolumn{1}{|l|}{DQ2} &
		\multicolumn{1}{l|}{\begin{tabular}[c]{@{}l@{}}How many years have you used formal methods in your daily work?\end{tabular}}
		&
		\multicolumn{1}{l|}{O} &
		LS2 \\ \hline
		\multicolumn{4}{|l|}{\textbf{Task Questions}}                                                                                                                                                                                                                                           \\ \hline
		TQ1                                  & \begin{tabular}[c]{@{}l@{}}Do  you think this use case is difficult?                    \end{tabular}                                                                                        & N \& O                 & LS5                                                                        \\ \hline
		TQ2                                  & \begin{tabular}[c]{@{}l@{}}How difficult was this use case for you to understand?         \end{tabular}                                                                                                    & N \& O                 & LS5                                                                        \\ \hline
		TQ3                                  & \begin{tabular}[c]{@{}l@{}}Do  you think you have understood  results   from the model checker?\\ (This question is only for pretest)   \end{tabular}                                       & N \& O                 & LS5                                                                        \\ \hline
		TQ4                                  & \begin{tabular}[c]{@{}l@{}}Do   you think you have understood the    explanations? \\(This question is only for the posttest)         \end{tabular}                                              & N \& O                 & LS5                                                                        \\ \hline
		TQ5                                  & \begin{tabular}[c]{@{}l@{}}Of the following list, please select the inconsistent components.   \end{tabular}    & N                             & NA                                                             \\ \hline
		TQ6                                  & \begin{tabular}[c]{@{}l@{}}Of the following list, please select the inconsistent specifications.\end{tabular}                                                                                                               & N                             & NA                                                             \\ \hline
		TQ7                                  & \begin{tabular}[c]{@{}l@{}}Please explain the reason that makes the specifications inconsistent \\from your understanding.     \end{tabular}                                                                                                                & N                             & NA                                                             \\ \hline		TQ8                                  & \begin{tabular}[c]{@{}l@{}}Please provide a solution to fix the inconsistency from your \\understanding.   \end{tabular}                                                                                                    & N                             & NA                                                             \\ \hline
		TQ9                                  & \begin{tabular}[c]{@{}l@{}}Please provide a nominal behavior that is expected in the \\counterexample's erroneous states from your understanding.        \end{tabular}                                               & N                            & NA                                                             \\ \hline
		\multicolumn{4}{|l|}{\textbf{Understanding Model   Checker Outputs (Pretest)}}                                                                                                                                                                                                                                           \\ \hline
		PRQ1                                 & \begin{tabular}[c]{@{}l@{}}The results from the model checker allow me to understand the\\ inconsistencies.  \end{tabular}                                                                   & N \& O                             & LS6                                                                        \\ \hline
		PRQ2                                 & \begin{tabular}[c]{@{}l@{}}Such a result from   the model checker could save me time.                       \end{tabular}           & N \& O                             & LS6                                                                        \\ \hline
		PRQ3  & \begin{tabular}[c]{@{}l@{}}The results from the model checker make me confident that I\\ really understand the inconsistencies that I am investigating.    \end{tabular}                               &                                                                     N \& O                             & LS6                                                                        \\ \hline
		PRQ4                                 &  \begin{tabular}[c]{@{}l@{}}The value added by   such a result from the model checker will be \\minimal.                  \end{tabular}                                                                              & N \& O                             & LS6                                                                        \\ \hline 
		\multicolumn{4}{|l|}{\textbf{Understanding Counterexample Explanation (Posttest)}}                                                                                                                                                                                                                \\ \hline
		
		POQ1                                 & \begin{tabular}[c]{@{}l@{}}An approach like   counterexample explanation allows me to better\\ understand inconsistencies.                   \end{tabular}                                                      & N \& O                             & LS6                                                                        \\ \hline
		POQ2                                 & \begin{tabular}[c]{@{}l@{}}An approach like   counterexample explanation saves me time.                                                  \end{tabular}                                                        & N \& O                             & LS6                                                                        \\ \hline
		POQ3                                 &   \begin{tabular}[c]{@{}l@{}}An approach like   counterexample explanation makes me more \\confident that I really understand  the inconsistencies that I am \\investigating.   \end{tabular}                                                 & N \& O                             & LS6                                                                        \\ \hline
		POQ4                                 &       \begin{tabular}[c]{@{}l@{}}The value added by an   approach like counterexample explanation \\is minimal.                                       \end{tabular}                 & N \& O                             & LS6                                                                        \\ \hline
		\multicolumn{4}{|l|}{\textbf{Counterexample   Explanation Features (Ratings)}}                                                                                                                                                                                                                                          \\ \hline
		FQ1                                  & \begin{tabular}[c]{@{}l@{}}Translation of   specifications from formal temporal format to \\natural language-like format.                \end{tabular}                                                        & N \& O                             & LS7                                                                        \\ \hline
		FQ2                                  & Listing inconsistent   specification.                                                                                                                          & N \& O                             & LS7                                                                        \\ \hline
		FQ3                                  & \begin{tabular}[c]{@{}l@{}}Highlighting   sub-parts of the inconsistent specifications that leads\\ to an inconsistency.     \end{tabular}                                                                       & N \& O                             & LS7                                                                        \\ \hline
		FQ4                                  & \begin{tabular}[c]{@{}l@{}}Providing the component   name that belongs to the\\ inconsistent specifications.       \end{tabular}                                                                             & N \& O                             & LS7                                                                        \\ \hline
		FQ5                                  & \begin{tabular}[c]{@{}l@{}}Providing an expected  nominal behavior in the explanation for \\the corresponding erroneous states and variables of the \\counterexample.       \end{tabular}                                 & N \& O                             & LS7                                                                        \\ \hline
		FQ6                                  & \begin{tabular}[c]{@{}l@{}}Highlighting the erroneous states and variables in the \\counterexample.               \end{tabular}                                                                                 & N \& O                             & LS7                                                                        \\ \hline
		\multicolumn{4}{|l|}{\textbf{Feedback (After   completion of the experiment)}}                                                                                                                                                                                                                                              \\ \hline
		FE1                                  & \begin{tabular}[c]{@{}l@{}}Are inconsistencies easier to understand with the results  created \\by the counterexample explanation approach in comparison those \\of the original model checker?\end{tabular}   & N \& O                 & LS5                                                                        \\ \hline
		FE2                                  & \begin{tabular}[c]{@{}l@{}}What challenges did you face while analyzing inconsistencies\\ in a specification with the proposed approach?        \end{tabular}         & Nominal                             & NA                                                             \\ \hline
		FE3                                  & \begin{tabular}[c]{@{}l@{}}Do you think it is easy to maintain consistency with the proposed \\counterexample explanation while refining requirements into\\ requirements for sub-components? \end{tabular}  
		& N \& O                 & LS3                                                                        \\ \hline
		FE4                                  & \begin{tabular}[c]{@{}l@{}}Do you think the  proposed counterexample explanation \\approach is usable in real-world development   processes?                         \end{tabular}                             & N \& O                 & LS5                                                                        \\ \hline
		FE5                                  & \begin{tabular}[c]{@{}l@{}}Would  you consider using formal methods with our approach\\ in real-world projects?       \end{tabular}                     & N \& O                 & LS5                                                                        \\ \hline
		FE6                                  & \begin{tabular}[c]{@{}l@{}}Would   you consider using the presented approach in your\\ project? If so, please name the project and a contact person?     \end{tabular}                                               & N \& O                 & LS5                                                                        \\ \hline
		FE7                                  & \begin{tabular}[c]{@{}l@{}}Do you think presenting a list of possible suggestions/fixes\\ would be helpful to understand and fix inconsistencies?                          \end{tabular}                            & N \& O                 & LS5                                                                       \\ \hline
		FE8                                  & Suggestions   for further improvements.                                                                                                                        & Nominal                             & NA                                                             \\ \hline
	\end{tabular}}
\end{table}

\subsubsection{Part\,2: One-Group Pretest-Posttest Experiment}
\label{sec:part2}

\emph{Part\,2} of our study is an exploratory pre-experimental study following a \emph{one-group pretest-posttest experiment} design to attain goal \goal{2}. We follow the guidelines by \cite{CampbellS63} to conduct this part of our study.

One of the main drawbacks of using a one-group pretest-posttest design is that it does not meet the scientific standards of an experimental design. 
The pre-experimental study designs does not have a control group like a true experiment~\citep{WohlinRHO12}. Thus, comparison and generalization of the results based on the provided intervention/stimulus may not be possible. However, we intend to use this pre-experimental study design because of the scarcity of participants. To find a considerable number of participants (30 to 40) with knowledge of formal methods and model checkers inside an industrial organization is way too ambitious. Performing a true experiment with a lower number of participants raises the threat to external validity.
Therefore, we intend to perform a one-group pretest-posttest experiment with Bosch automotive engineers that allows us to capture results from real-world user behavior, even with a limited number of participants. However, the pre-experimental study has several internal and external threats to be considered. In \sect{ttv}, we discuss corresponding threats raised by \cite[Table~1]{CampbellS63}.

Along with the guidelines by \cite{CampbellS63}, we refer to the protocol by \cite{ZaidmanMSD13} for a one-group pretest-posttest experiment. They evaluate a tool called \emph{FireDetective} that supports understanding of Ajax applications at both the client-side (browser) and server-side. Their evaluation is performed using two user study variants (i)~pretest-posttest user study, and (ii)~a field user study, where the former is performed with eight participants and the latter is performed with two participants. In our study, we perform the one-group pretest-posttest experiment with Bosch automotive engineers and discard the field user study for our evaluation.
The questionnaire shown in \tab{onegroup} is used for the one-group pretest-posttest study (\emph{Part\,2} of our overall study). Similar to \emph{Part\,1}, responses to each question are either qualitative statements, selections of options on predefined eight-point Likert scales (\tab{likert_scale}), or a combination of both.

\subsection{Tools Used for the Study}
\label{sec:survey_tool}

\change{Both studies are performed remotely due to the COVID-19 pandemic. In such a setting, it would not be easy to capture the time that participants spent on the study. For example, there could be a situation where participants could have taken a break or could respond to some urgent emails while during the study. Thus, we asked separately a question for the time taken by the participants to conduct the pretest and posttest.
To perform the studies and collect the answers by the participants, we use \textit{Microsoft Forms for Excel}. that} is easily accessible within the company and already familiar to the participants. 
The results are stored in a Microsoft Excel file, which we use to perform the analysis.
All content-wise explanations for this study are provided as a video that are accessible Bosch internally via an online platform called \emph{BoschTube}.

\subsection{Participants}
\label{sec:participants}

Our counterexample explanation approach focuses on enhancing safety analysis for automotive systems~\citep{KaleeswaranNVG20}. Thus, we are interested in performing this user study only with automotive engineers have at least basic knowledge of formal methods, particularly engineers working on system development, requirement elicitation, and safety analysis. 

The target population for our study is very specific and thus, it is hard to make a finite list of participants by applying probabilistic sampling. According to \cite{KitchenhamP08}, when a target population is very specific and limited, non-probabilistic sampling can be used to identify the participants. Therefore, we intend to use two non-probabilistic sampling methods for \emph{Part\,1} of our study, namely, \emph{convenience sampling} and \emph{snowball sampling}. Further, we invite participants with knowledge on formal methods for \emph{Part\,2} of our study by filtering the participants of \emph{Part\,1} based on the responses to the demographic questions Q1 to Q3 listed in Table~\ref{tab:survey_questions}.

First, we start with the convenience sampling for \emph{Part\,1}. We send e-mails with the survey link to participants collected through department mailing lists and community mailing lists of all relevant Bosch business units. We perform snowball sampling with the accepted participants by asking for further potential participants at the end of the survey. In the e-mail invitation, we explicitly mention that the anonymity of results will be preserved. So, while summarizing analysis results, we remove all personal, product- and project-related information. For both the studies, the reminder mail is sent three times in the interval of one week.

\begin{figure}[!tbh]
	\centering
	\includegraphics[width=0.6\linewidth]{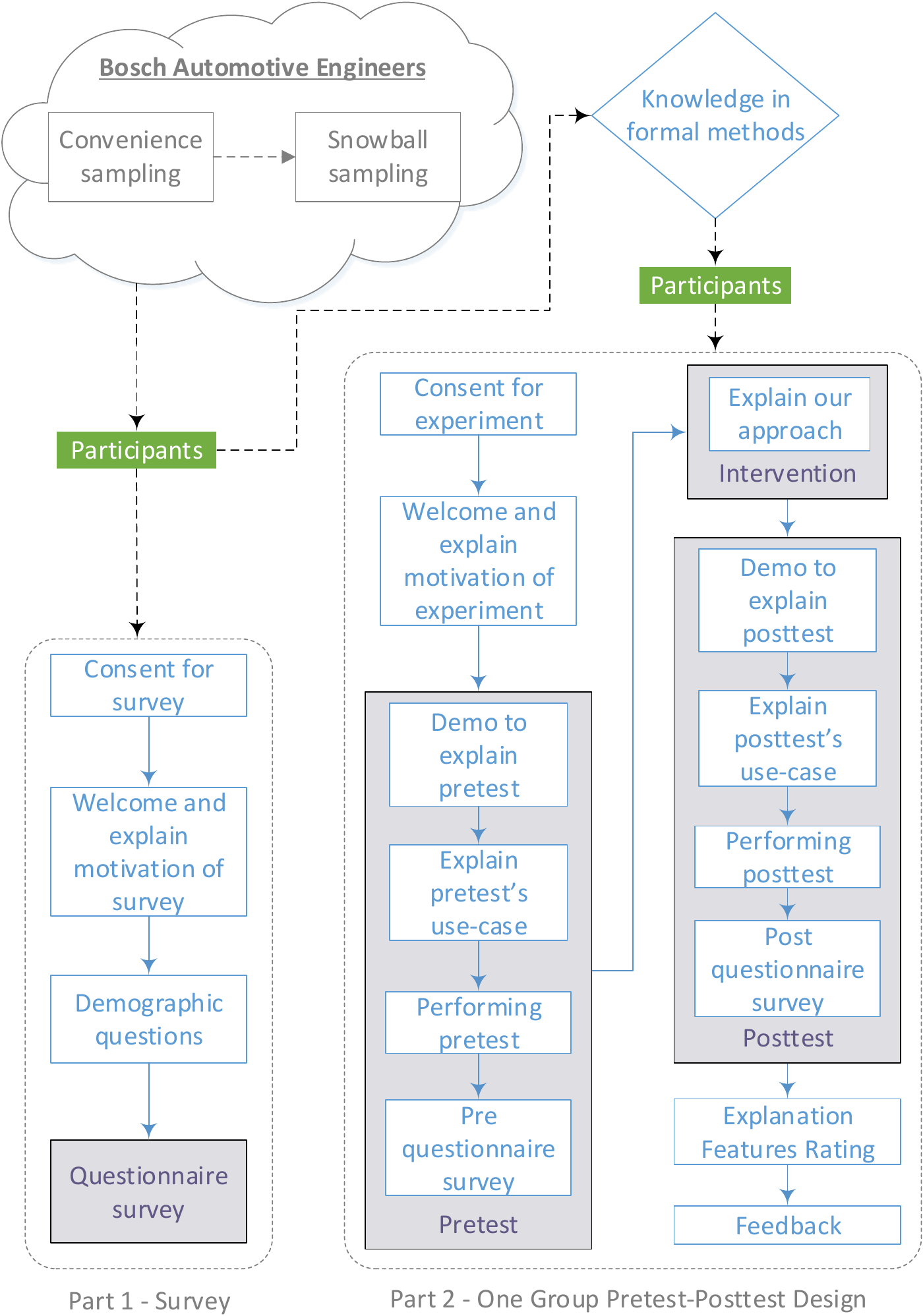}
	\caption{Overview of the study. \emph{Part\,1} is a user survey performed with a wide range of participants. \emph{Part\,2} is a one-group pretest-posttest experiment performed with engineers having knowledge in formal methods. Gray color boxes indicate the main tasks, \ie questionnaire survey, pretest and posttest.}
	\label{fig:steps}
\end{figure}

\subsection{Execution Plan}
\label{sec:execution}

In this section, we describe the execution plan of the \emph{user survey (Part\,1)} and the \emph{one-group pretest-posttest experiment (Part\,2)} depicted in \fig{steps}. 

\subsubsection{Execution Plan of Part\,1}
\label{sec:plan_part1}

The \emph{user survey (Part\,1)} comprises four steps (\fig{steps}).
First, we notify participants regarding the data processing agreement. Additionally, we also state explicitly that their names, project- and product-related information will be removed while results are shared for evaluation. Then we show a video, welcoming the participant and explaining the background and motivation of this survey. Then, we ask participant to answer the demographic questions (Q1 to Q6 in \tab{survey_questions}), and further the main survey questions (Q7 to Q20 in \tab{survey_questions}). Finally we conclude the survey with a thanks note.

\subsubsection{Execution Plan of Part\,2}
\label{sec:plan_part2}

For the one-group pretest-posttest experiment, we invite participants from \emph{Part\,1} who have knowledge in formal methods. Similar to \emph{Part\,1}, \emph{Part\,2} starts with a data processing agreement, followed by a background and motivation video. Our one-group pretest-posttest experiment is executed with the invited participants as follows: a pretest experiment, then intervention, and finally the posttest experiment. 

\paragraph{Pretest.}
The pretest experiment starts with a video demonstrating the experiment with a simple example of an OR-gate and the behavior of the OR-gate. After that, another video introduces an airbag system with the corresponding system model and specification, that serves as a use case for the actual pretest experiment.
During the actual experiment, the participant analyzes the violated specification and the counterexample returned by the model checker to understand the inconsistent parts of the specification. Furthermore, based on the understanding, the participant answers the task questions (TQ1 to TQ9 except of TQ4 in \tab{onegroup}).
Finally, the pretest is concluded by answering the pre-questionnaire survey questions PRQ1 to PRQ4 listed in \tab{onegroup}.

\paragraph{Intervention.}
After the pretest experiment, a video introducing the counterexample explanation approach~\cite{KaleeswaranNVG20} is shown to the participants. This serves as an intervention in our study.   

\paragraph{Posttest.}
Like the steps followed for the pretest experiment, the posttest experiment starts with a demonstration video with the same use case of the OR-gate, but this time with the counterexample explanation approach. This is followed by a video that introduces the electronic power steering system (EPS), a commercial Bosch product, with the corresponding system model and specification. Then the participants interpret the explanation provided by the counterexample explanation approach to understand the inconsistency. Based on the explanation, participants answer the task questions (TQ1 to TQ9 except of TQ3 in \tab{onegroup}). Subsequently, they answer the post-questionnaire survey questions POQ1 to POQ4 listed in \tab{onegroup}.
After completing the posttest experiment, participants rate the features (FQ1 to FQ6 in \tab{onegroup}) provided by the counterexample explanation approach and respond to the feedback questions (FE1 to FE8 in \tab{onegroup}). Finally, \emph{Part\,2} of our study concludes with a thanks note to the participants.

\subsection{Presentation of the Analysis Results}
\label{sec:analysis}

To obtain the results from the study, we follow the recommendation by \cite{Robbins2011}. We use normal, grouped and stacked bar charts to plot the results. 
Qualitative statements received from participants are gathered, organized, and summarized individually for every question. We summarize the qualitative statements through the following three steps: \textit{(i)~Microanalysis:} The first author goes through the individual answers from the participants and assigns labels to the statements. The rest of the authors validate the initial labels and provide feedback for improvement. At the end of this step, all authors come to a mutual agreement on the initial labels. \textit{(ii)~Categorization:} Based on the feedback for improvement, the first author performs second iteration. As a result, a set of themes are extracted which are deemed to be essential. \textit{(iii)~Saturation:} This is the final step where all the authors come to the final agreement on labels, themes, and summarized statements. Since the qualitative statement is a medium to express an individual opinion, the categorization of labels are associated with the demographic answers. For example: \emph{``an engineer who has seven years of experience states that the counterexample explanation approach can promote the usage of model checkers among system engineers''}.

\section{User Survey (Part\,1): Results and Analysis}
\label{sec:phase1}

In this section, we present and analyze the results of the user survey (\textit{Part\,1}). We gathered answers to the questionnaire shown in \tab{survey_questions} from 41\,participants.

\subsection{Participants}
\label{sec:phase1:participants}
 
We first present demographic information of the participants that we obtained from the first six questions of the questionnaire (\qsone to \qssix in \tab{survey_questions}). 

\subsubsection{Experience in Formal Methods and Safety}
\label{sec:phase1:participants:experience}

\begin{figure}[t]
	\centering
	\includegraphics[width=0.8\textwidth]{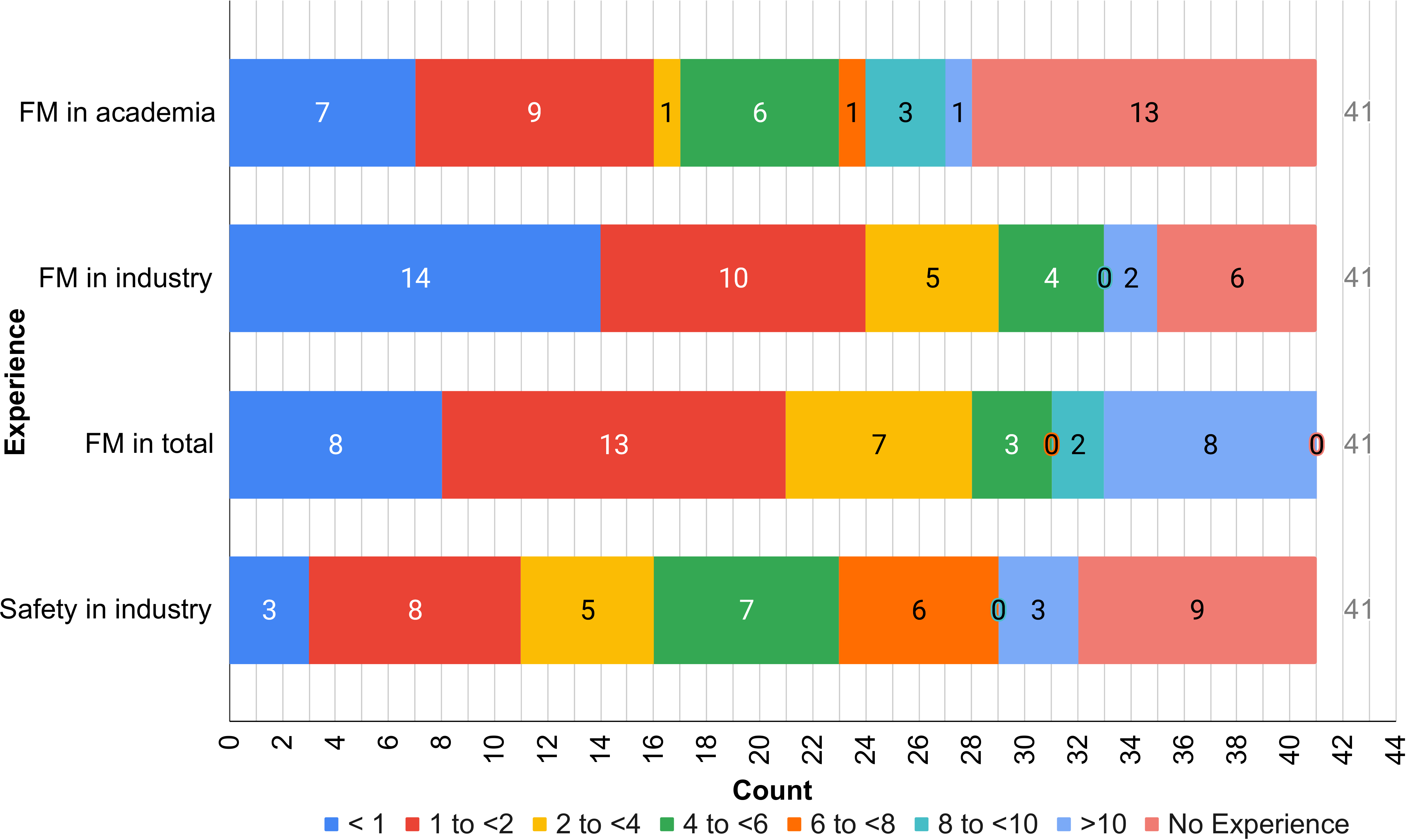}
	\caption{Experience of participants in formal methods gained in academia, industry, and the overall experience, along with industrial experience on safety.}
	\label{fig:experience_survey}
\end{figure}

The experience of the participants in using formal methods and their experience focusing on safety are collected through the questions \qsthree and \qsfour (\tab{survey_questions}), which are answered in scale \lstwo (\tab{likert_scale}). Participants are asked to fill in their experience in formal methods gained individually in academia, and in industry, and their overall experience in academia and industry combined together with \qsthree, while \qsfour collects industrial experience in safety. \fig{experience_survey} represents the responses for experience in formal methods and safety.

From \fig{experience_survey}, it is clear that all the 41\,participants have gained knowledge in formal methods in academia or industry \emph{to some extent}. When excluding 13\,participants who have no academic experience in formal methods, the predominant number of participants (82\% of all participants who have academic experience) have experience of $<$6 years. Likewise, excluding 6\,participants who do not have industrial experience in formal methods, the number of participants who have experience of $<$2 years is dominant (40\% of all participants who have industrial experience). 29\% of all participants with some experience in industry only have 1 to $<$2 years of experience. 

In summary, 52\% of all participants have $<$2 years, 24\% have 2 to $<$6 years, no participants with 6 to $<$8 years, and 24\% have $>$8 years of experience in formal methods. Looking at the participants' experience on safety, the results are scattered, with no clear majority. Excluding the 9\,participants who do not have any experience in safety, 41\% of all participants are experienced 1 to $<$4 years, further 41\% are experience 4 to $<$8 years, 9\% of participants have $<$1 year of experience, and the final 9\% have $>$10 years of experience.

\subsubsection{Knowledge of Formal Methods}
\label{sec:phase1:participants:knowledge}

\begin{figure}[t]
	\centering
	\includegraphics[width=0.8\textwidth]{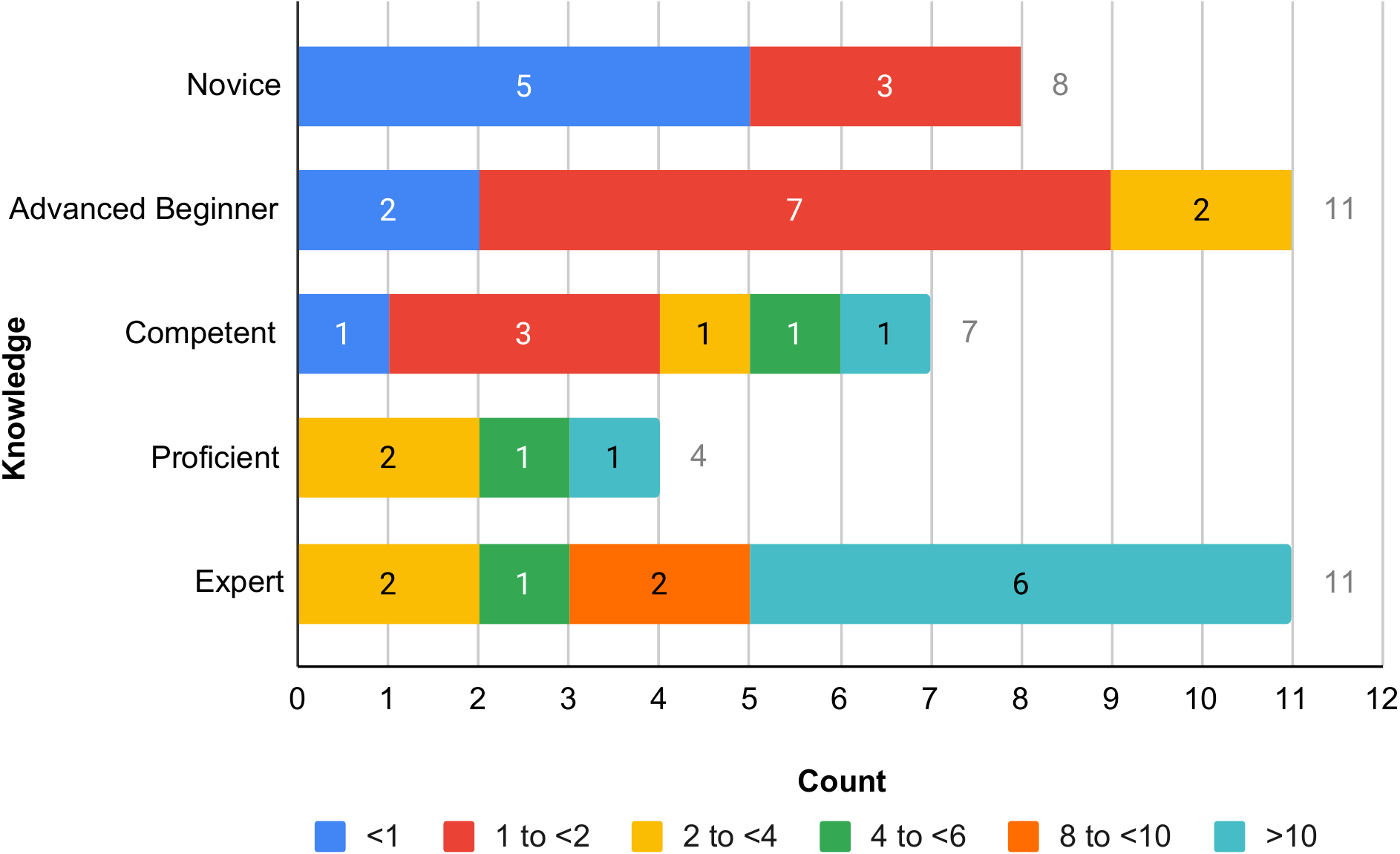}
	\caption{Knowledge of the participants categorized based on their overall experience in formal methods.}
	\label{fig:knowledge_exp_total}
\end{figure}

Question \qsone (\tab{survey_questions}) asks participants to rate their knowledge in formal methods according to the scale \lsone (\tab{likert_scale}). \fig{knowledge_exp_total} shows the results. All participants have rated their knowledge within the scale \textit{novice} to \textit{expert}, while no participant provided a rating of \textit{mastery} and \textit{practical wisdom}. The majority of participants rate their knowledge in formal methods as \textit{advanced beginner} and \textit{expert} with 27\% of all participants each. Further, eight participants (20\%) rate themselves as a \textit{novice}, seven participants (17\%) as \textit{competent}, and four participants rate themselves to be \textit{proficient}.

\fig{knowledge_exp_total} also presents participants' knowledge together with their years of experience in formal methods. The majority of participants with an experience of $<$1 year rated themselves as \textit{novice}, with 1 to $<$2 years of experience as \textit{advanced beginner}, and those with 8 to $<$10 years and $>$10 years of experience rated themselves as an \textit{expert}. However, the participants who have experience of 2 to $<$4 years and 4 to $<$6 years have distributed their rating among various classes. Participants with experience of 2 to $<$4 years rated themselves from \textit{advanced beginner} to \textit{expert}, participants with 4 to $<$6 years of experience rated themselves from \textit{competent} to \textit{expert}.

\subsubsection{Designation}
\label{sec:phase1:participants:designation}

\begin{figure}[t]
\begin{subfigure}{0.48\textwidth}
	\includegraphics[width=\textwidth]{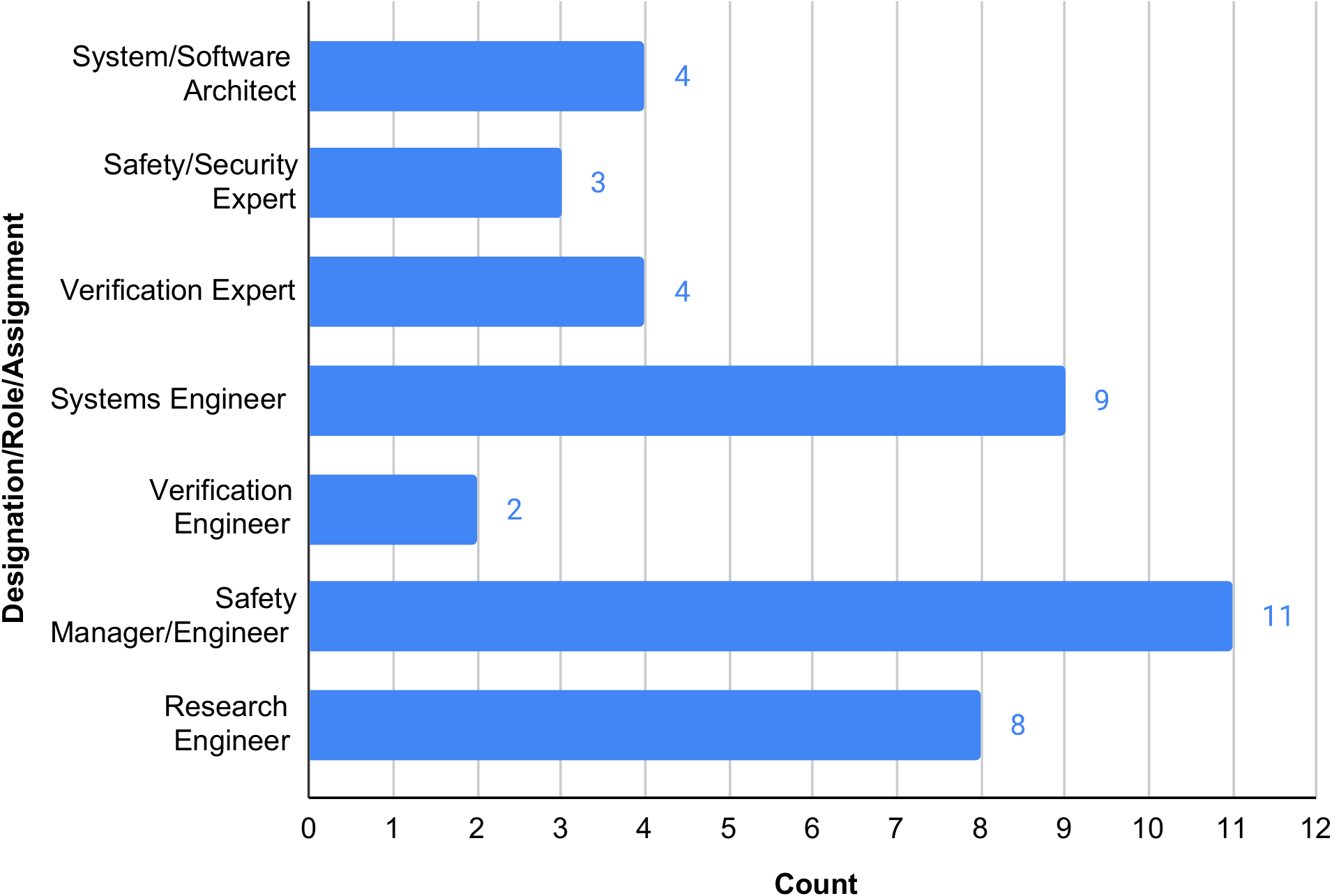}
\caption{Designations of participants.}
\label{fig:designation}
\end{subfigure}
\hspace{0.3cm}
\begin{subfigure}{0.48\textwidth}
		\includegraphics[width=\textwidth]{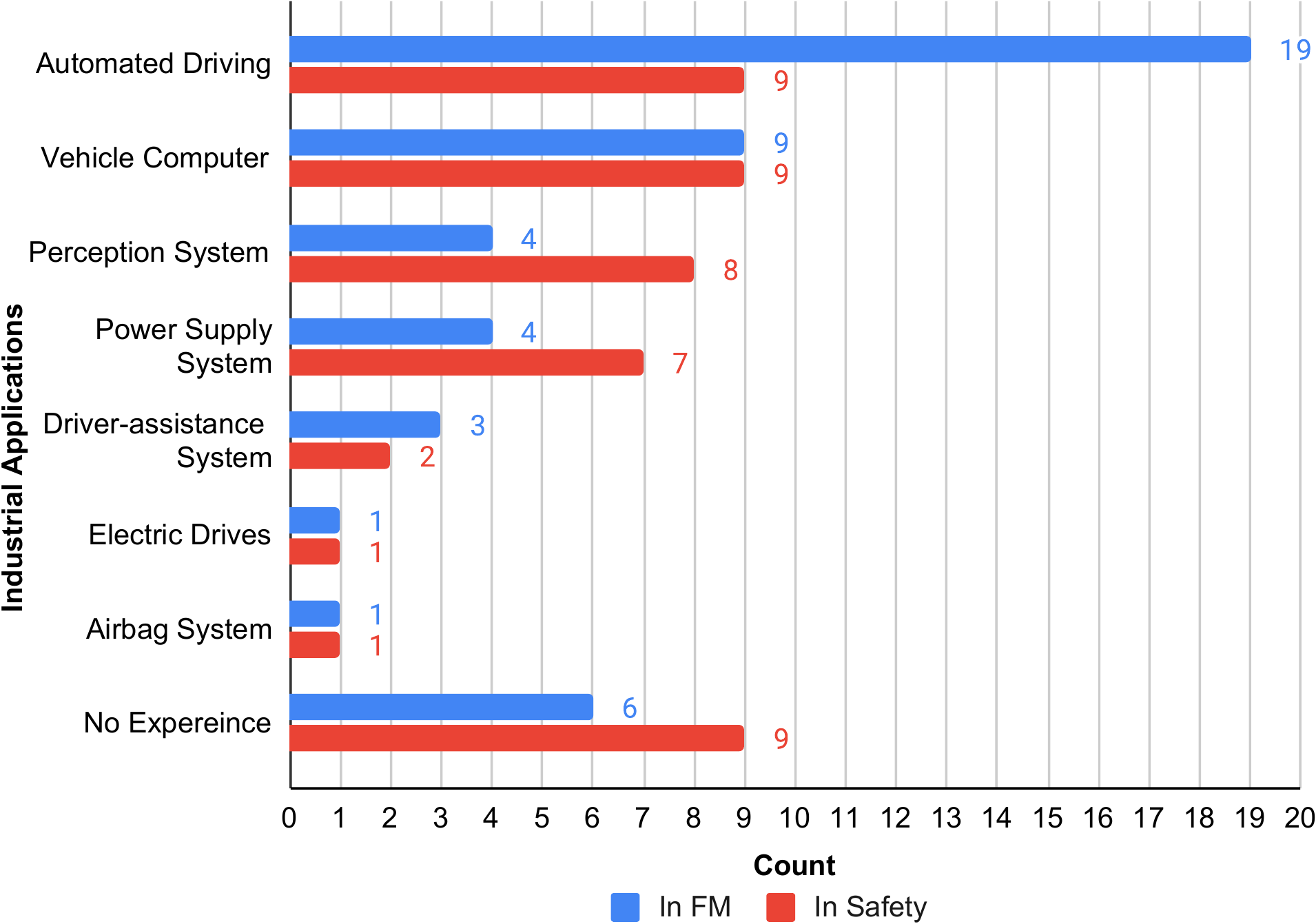}
		\caption{Industrial applications for formal methods and safety-related activities.}
		\label{fig:applications}
\end{subfigure}
\caption{Designations and industrial applications of participants.}
\end{figure}

Designations of the participants are collected with a free text field for question \qstwo (\tab{survey_questions}), depicted in \fig{designation}. The majority of the participants (11\,participants, 27\%) consider themselves as \textit{safety manager/engineer}, 22\% as \textit{systems engineer}, and 20\% as a \textit{research engineer}. Further, 11 participants (27\%) are either an expert or an architect in either safety, system, software, or verification.

\subsubsection{Industrial Application}
\label{sec:phase1:participants:application}

Questions \qsfour and \qssix in \tab{survey_questions} assess the industrial applications, to which participants applied formal methods or worked on safety aspects. Identified clusters of applications are shown in \fig{applications}. Formal methods are used mostly for \textit{automated driving} applications (19\,participants) and \textit{vehicle computer} (9\,participants). These are also the most-called applications for safety aspects (9\,participants each), followed by \textit{power supply system} (8\,participants) and \textit{driver-assistance system} (7\,participants). 

\subsection{Understanding Formal Notations}
\label{sec:phase1:notation}

With question \qsseven (\tab{survey_questions}), we collect the participants' opinions on understanding formal notations. The possible answers follow the scale \lsthree (\tab{likert_scale}) and further allow comments as free text. \tab{formal_notation} shows the results according to the scale. The majority (80\% of all participants) is almost equally distributed between answers \textit{hard} and \textit{slightly easy}. 44\% find formal notations between \textit{slightly hard} and \textit{extremely hard}, 34\% find  formal notations between \textit{slightly easy} and \textit{extremely easy}.
In the received comments, the majority of participants agree that understanding formal notations gets easier with more usage and experience, and that it is highly dependent on the focus of the system domain such as automotive and railway. A supporting statement from a participant answering \textit{slightly easy} states that \textit{\enquote{If I am familiar with the formal language in which the formal notations are written (e.g., first-order language), typically it is easy.}}

\begin{table}[b]
	\centering
	\caption{Understanding formal notations.}
	\label{tab:formal_notation}
	\resizebox{0.8\textwidth}{!}{%
		\begin{tabular}{|l|l|l|l|l|l|l|l|l|}
			\hline
			\textbf{Agreement} &
			\begin{tabular}[c]{@{}l@{}}Extremely\\ Hard\end{tabular} &
			Hard &
			\begin{tabular}[c]{@{}l@{}}Slightly \\ Hard\end{tabular} &
			\begin{tabular}[c]{@{}l@{}}Neither Hard \\ nor Easy\end{tabular} &
			\begin{tabular}[c]{@{}l@{}}Slightly \\ Easy\end{tabular} &
			Easy &
			\begin{tabular}[c]{@{}l@{}}Extremely \\ Easy\end{tabular} &
			\begin{tabular}[c]{@{}l@{}}No \\ Opinion\end{tabular} \\ \hline
			\textbf{Count} &
			2 (4\%) &
			8 (20\%) &
			8 (20\%) &
			9 (22\%)&
			8 (20\%)&
			5 (12\%)&
			1 (2\%)&
			0 \\ \hline
		\end{tabular}%
	}
\end{table}

\begin{figure}[t]
\begin{subfigure}{0.48\textwidth}
	\includegraphics[width=\textwidth]{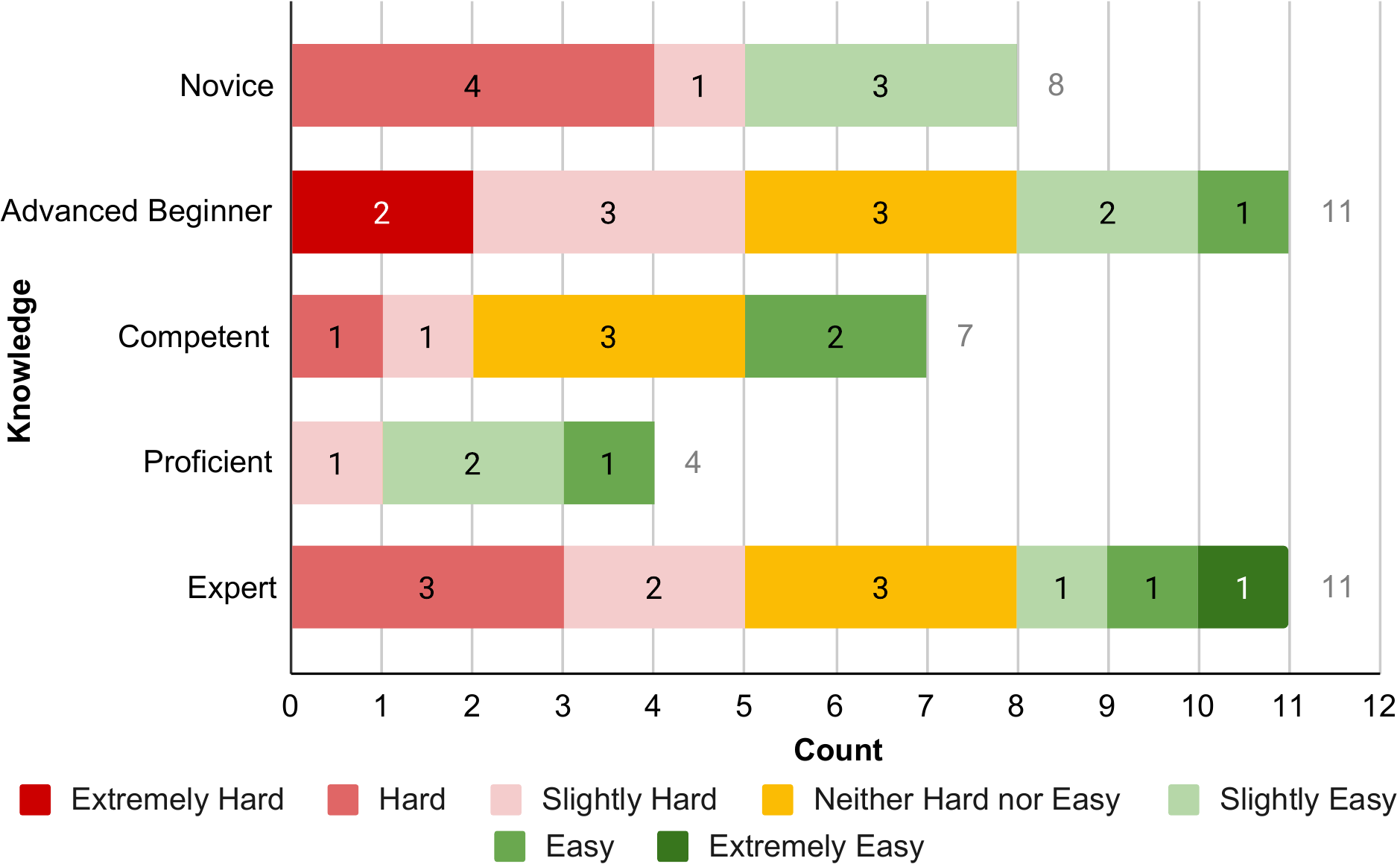}
	\caption{}
	\label{fig:formal_knowledge}
\end{subfigure}
\hspace{0.3cm}
\begin{subfigure}{0.48\textwidth}
	\includegraphics[width=\textwidth]{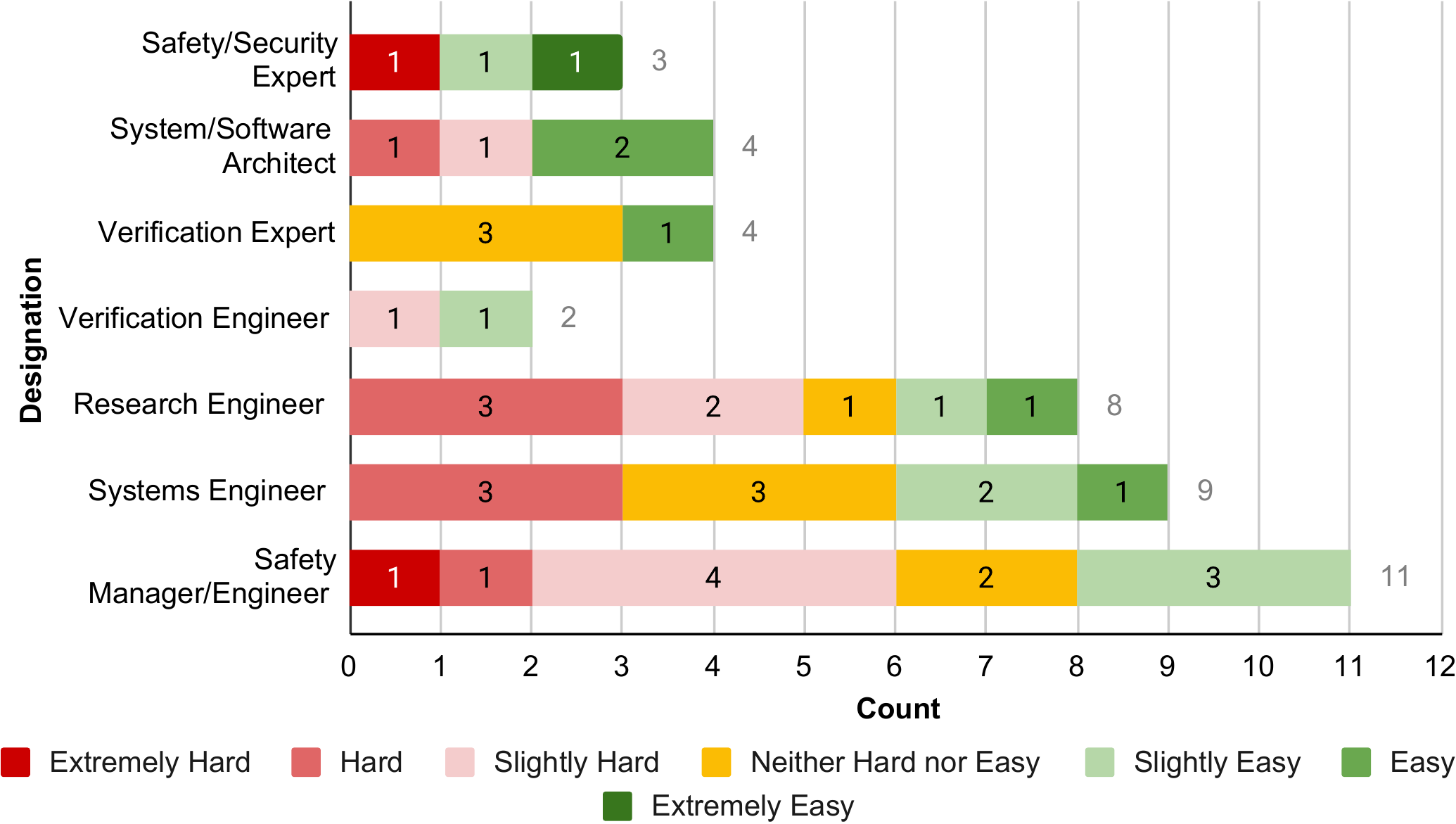}
	\caption{}
	\label{fig:formal_designation}
\end{subfigure}
\caption{Results for understanding formal notations grouped by (a) the knowledge of participants in formal methods and (b) the designation of participants.}
\label{fig:formal_knowledge+formal_designation}
\end{figure}

\fig{formal_knowledge+formal_designation}
shows responses for understanding formal notations based on the participants' knowledge in formal methods (\cf \sect{phase1:participants:knowledge}) and designation (\cf \sect{phase1:participants:designation}).
For the following discussion, we cluster answers among \textit{extremely hard}, \textit{hard}, and \textit{slightly hard} as {``harder''} and the ones among \textit{extremely easy}, \textit{easy}, and \textit{slightly easy} as {``easier''}.

\paragraph{Understanding of formal notations based on knowledge in formal methods.}

Looking at \textit{novice}s and \textit{advanced beginner}s in \fig{formal_knowledge}, the predominant number of participants perceives understanding of formal notations as \textit{harder}. Among the 11\,\textit{expert}s, five participants perceive understanding formal notations to be \textit{harder}, while three participants perceive understanding formal notation as \textit{easier}. This is notable because even the majority of \textit{experts} perceive understanding of formal notation to be \textit{harder}. A participant rated as being an \textit{expert} and answered \textit{hard} mentions understanding notations used in logics like LTL, CTL, or TCTL are pretty straightforward, but using them to formalize real-world requirements is harder. Another \textit{expert} who answers \textit{extremely easy} states that using approaches like pattern-based languages \citep{DwyerAC99} helps to ease understanding formal notations. 

\paragraph{Understanding of formal notations based on participants' designation.}

Participants with designations of a \textit{system/software architect}, \textit{systems engineer}, and \textit{verification engineer} perceive formal notations to be \textit{harder} to understand while the same number of participants from each of the three mentioned categories vote for formal notations to be \textit{easier} for understanding. A verification expert highlights: \textit{\enquote{In general, formal notations are clear and precise. However, even for very experienced engineers, some formal notations (for example in temporal logics) are hard to understand on the semantic level, i.e., really telling what the formula means for the system at hand.}} Considering the designations of a \textit{safety/security expert} and \textit{verification expert}, the majority of the answers (excluding \textit{neither hard nor easy}) opt for understanding formal notations to be \textit{easier} with two out of three participants and one participant in each designation. For the remaining designations of a \textit{safety manager/engineer} and \textit{research engineer}, the majority of participants answer that understanding notations are \textit{harder} with six out of 11\,participants and five out of eight participants, respectively. A \textit{systems engineer} state that \textit{\enquote{I have never really learned formal notation; thus, I learned it on the job. A real introduction might have turned out helpful.}}

\paragraph{Summary.}
44\% of all participants answer understanding formal notations to be hard, only 34\% of participants perceive understanding of formal notations as easy. From the answers received as free text, it is clear that experience plays a major role in understanding formal notations.

\subsection{Inconsistent Formal Specifications}
\label{sec:phas1:inconsistent_spec}

In this section, we discuss difficulties in identifying inconsistent specifications, understanding inconsistencies, the time taken to identify inconsistent specifications, as well as challenges and different methods to identify inconsistent specifications based on the answers to questions \qsnine to \qsthirteen (\tab{survey_questions}).

\subsubsection{Identifying Inconsistent Formal Specification}
\label{sec:phase1:inconsistent}

With question \qsnine, we assess difficulties of identifying inconsistent formal specifications. Answers are given according to the scale \lsthree (\tab{likert_scale}) and as free text. The results of the answering 41\,participants are shown in \tab{inconsistent}. Notably, a majority (51\%) perceive the identification of inconsistent formal specifications as \textit{hard}, with a total of 73\% perceiving it as at least \textit{slightly hard}.

\begin{table}[b]
	\centering
	\caption{Results for the difficulty of identifying inconsistent specifications.}
	\label{tab:inconsistent}
	\resizebox{0.8\textwidth}{!}{%
		\begin{tabular}{|l|l|l|l|l|l|l|l|l|}
			\hline
			\textbf{Agreement} & \begin{tabular}[c]{@{}l@{}}Extremely\\  Hard\end{tabular} & Hard & \begin{tabular}[c]{@{}l@{}}Slightly \\ Hard\end{tabular} & \begin{tabular}[c]{@{}l@{}}Neither Hard\\  nor Easy\end{tabular} & \begin{tabular}[c]{@{}l@{}}Slightly \\ Easy\end{tabular} & Easy & \begin{tabular}[c]{@{}l@{}}Extremely \\ Easy\end{tabular} & \begin{tabular}[c]{@{}l@{}}No \\ Opinion\end{tabular} \\ \hline
			\textbf{Count}     & 4  (10\%)                                                       & 21  (51\%) & 5  (12\%)                                                      & 7   (17\%)                                                             & 0                                                        & 3  (7\%)  & 1   (3\%)                                                      & 0                                                     \\ \hline
		\end{tabular}%
	}
\end{table}

From the free-text responses, 
13 participants state that the effort of identifying inconsistent specifications highly depends on the kind of specification, as well as the complexity of the specification and system. For example, a \textit{verification expert} who answers \textit{neither hard nor easy} states that identifying inconsistent specification \textit{\enquote{depends on the size and kind of specification, \eg debugging temporal logic is not trivial.}} Regarding complexity of the specification, a \textit{systems engineer} highlights an example that it \textit{\enquote{depends on the complexity of the formal specification itself and how well the inconsistency is hidden, \eg $x<10$ and $x>20$ is easy to stop, but if you replace one x with y and link them somewhere else, it’s already hard to find.}} However, on contrary, an interesting point is highlighted by a \textit{verification expert}, stating that one \textit{\enquote{can’t really say, [as it] highly depends on the complexity of the application. With respect to model checking activation/deactivation of autonomous driving functions we realized that even small state machines can contain possibly critical errors. Also, these errors would have never been identified by classic testing (test runs, simulation).}}
\begin{figure}[t]
	\centering
	\includegraphics[width=0.8\textwidth]{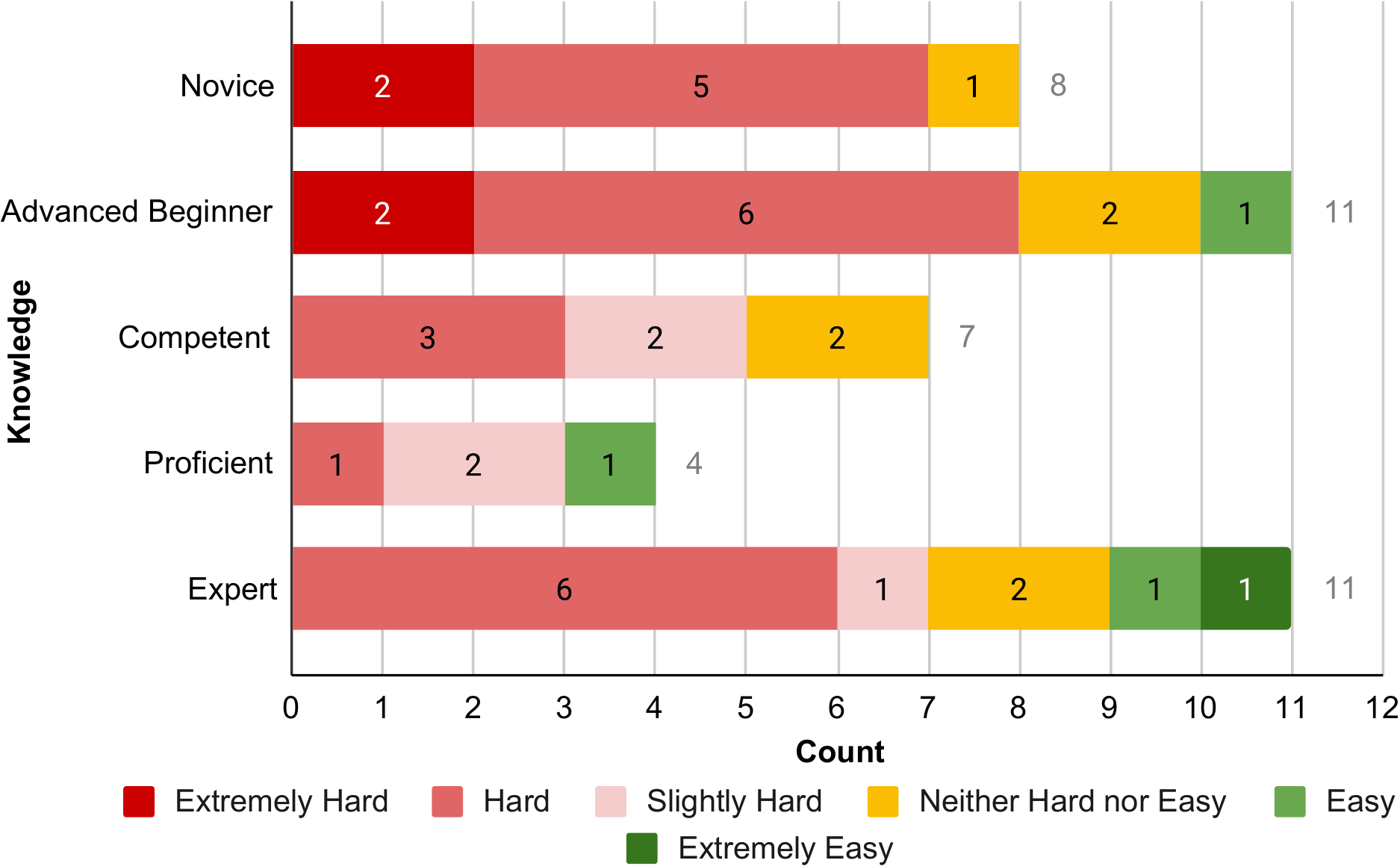}
	\caption{Results for identifying inconsistent formal specifications grouped by the participants' knowledge in formal methods.}
	\label{fig:inconsistent_knowledge}
\end{figure}

\paragraph{Identification of inconsistent formal specifications based on participants knowledge in formal methods.}

\fig{inconsistent_knowledge} depicts the responses gathered for identifying inconsistent specifications together with the participants' knowledge in formal methods. The count of responses as \textit{harder} is dominating all the classes ranging from \textit{novice} to \textit{expert}.
Among all participants who answer identifying inconsistent specifications is \textit{easier}, the majority is either \textit{proficient} or an \textit{expert} in formal methods.
An \textit{expert} safety engineer states that \textit{\enquote{the degree of difficulty heavily depends on the nature of the formalism: Spotting an error in a boolean formula is significantly easier than identifying erroneous temporal specification, which is again significantly easier compared to spotting erroneous specification for continuous behavior}}.

\paragraph{Summary.} Predominantly, participants answer that identification of inconsistent specifications is \textit{hard}. Free-text answers agree that the effort for identifying inconsistent specifications highly depends on the complexity of specifications and systems. 

\subsubsection{Understanding inconsistency in formal specifications}
\label{sec:phase1:inconsistency}

The question \qsten (\tab{survey_questions}) is used to collect whether understanding of the actual inconsistency in the identified inconsistent specifications is \textit{easier} or \textit{harder}. The answer scale used to collect the response is again \lsthree and a free text field. Responses are aggregated in \tab{inconsistency}, where 39\,participants provided a rating, while two participants indicated \textit{no opinion}.

\begin{table}[b]
	\centering
	\caption{Understanding inconsistent formal specification.}
	\label{tab:inconsistency}
	\resizebox{0.8\textwidth}{!}{%
		\begin{tabular}{|l|l|l|l|l|l|l|l|l|}
			\hline
			\textbf{Agreement} & \begin{tabular}[c]{@{}l@{}}Extremely\\  Hard\end{tabular} & Hard & \begin{tabular}[c]{@{}l@{}}Slightly \\ Hard\end{tabular} & \begin{tabular}[c]{@{}l@{}}Neither Hard\\  nor Easy\end{tabular} & \begin{tabular}[c]{@{}l@{}}Slightly \\ Easy\end{tabular} & Easy & \begin{tabular}[c]{@{}l@{}}Extremely \\ Easy\end{tabular} & \begin{tabular}[c]{@{}l@{}}No \\ Opinion\end{tabular} \\ \hline
			\textbf{Count}     & 6  (15\%)                                                       & 18 (44\%)  & 6  (15\%)                                                         & 5   (12\%)                                                             & 2  (5\%)                                                      & 1   (2\%)  & 1    (2\%)                                                     & 2    (5\%)                                                 \\ \hline
		\end{tabular}%
	}
\end{table}

The result is comparable to the result obtained for identifying inconsistent specification in \sect{phase1:inconsistent}. The predominant number of 18\,participants answers understanding inconsistency in the formal specifications is \textit{hard}. Overall 77\% of participants responded at least \textit{slightly hard}. 13\% answer \textit{neither hard nor easy} and perceive it \textit{easier}.

Three participants who answer \textit{neither hard nor easy} highlight that the effort of understanding the inconsistency in formal specifications depends on the complexity of specification and the system. A \textit{verification expert} with the response \textit{hard} states that \textit{\enquote{the key question is: whether the model or the specification is wrong \ie finding inconsistencies can often not only be done on the level of just the specification}.} 

\begin{figure}[b]
	\centering
	\includegraphics[width=0.8\textwidth]{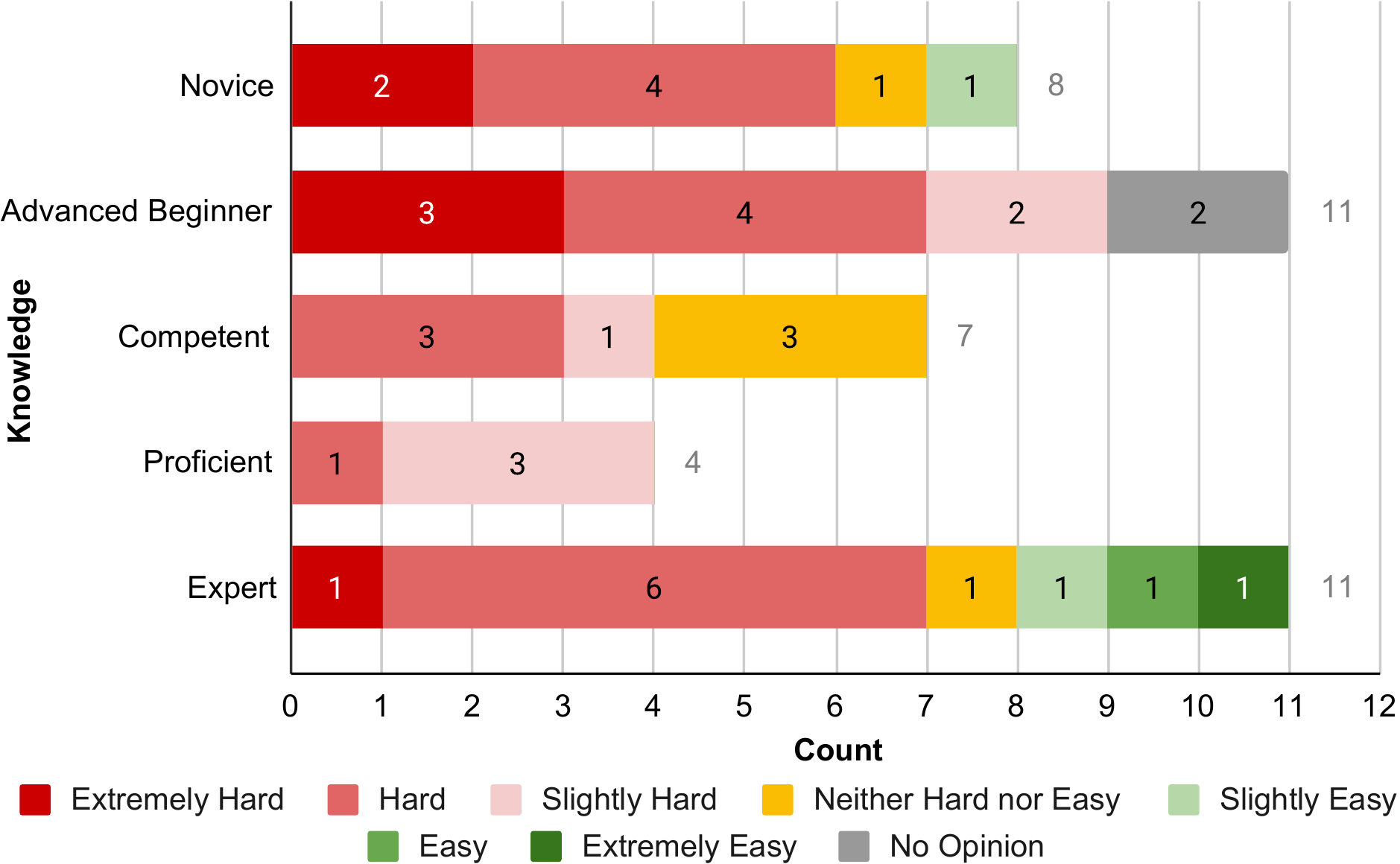}
	\caption{Results for the complexity in identifying inconsistency in formal specifications grouped by the participants' knowledge in formal methods.}
	\label{fig:inconsistency_knowledge}
\end{figure}

\paragraph{Understanding inconsistency in formal specifications based on the participants' knowledge in formal methods.}

\fig{inconsistency_knowledge} depicts the responses gathered for understanding inconsistency in formal specifications together with the participants knowledge in formal methods (\cf \sect{phase1:participants:knowledge}). The majority of participants from \textit{novices} to \textit{experts} perceive understanding of inconsistencies to be \textit{harder}. Particularly, this view is shared by six of eight \textit{novices}, seven of 11 \textit{advance beginners}, four of seven \textit{competents}, all four \textit{proficients}, and seven of 11 \textit{experts}.

One of the \textit{advanced beginners} explains a common industrial issue: \textit{\enquote{understanding can be hard especially when the specifications have multiple authors, each holding a slightly different view on the object to be specified}.} 
A \textit{system/software architect} who rates their knowledge in formal methods as \textit{expert}, states that \textit{\enquote{formal proofs typically lead to the source of the error. Counterexamples in model checkers also help a lot}}.
On contrary to this statement a \textit{systems engineer} who rates their knowledge in formal methods as \textit{expert} highlights that it depends on the particular inconsistency, \eg if it is real-time inconsistency that only occurs after several cycles, then it is difficult to find even with the help of counterexamples.

\paragraph{Summary.}
Similar to the result of identifying inconsistent specification in \sect{phase1:inconsistency}, a majority of participants (77\%) value understanding of the inconsistency as harder. In the free-text responses, participants indicate that the effort to understand an inconsistency depends on the complexity of the specification and system. Further, some of the responses highlight the use of tools like model checkers to support identification and understanding of inconsistencies.

\subsubsection{Time Taken to Identify Inconsistent Formal Specifications}
\label{sec:phase1:timetaken}

With question \qseleven (\tab{survey_questions}), we consider the time that it takes to identify inconsistent specifications. The answers follow scale \lsfour (\tab{likert_scale}) and allow for a free-text comment. \tab{identify_inconsistentSpec} represents the response collected from 37\,participants as four participants responded \textit{no opinion}.
\begin{table}[b]
	\centering
	\caption{Time taken to identify inconsistent formal specifications.}
	\label{tab:identify_inconsistentSpec}
	\resizebox{0.8\textwidth}{!}{%
		\begin{tabular}{|l|l|l|l|l|l|l|l|l|}
			\hline
			Agreement &
			\begin{tabular}[c]{@{}l@{}}Extremely\\ Fast\end{tabular} &
			Fast &
			\begin{tabular}[c]{@{}l@{}}Slightly\\ Fast\end{tabular} &
			\begin{tabular}[c]{@{}l@{}}Neither Fast\\ nor Slow\end{tabular} &
			\begin{tabular}[c]{@{}l@{}}Slightly\\ Slow\end{tabular} &
			Slow &
			\begin{tabular}[c]{@{}l@{}}Extremely\\ Slow\end{tabular} &
			\begin{tabular}[c]{@{}l@{}}No\\ Opinion\end{tabular} \\ \hline
			Count &
			0 &
			3 (7\%) &
			4 (10\%)&
			12 (29\%)&
			4 (10\%)&
			11 (27\%)&
			3 (7\%)&
			4 (10\%)\\ \hline
		\end{tabular}%
	}
\end{table}

A majority of participants, 18 of them (44\% of all participants) answer that the time taken to identify inconsistent specifications is \textit{slower} (\textit{slightly slow}, \textit{slow}, and \textit{extremely slow}).
Further, 12 participants (29\%) answer that the time taken to identify inconsistent specifications is \textit{neither fast nor slow}, with six participants stating that it depends on the complexity and number of specifications.

A \textit{system/software architect} answering with \textit{slightly slow} highlights that \textit{\enquote{the time it takes is the time taken to understand requirements}}. Further, a \textit{verification expert} states that in one of their projects it took a couple of hours to formalize the system model and specifications, but only a few seconds to then run the model checking and perform verification and further optimization.

Seven participants (17\%) answer that the time taken to identify inconsistent specifications is \textit{faster} (\textit{slightly fast}, \textit{fast}, and \textit{extremely fast}). Three of them answer that by using a model checker, the time taken to identify inconsistent specifications is reduced. 

\begin{figure}[t]
	\centering
	\includegraphics[width=0.8\textwidth]{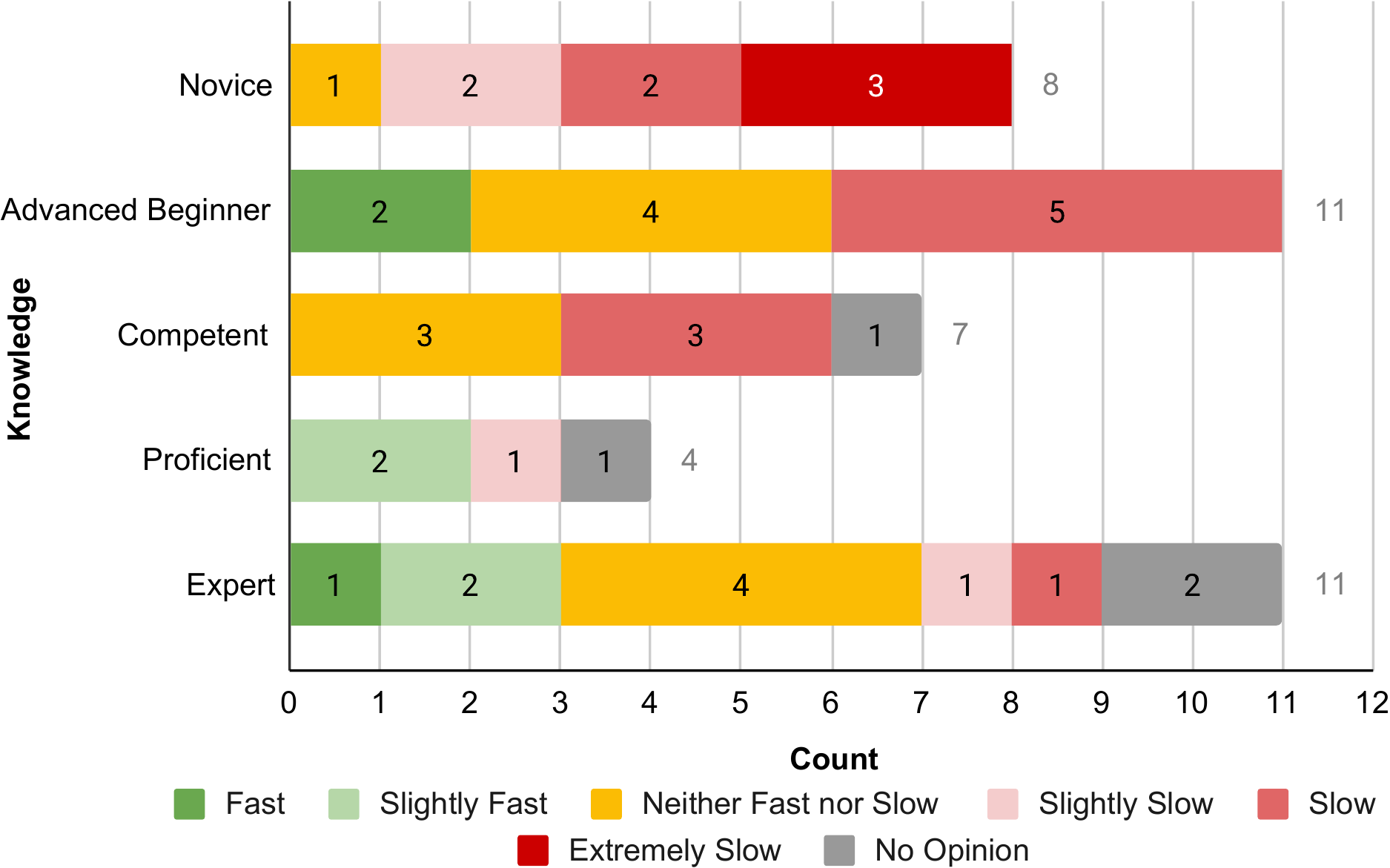}
	\caption{Results for the time taken to identify inconsistent formal specifications grouped by the participants' knowledge in formal methods.}
	\label{fig:identify_inconsistentSpec_knowledge}
\end{figure}

\paragraph{Time taken to identify inconsistent formal specifications grouped by the participants' knowledge in formal methods.}

\fig{identify_inconsistentSpec_knowledge} presents the results of time taken to identify inconsistent formal specifications grouped by the participants' knowledge (\cf \sect{phase1:participants:knowledge}). The majority of participants who rate their knowledge as \textit{novice}, \textit{advanced beginner}, \textit{competent}, and \textit{expert} answer that the time taken to identify inconsistent specifications is \textit{slower} or \textit{neither fast nor slow}. Five of seven participants who answer \textit{faster} rated themselves as \textit{proficient} and \textit{experts}. The majority answers of about 37\% of all participants respond that using formal verification makes \textit{very probably} a system safer while 34\% vote for \textit{definitely} safer.

\paragraph{Summary.} \change{In total, 44\% of the participants answer that identifying inconsistent formal specifications is \textit{slower}. In fact, 44\% is predominant here because 29\% of them have the neutral response answering \textit{neither fast nor slow} and only 17\% of the participants answer to be \textit{faster}.} Free-text responses indicate that the time it takes depends on the kind of specifications and the complexity of the system and specifications. This is similar to the responses received for identifying inconsistent specifications in \sect{phase1:inconsistent} and \sect{phase1:inconsistency}.

\subsubsection{Challenges to Identify Inconsistent Specifications}
\label{sec:phase1:challanges_inconsistent}

With question \qstwelve (\tab{survey_questions}), we collect challenges in identifying inconsistent specification based on free-text responses. The responses are summarized based on the participants' designations. 

\paragraph{Architects and Experts.}

Most of the \textit{system/software architects}, \textit{safety/security experts}, and \textit{verification experts} highlight two general challenges: (1)~architectural models and system requirements are often incomplete in industry, and (2)~understanding the semantics and formal notations. A notable statement from a \textit{system/software architect} is that \textit{\enquote{if the specification has many items (which is very common in the industry), checking them one-by-one by human is time-consuming and error-prone. Furthermore, if the specification is written in natural language, how to interpret it in an objective way can also be a question}}. Further, a \textit{verification expert} states that it is challenge to \textit{\enquote{understand the intended semantics of the specifications at scale}}. Another notable statement from a \textit{safety/security expert} is that \textit{\enquote{the prime challenge is to identify inconsistency and then to get acceptance to the amount of inconsistencies [that can be tolerated]}}.

\paragraph{Safety Managers/Engineers.}

Two main challenges listed by \textit{safety managers and engineer} are: (1)~understanding formal specification and (2)~understanding verification results. A \textit{safety engineer} rated as \textit{expert} states that \textit{\enquote{some notions (\eg LTL formulas) are hard to understand. Often you just think you understood correctly what they mean while using them the wrong way. Furthermore, there might be corner cases in the processes to be described which make your life particularly hard, \eg specific startup behavior of systems. Even simple formulas tend to become huge quickly, which makes it pretty hard to focus}}. Furthermore, a statement from a \textit{novice} is, that \textit{\enquote{it is not really intuitive to understand the dependencies between the formal specifications – especially if you investigate a larger set of specifications}}.

\paragraph{Systems Engineers.}
 
The main challenge mentioned by \textit{systems engineers} are formalizing specifications. A \textit{systems engineer} rated as \textit{expert} states that \textit{\enquote{the main challenge to identify inconsistent formal specifications is that specifications have to be formalized first. There are only few engineers at Bosch who want to do that, even semi formal requirements patterns are seldom used}}. Further, a challenge mentioned by a \textit{systems engineer} rated as \textit{competent} is \textit{\enquote{insufficient granularity of high-level system behavior, thus difficult to perform verification}}. In addition, a general challenge to identify inconsistent specification is that \textit{\enquote{often hundreds of cases/situations have to be considered while only a few are inconsistent}}.

\paragraph{Research Engineers and Verification Engineers.}
 
Two challenges identified by \textit{research engineers} and \textit{verification engineers} are: (1)~complexity in using appropriate tools, and (2)~understanding the verification results. A \textit{research engineer} rated as \textit{proficient} states that the \textit{\enquote{large size and complexity of the specification might not allow one to debug it manually. When the formal specification is written in an expressive language and is very complex in general, automatic method for formal analysis might not scale either}}. Further, a \textit{verification engineer} highlights that even if a model checker supports to identify an inconsistent specification, the time taken to perform verification for large system is huge. Regarding explainability, a \textit{research engineer} states that inconsistencies detected by the verification tools are cryptic and thus, require additional support to derive a useful explanation.

\paragraph{Summary.} 

From the collected responses, the four different challenges to identify inconsistent specification are: (1)~verification performed with incomplete 
models, (2)~understanding formal semantics, notations, and specifications, (3)~complexity in using verification tools, and (4)~understanding of verification results by domain experts.

\subsubsection{Methods used for Verification}
\label{sec:phase1:methods}

As discussed in \sect{phase1:inconsistency}, several participants mentioned the use of verification methods to identify inconsistent specifications and understanding the inconsistencies. With \qsthirteen (\tab{survey_questions}), we collect the used verification methods with a free-text field. Several participants mentioned multiple methods.
All responses are clustered into five different methods shown in \tab{methods}.

\begin{table}[b]
	\centering
	\caption{Results for the used verification methods.}
	\label{tab:methods}
	\resizebox{0.8\textwidth}{!}{%
		\begin{tabular}{|l|l|l|l|l|l|l|}
			\hline
			Methods &
			\begin{tabular}[c]{@{}l@{}}Model\\ Checking\end{tabular} &
			\begin{tabular}[c]{@{}l@{}}Manual \\ Inspection/Review\end{tabular} &
			Simulation &
			Reasoner &
			\begin{tabular}[c]{@{}l@{}}Contract-Based\\  Design\end{tabular} &
			\begin{tabular}[c]{@{}l@{}}No\\  Opinion\end{tabular} \\ \hline
			Count &
			18 (32\%) &
			10 (18\%)&
			6 (11\%)&
			5 (9\%)&
			4 (7\%)&
			13 (23\%)\\ \hline
		\end{tabular}%
	}
\end{table}

The predominantly used methods are \textit{model checking} with 18\,participants and \textit{manual inspection/review} with ten participants. \textit{Manual inspection/review} is the only purely manual method, the other methods like \textit{model checking}, \textit{simulation}, \textit{reasoner}, \textit{contract-based design} are (semi-)automated methods, supported by appropriate tools. Five participants indicate that model checkers support overcoming the manual inspection to find inconsistency. 

\begin{figure}[b]
	\centering
	\includegraphics[width=0.8\textwidth]{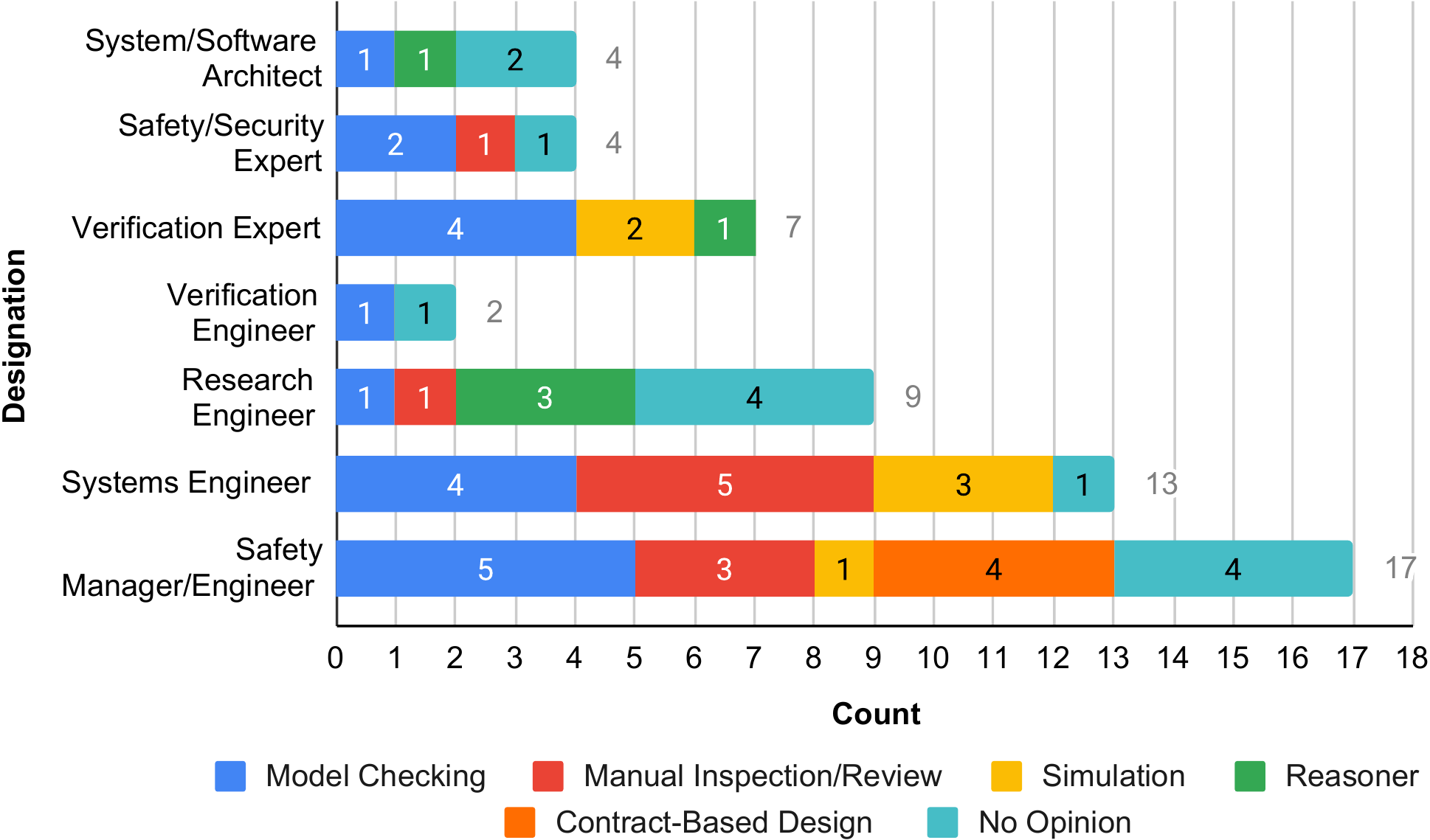}
	\caption{Methods used by participants to identify inconsistent formal specifications grouped by designation.}
	\label{fig:method_designation}
\end{figure}

\paragraph{Methods used for verification based on designation.} 

\fig{method_designation} represents the different verification methods together with participants' designation. \textit{Model checking} is used by at least one participant of each designation. \textit{Manual inspection/review} is mostly mentioned by \textit{systems engineers} and \textit{safety managers/engineers}, \textit{simulation} mostly by \textit{systems engineers} and \textit{verification experts}, \textit{reasoners} by \textit{research engineers}, and \textit{contract-based design} mostly by \textit{safety managers/engineers}.

\paragraph{Verification methods and the complexity of identifying inconsistent specifications.}

\begin{figure}[t]
	\centering
		\includegraphics[width=0.8\textwidth]{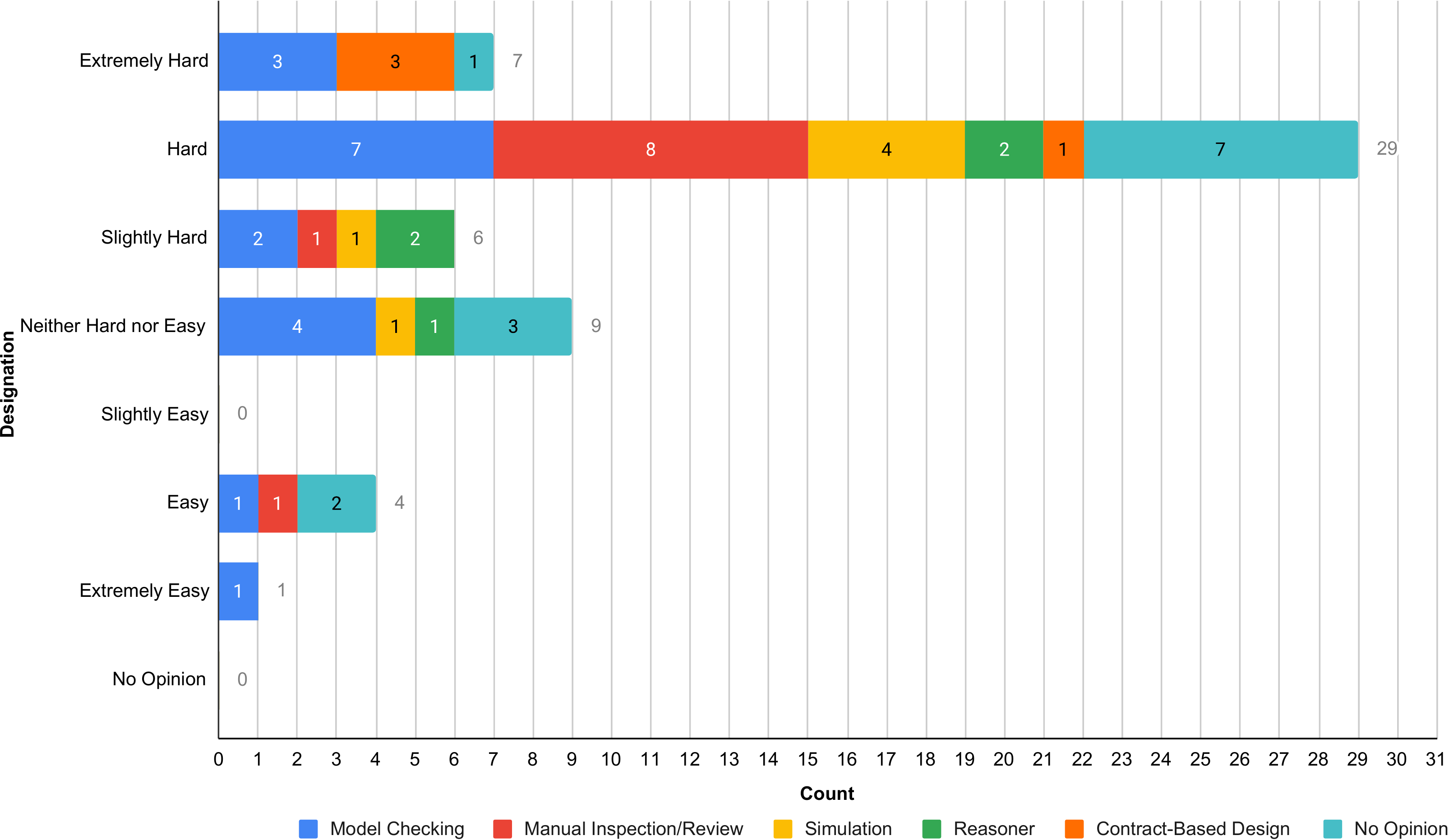}
		\caption{Methods used to identify inconsistent formal specifications along with complexity of identifying inconsistent specifications.}
		\label{fig:method_inconsistent}
\end{figure}

\fig{method_inconsistent} depicts the different verification methods together with the results obtained for identifying inconsistent formal specifications in \sect{phase1:inconsistent}. Since the results for both identifying inconsistent specification (\cf \sect{phase1:inconsistent}) and understanding inconsistent specification (\cf \sect{phase1:inconsistency}) are similar, we choose to discuss the relation only with results of identifying inconsistent specification.

As a majority of the participants (21 of 41\,participants) have answered that identifying inconsistent formal specifications is \textit{hard}, they mentioned each of the listed methods at least once. To be more precise, among 21\,participants who answer \textit{hard} for identifying inconsistent specification, eight participants use \textit{manual inspection/review} while seven of them use \textit{model checking}.
Addition\-ally, participants who answer \textit{extremely hard} are found to use \textit{model checking} and \textit{contract-based design}. 
A participant who rated herself/himself as \textit{proficient} and classify the problem as \textit{hard} state that the correct usage of verification tools is mostly an error-prone way. Furthermore, an \textit{expert} states: \textit{\enquote{If you see UML Models Rhapsody/EA as formal models: typically, they are not complete and therefore model checker do not support well}}. 
On other hand, participants classifying the problem as \textit{easier} are found to use \textit{model checking} and \textit{manual inspection/review}. 
An \textit{advanced beginner} classifying the problem as \textit{easy} by using \textit{manual inspection/review} highlights that \textit{\enquote{by using the modern requirement engineering tools traceability can be maintained within dependent specification and corresponding architectural model}}.  

\paragraph{Summary.} From the collected responses, five different verification methods are used to identify inconsistent specifications: \textit{model checking}, \textit{manual inspection/review}, \textit{simulation}, \textit{reasoner}, and \textit{contract-based design}. Among these, the predominantly used methods are \textit{model checking} and \textit{manual inspection/review}, mentioned by 18\,participants and ten participants respectively.

\subsection{Refinement of Specifications}
\label{sec:phase1:refinement}

The early stages in the V-model~\citep{Weber09} are about refining the top-level requirements and associate them to system components and sub-components. With questions \qsfourteen and \qsfifeteen (\tab{survey_questions}), we investigate whether maintaining the consistency of refined specifications with a system architecture and verifying the refinement consistency are \textit{harder} or \textit{easier}. 

\begin{table}[b]
	\centering
	\caption{Results for the difficulty of maintaining the consistency when refining formal requirements for sub-components of a system architecture.}
	\label{tab:consistency}
	\resizebox{0.8\textwidth}{!}{%
		\begin{tabular}{|l|l|l|l|l|l|l|l|l|}
			\hline
			Agreement & \begin{tabular}[c]{@{}l@{}}Extremely\\  Hard\end{tabular} & Hard & \begin{tabular}[c]{@{}l@{}}Slightly \\ Hard\end{tabular} & \begin{tabular}[c]{@{}l@{}}Neither Hard\\ nor Easy\end{tabular} & \begin{tabular}[c]{@{}l@{}}Slightly \\ Easy\end{tabular} & Easy & \begin{tabular}[c]{@{}l@{}}Extremely \\ Easy\end{tabular} & \begin{tabular}[c]{@{}l@{}}No \\ Opinion\end{tabular} \\ \hline
			Count     & 2 (5\%)                                                        & 9 (22\%)   & 9    (22\%)                                                     & 1   (2\%)                                                            & 5  (13\%)                                                      & 2   (5\%) & 1     (2\%)                                                    & 12   (29\%)                                                 \\ \hline
		\end{tabular}%
	}
\end{table}

\subsubsection{Consistency of Refined Specifications and System Architectures}
\label{sec:phase1:refinement_consistency}

With question \qsfourteen (\tab{survey_questions}), we collects the responses whether maintaining consistency of the refined specifications is \textit{harder} or \textit{easier} following the answer scale \lsthree in \tab{likert_scale}. Twelve out of 41\, participants state that they have \textit{no opinion} and thus, the responses received from 29\,participants are shown in \tab{consistency}. The predominant number of participants answer that maintaining consistency is \textit{hard} and \textit{slightly hard}, with an exact count of nine participants (31\% of responded participants) each. 
Overall, most participants (20\,participants, 69\% of all participants) answer that maintaining consistency is \textit{harder}, eight participants (28\%) answer with \textit{easier}, and one participant (3\%) with \textit{neither hard nor easy}.

\paragraph{Maintaining refinement consistency grouped by the participants' experience in formal methods.}

\begin{figure}[t]
	\centering
	\includegraphics[width=0.8\textwidth]{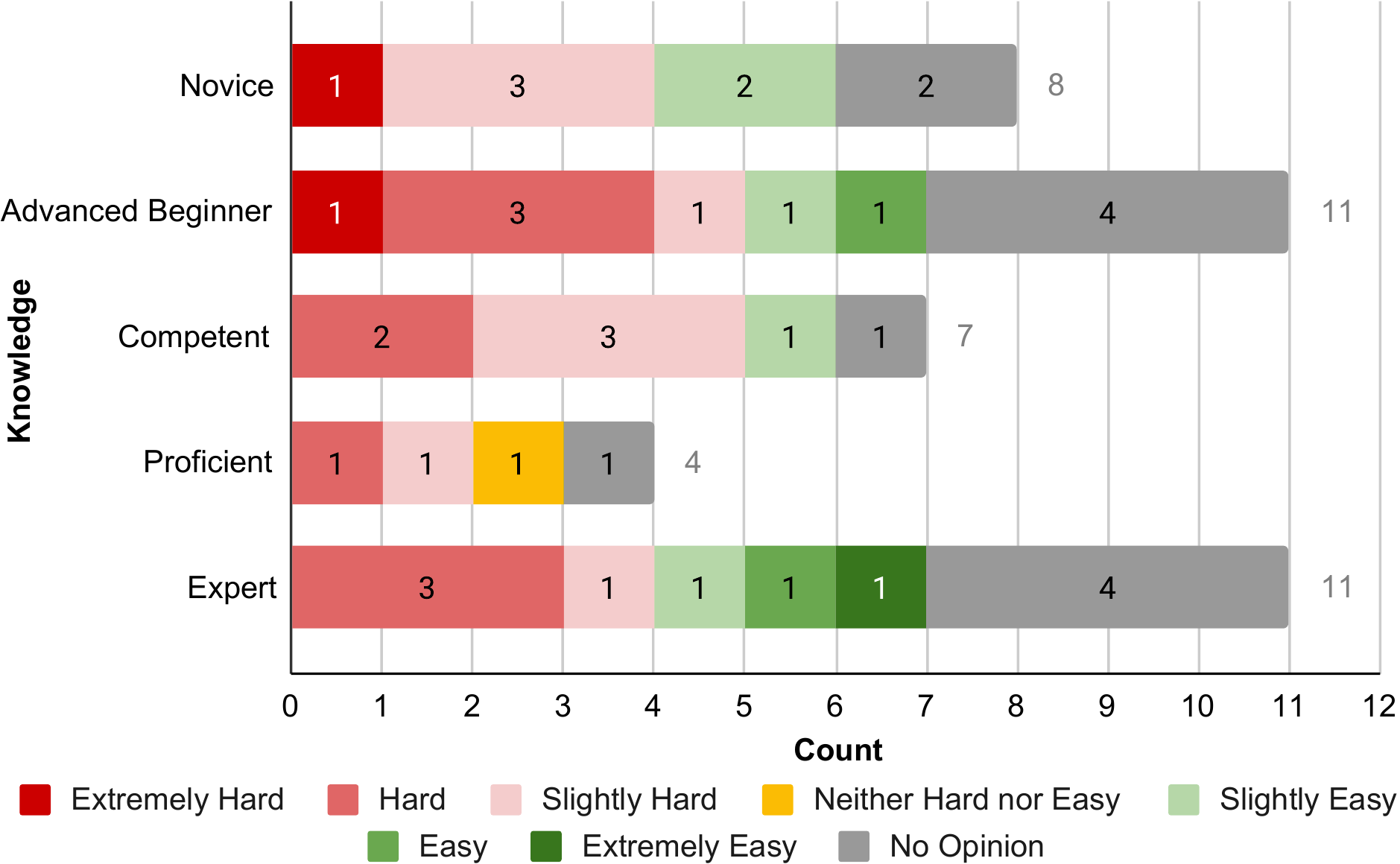}
	\caption{Results for the difficulty of maintaining refinement consistency grouped by the participants' knowledge in formal methods.}
	\label{fig:consistency_knowledge}
\end{figure}

\fig{consistency_knowledge} depicts the results of maintaining refinement consistency between the specifications and architecture considering the participants' knowledge in formal methods (\cf  \sect{phase1:participants:knowledge}).
Participants who rate the difficulty as \textit{extremely hard} are \textit{novices} and \textit{advanced beginners} while those who answer \textit{extremely easy} are designated as an \textit{expert}. Since the majority of participants rate the difficulty as \textit{harder} (\cf \tab{consistency}), all the categories from \textit{novice} to \textit{expert} have higher count for \textit{harder} than \textit{easier}. An \textit{advance beginner} who rates the problem as \textit{hard} highlights: \textit{\enquote{It depends on the size and complexity of the System-Architecture; in L4 Sensor-set architectures it[']s hard}}. Additionally, an \textit{expert} states that performing a manual review with natural language requirements against the requirements of the higher abstraction level in a contract-based design is \textit{hard}. From all of the categories except of \textit{proficient}, at least one participant answers that maintaining inconsistent specifications is \textit{easier}. 
An \textit{advanced beginner} whose rates the problem as \textit{easy} mentions: \textit{\enquote{I assume I formulated a formal requirement for the parent component, and I have to define formal requirements for sub-components, I think that is straight forward. If somebody else wrote the formal requirement of the parent component, it becomes harder since I need to understand what is expressed there first}}. 

\begin{figure}[t]
	\centering
	\includegraphics[width=0.8\textwidth]{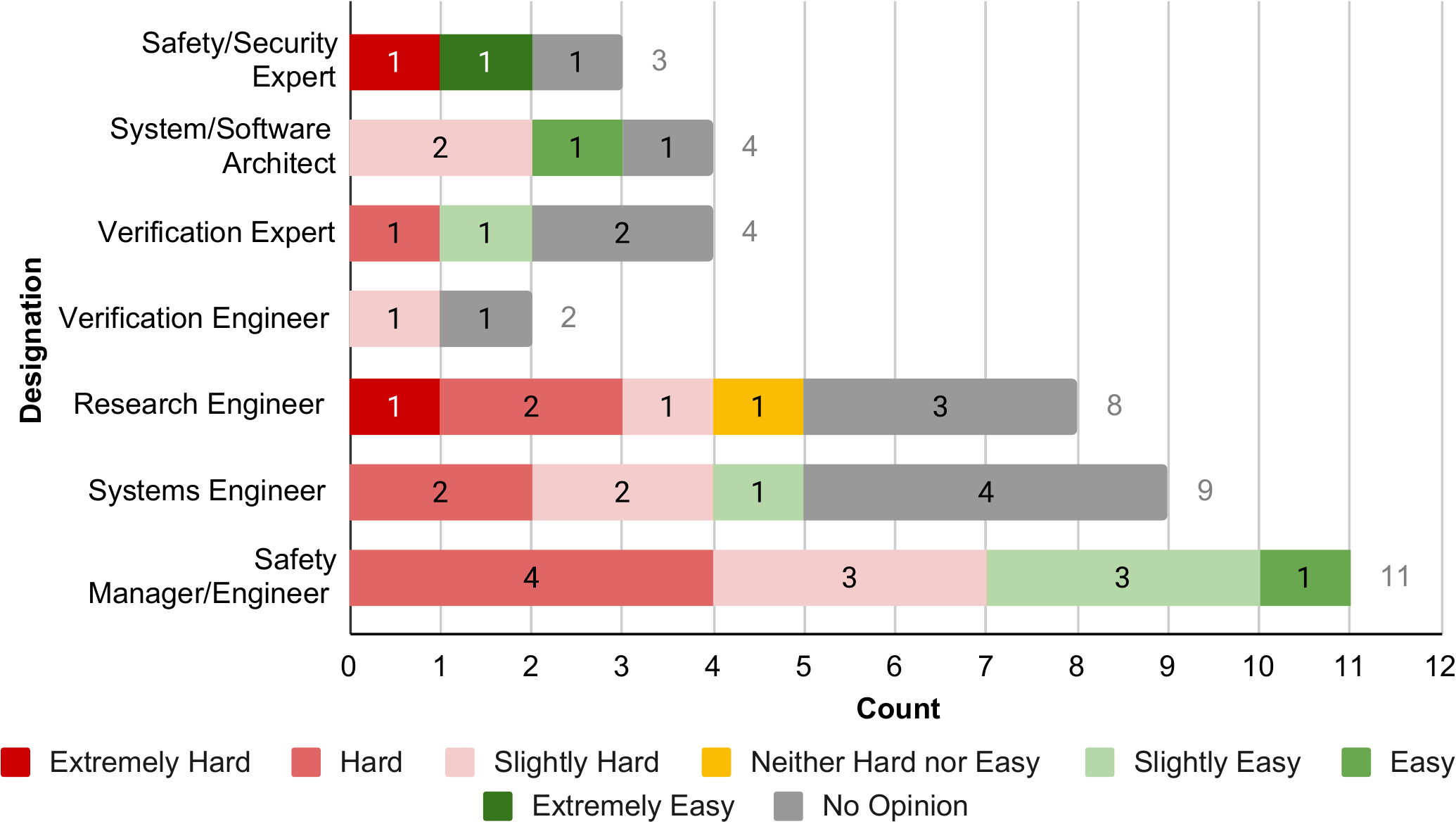}
	\caption{Results for the difficulty of maintaining refinement consistency grouped by the designations of the participants.}
	\label{fig:consistency_designation}
\end{figure}

\paragraph{Maintaining refinement consistency grouped by the participants' designations.}

\fig{consistency_designation} depicts the results for the difficulty of maintaining refinement consistency between the specifications and architecture grouped by the participants' designations (\cf \sect{phase1:participants:designation}). Focusing on valid responses (excluding \textit{no opinion}) from \textit{system/software architects}, \textit{safety/security expert}, and \textit{verification expert}, four participants answer \textit{harder} and three participants answer  \textit{easier}. Notably, among two \textit{system/software architects}, one answers \textit{extremely easy} and the another one answers \textit{extremely hard}. The participants who answer \textit{extremely easy} state that using model-based development it is easy to maintain the consistency with natural language requirements. Additionally, according to a \textit{system/software architect} who answers \textit{slightly hard}, maintaining refinement consistency \textit{\enquote{[s]cales with number of requirements. With graphical models like state machines one can keep complexity under control}}. Participants from the  remaining designations like \textit{systems engineer}, \textit{verification engineer}, \textit{safety manager/engineer}, and \textit{research engineer}, predominantly rated the problem as \textit{harder}. A \textit{safety manager/engineer} classifying the problem as \textit{slightly hard} states: \textit{\enquote{To maintain the consistency, we need to refine all software development phases results from requirements till test}}.

\paragraph{Summary.} Among all of the 29\,participants who answered this question, 20\,parti\-cipants (69\%) vote that maintaining consistency is \textit{harder}. From the free-text field responses, the majority of the participants finds that maintaining consistency is \textit{harder} when the system gets complex. Furthermore, a notable response is that model-based system development could ease maintaining consistency. 

\subsubsection{Verification of the Refinement Consistency}
\label{sec:phase1:refinement_check}

With question \qsfifeteen (\tab{survey_questions}), we collect the responses whether verifying refined specifications is \textit{harder} or \textit{easier}. Possible responses follow answer scale \lsthree in \tab{likert_scale}. Among the 41\,participants, eleven answer to have \textit{no opinion} while the remaining 30\,participants rate the difficulty of the verification problem. The overall result shown in \tab{consistency_check} is similar to the responses of \qsfourteen about maintaining consistency discussed in \sect{phase1:refinement_consistency}. The majority of 24 participants (80\% of responded participants) answer that verifying refined specifications is \textit{harder}. Further four participants answer it to be \textit{easier} and two participants answer, \textit{neither hard nor easy}. Notably, among the 24\,participants who answer \textit{harder}, 16 (66\%) answer \textit{hard} which takes the predominant count of all the given options.  

\begin{table}[b]
	\centering
	\caption{Difficulty of checking consistency when refining formal requirements for sub-components of a system architecture}
	\label{tab:consistency_check}
	\resizebox{0.8\textwidth}{!}{%
		\begin{tabular}{|l|l|l|l|l|l|l|l|l|}
			\hline
			Agreement & \begin{tabular}[c]{@{}l@{}}Extremely\\  Hard\end{tabular} & Hard & \begin{tabular}[c]{@{}l@{}}Slightly \\ Hard\end{tabular} & \begin{tabular}[c]{@{}l@{}}Neither Hard\\ nor Easy\end{tabular} & \begin{tabular}[c]{@{}l@{}}Slightly \\ Easy\end{tabular} & Easy & \begin{tabular}[c]{@{}l@{}}Extremely \\ Easy\end{tabular} & \begin{tabular}[c]{@{}l@{}}No \\ Opinion\end{tabular} \\ \hline
			Count     & 2 (5\%)                                                        & 16 (39\%)   & 6   (15\%)                                                     & 2 (5\%)                                                              & 2  (5\%)                                                       & 1 (2\%)    & 1     (2\%)                                                     & 11  (27\%)                                                   \\ \hline
		\end{tabular}%
	}
\end{table}

\begin{figure}[t]
	\centering
	\includegraphics[width=0.8\textwidth]{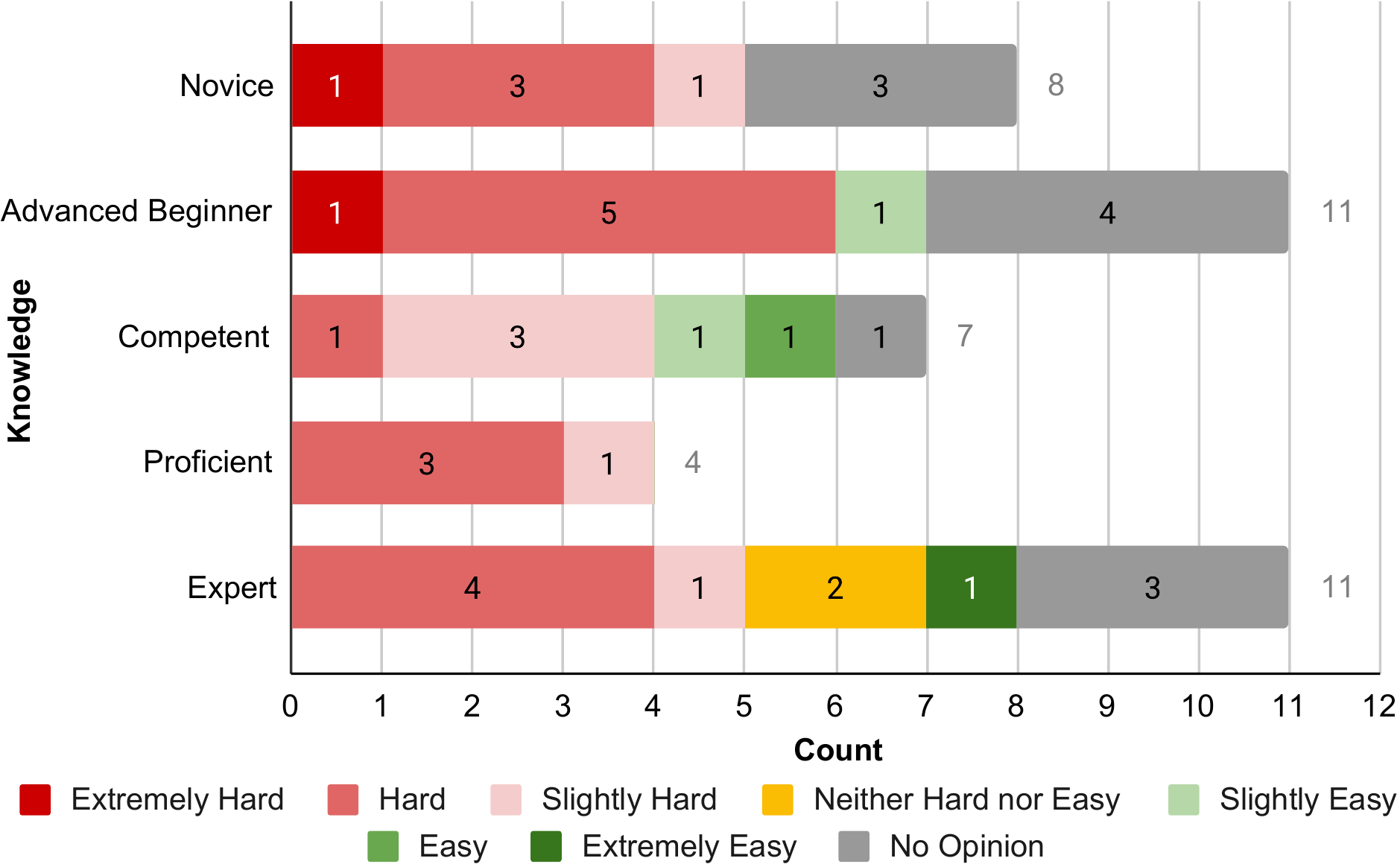}
	\caption{Results for the difficulty of verifying refinement consistency grouped by the participants' experience in formal methods.}
	\label{fig:consistency_check_knowledge}
\end{figure}

\paragraph{Verifying refinement consistency grouped by the participants' knowledge in formal methods.}

\fig{consistency_check_knowledge} depicts the results of rating the difficulty of verifying refinement consistency grouped by the participants' knowledge in formal methods (\sect{phase1:participants:knowledge}). None of the \textit{novice} and \textit{proficient} participants rate the problem of verifying refinement consistency as \textit{easier}. A \textit{proficient} participant mentions: \textit{\enquote{You can only rely on provisioned tools (model checker)}}. Among the rest, at least one participant answers that verifying refinement consistency is \textit{easier}, however, the majority of the answers rate it to be \textit{harder}. According to an \textit{advanced beginner} rating the problem as \textit{hard}, the difficulty depends on the formalization of requirements: \textit{\enquote{If you mean non-formal requirements, I think checking the consistency is extremely hard. If you mean formal requirements but linked to components, I think it is slightly hard. However, links to components help to understand the relationship between variables}}.

\begin{figure}[t]
	\centering
	\includegraphics[width=0.8\textwidth]{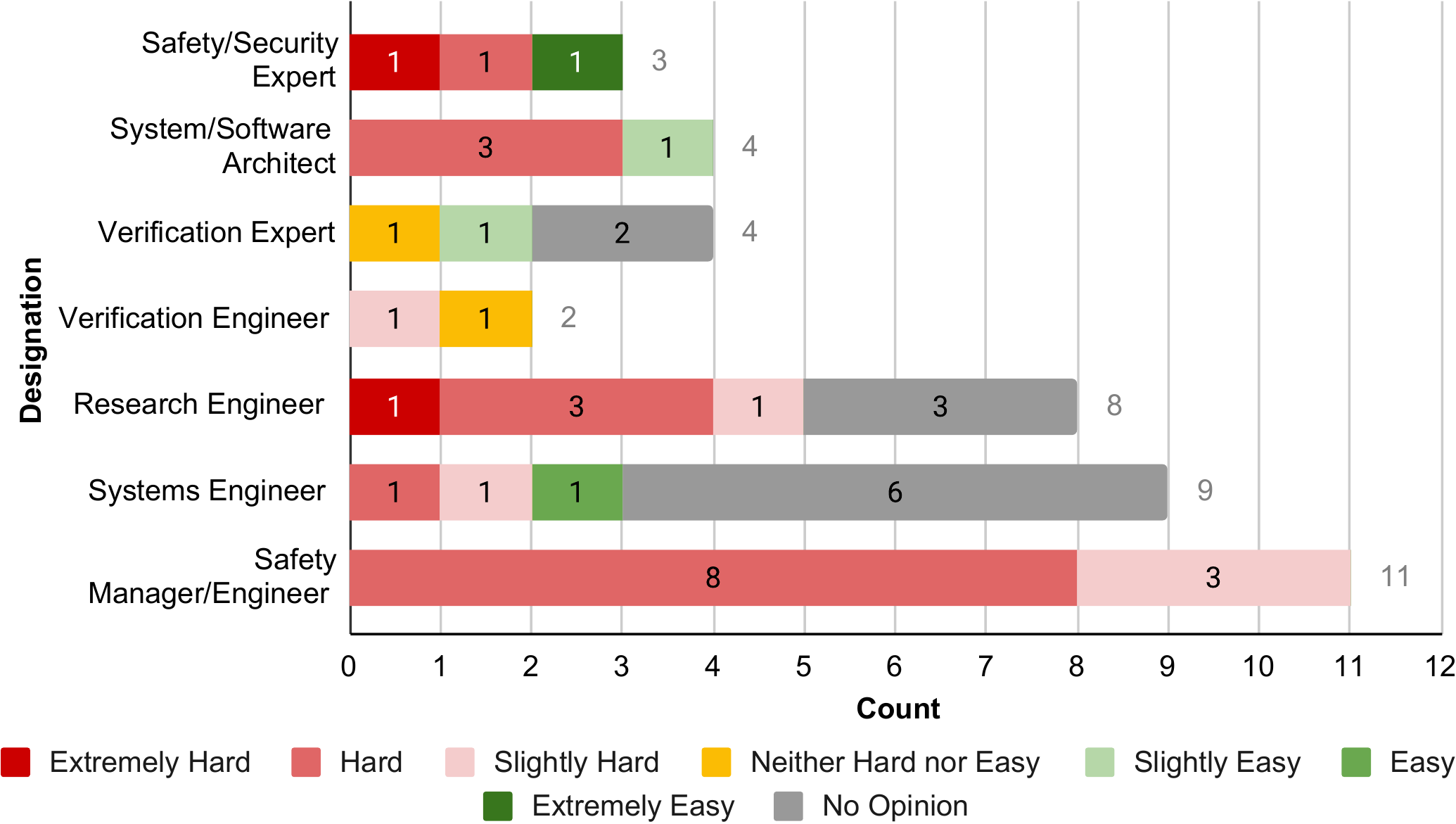}
	\caption{Results for the difficulty of verifying refinement consistency grouped by the participants' designations.}
	\label{fig:consistency_check_designation}
\end{figure}

\paragraph{Verifying refinement consistency grouped by the participants' designations.}

\fig{consistency_check_designation} depicts the results of rating the difficulty of verifying refinement consistency grouped by participants' designations (\sect{phase1:participants:designation}). Focusing on the designation of a \textit{safety/security expert}, one participant answers that verifying refinement consistency is \textit{extremely hard} while another participant answers it to be \textit{extremely easy}. One of the \textit{safety/security expert} states that \textit{\enquote{[e]specially when the requirements to be checked are distributed, without tool support, checking can be extremely tedious}}. With the designations of a verification engineer, safety manager/engineer, and research engineer as exceptions, all other designations have at least one participant rating the difficulty as \textit{easier}. A \textit{verification engineer} rating the problem as \textit{neither hard nor easy} mentions: \textit{\enquote{Ideally, the formal requirements should be specified in a way that they can be automatically checked by some tool – then it is easy}}. Additionally, a \textit{systems engineer} rating the problem as \textit{easy} states that \textit{\enquote{[i]t is easy as long as the requirements can be functionally separated. If different requirements have impact on a state behavior or contradicting safety goals, it is hard}}.

\paragraph{Summary.}

Among the 30\,participants who provide a response other than having \textit{no opinion}, 80\% perceive the verification of refinement consistency as a \textit{harder} problem. Furthermore, most participants who rated the problem as \textit{easier} highlight that the effort of verifying refinement consistency could be reduced by using well-formalized specifications and verification tools.

\subsection{Formal Verification Focusing on Safety}
\label{sec:phase1:fm_safety}

With questions \qssixteen to \qseighteen, we collect the participants' opinion on weather formal verification could support safety analysis and make a system safer.

\subsubsection{Using Formal Verification to Make Systems Safer}
\label{sec:phase1:fm_make_safer}

Particularly, with question \qssixteen we investigate the opinion of participants on whether using formal verification could make a system safer (response according to scale \lsfive in \tab{likert_scale} plus free-text comments). Aggregated responses are shown in \tab{fv}. A predominant  number of participants answers \textit{definitely} (14\,participants, 34\%) and \textit{very probably} (15\,participants, 37\%).

\begin{table}[b]
	\centering
	\caption{Results on whether using formal verification makes systems safer.}
	\label{tab:fv}
	\resizebox{0.8\textwidth}{!}{%
		\begin{tabular}{|l|l|l|l|l|l|l|l|l|}
			\hline
			Likelihood & Definitely & \begin{tabular}[c]{@{}l@{}}Very \\ Probably\end{tabular} & Probably & \begin{tabular}[c]{@{}l@{}}Neither Probably\\ nor Possibly\end{tabular} & Possibly & \begin{tabular}[c]{@{}l@{}}Probably \\ Not\end{tabular} & \begin{tabular}[c]{@{}l@{}}Definitely \\ Not\end{tabular} & \begin{tabular}[c]{@{}l@{}}No \\ Opinion\end{tabular} \\ \hline
			Count      & 14  (34\%)       & 15     (37\%)                                                  & 6  (15\%)      & 1       (2\%)                                                               & 4    (10\%)    & 0                                                       & 1  (2\%)                                                       & 0                                                     \\ \hline
		\end{tabular}%
	}
\end{table}

Most free-text comments collected by participants who answer \textit{definitely}, \textit{very probably}, or \textit{probably} highlight that formal verification is required to tackle the increasing system complexity and shorten development time. 
A \textit{research engineer}, whose answer is \textit{definitely}, states that \textit{\enquote{it can provide yet another level of surety about the safety of systems. The more sure we are the better it is. The human brain tries to save energy. As a human, we may overlook a lot of details when it comes to habitual/routine work. This is where formal methods can provide further surety about safer systems.}} A \textit{safety engineer}, whose answer is \textit{very probably}, highlights a further challenge in industry: \textit{\enquote{Personally, I think it is useful if the environment is open to this (\eg interested in the results and willing to spend resources on it) and you have the right well-trained people working on it. However, I’ve seen these ideas failing too many times in industry. I conclude that in most cases using formal verification is not yet ready for production.}} 

\begin{figure}[t]
	\centering
	\includegraphics[width=0.8\textwidth]{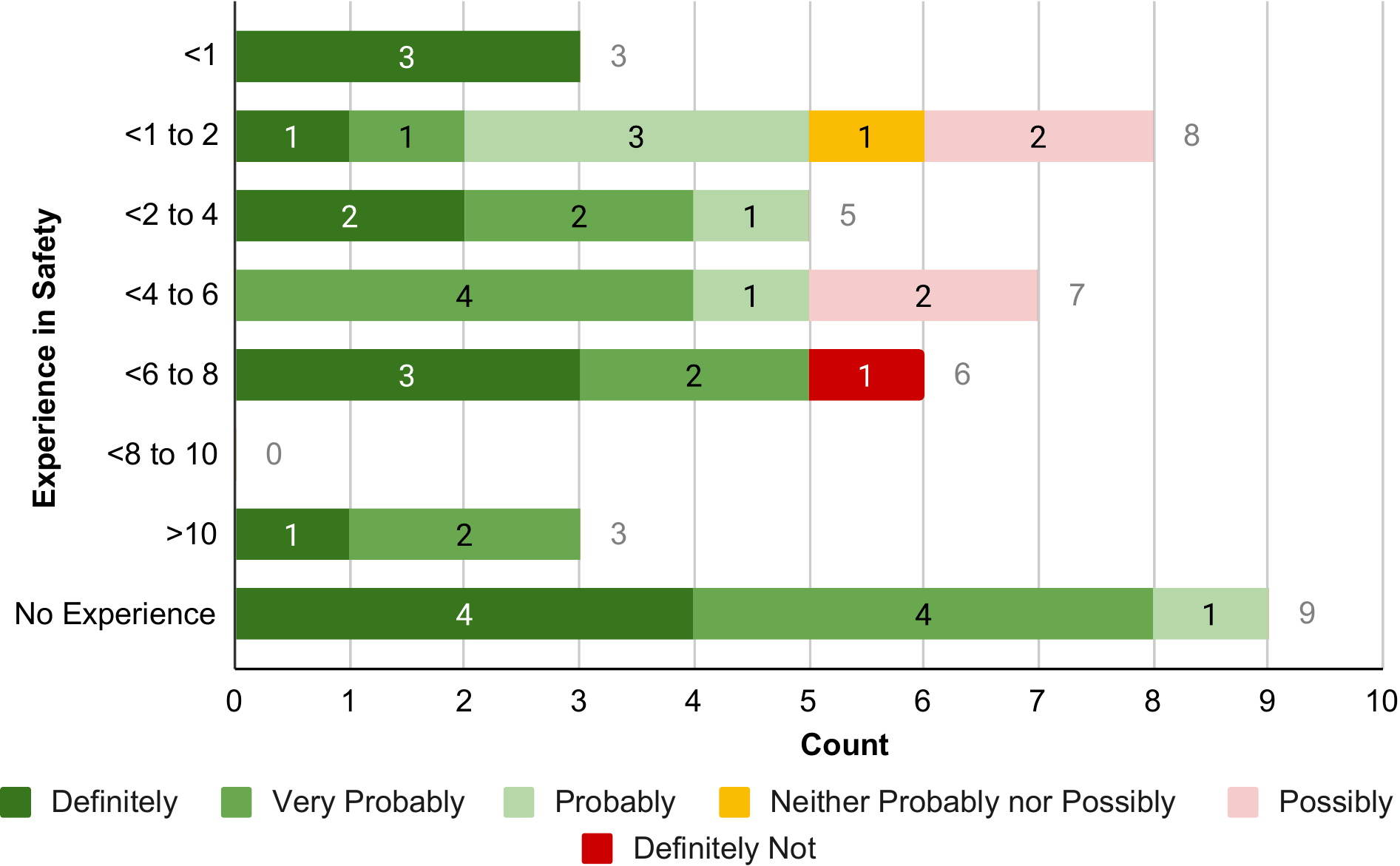}
	\caption{Results on whether using formal verification makes systems safer grouped by the participants' industrial experience in safety.}
	\label{fig:fv_experience}
\end{figure}

\paragraph{Relation to participants' experience in safety.} 

\fig{fv_experience} depicts the results of using formal verification to make systems safer based on the participants' industrial experience in safety.
Nine participants with \textit{no experience} present their view that using formal verification makes \textit{definitely}, \textit{very probably}, or \textit{probably} the systems safer. All of the participants with $<$1 year experience answer that using formal verification could make the system \textit{definitely} safer. An example statement by a \textit{systems engineer} without safety experience states that \textit{\enquote{[g]aps and errors can be found earlier, easier and with higher reliability.}} 
Considering the experience levels from 1 to $<$2 years and up to $>$10 years, the majority of participants answer with \textit{very probably}. A \textit{safety manager/engineer} with 2 to $<$4 years of experience who answered with \textit{very probably} mentions that formal verification \textit{\enquote{can help to make system consistent and reduce the human error.}} Furthermore, another \textit{safety manager/engineer} strengthens this view: \textit{\enquote{With higher system and organizational complexity combined with shorter development time, the need for more formal methods is increasing.}}

Focusing on the answers \textit{possibly} and \textit{definitely not}, two participants either with 1 to $<$2 or 4 to $<$6 years of experience answer \textit{possibly} while one participant with 6 to $<$8 years of experience answers \textit{definitely not}. A \textit{safety manager/engineer}, whose  answer is \textit{definitely not}, states the following: \textit{\enquote{We need to distinguish between two terms: Reliability and Safety. The formal verification are used to define the failures /Errors in software but not necessary that the all errors are safety-critical }}. Finally, a \textit{verification engineer}, whose answer is \textit{possibly}, highlights that formal verification \textit{\enquote{could potentially make them safer, but I think the effort is probably too high in relation to the benefit.}}

\paragraph{Summary.}
The majority answers of about 37\% of all participants respond that using formal verification makes \textit{very probably} a system safer while 34\% vote for \textit{definitely} safer. Most of the participants highlight that using automated formal verification methods could reduce a lot of manual work. However, a considerable number of participants also highlight that without proper training it is hard to imagine using formal verification in industry. 

\subsubsection{Formal Verification as an Add-on}
\label{sec:phase1:fm_addon}

With question \qsseventeen, we collect the participants' opinion whether formal verification could be a meaningful addition to the functional safety methods to ensure safety (possible answers follow scale \lsfive in \tab{likert_scale} and allow free-text comments). 
Among the 41\,participants, 39 have answered this question. Results are shown in \tab{fva}.
From these 39\,participants, 19 (49\%) answer that formal verification could \textit{definitely} be an add-on to functional safety methods, and eleven participants (28\%) consider this as \textit{very probably}. A \textit{safety manager/engineer}, whose answer is \textit{very probably}, states that \textit{\enquote{formal verification methods probably can [be] used to specify and verify the functional safety related properties of the system \eg time related issues such as Fault-Tolerance Time Intervals etc.}} The answers to question \qssixteen (\cf \sect{phase1:fm_make_safer}) correlates with the results shown in \tab{fva}.

\begin{table}[b]
	\centering
	\caption{Results for whether using formal verification could be a meaningful add-on to functional safety methods to ensure safety.}
	\label{tab:fva}
	\resizebox{0.8\textwidth}{!}{%
		\begin{tabular}{|l|l|l|l|l|l|l|l|l|}
			\hline
			Likelihood & Definitely & \begin{tabular}[c]{@{}l@{}}Very \\ Probably\end{tabular} & Probably & \begin{tabular}[c]{@{}l@{}}Neither Probably\\ nor Possibly\end{tabular} & Possibly & \begin{tabular}[c]{@{}l@{}}Probably \\ Not\end{tabular} & \begin{tabular}[c]{@{}l@{}}Definitely \\ Not\end{tabular} & \begin{tabular}[c]{@{}l@{}}No \\ Opinion\end{tabular} \\ \hline
			Count      & 19  (46\%)       & 11    (27\%)                                                   & 6  (15\%)      & 0                                                                     & 2   (5\%)     & 1   (2\%)                                                   & 0                                                         & 2  (5\%)                                                   \\ \hline
		\end{tabular}%
	}
\end{table}

\begin{figure}[b]
	\centering
	\includegraphics[width=0.8\textwidth]{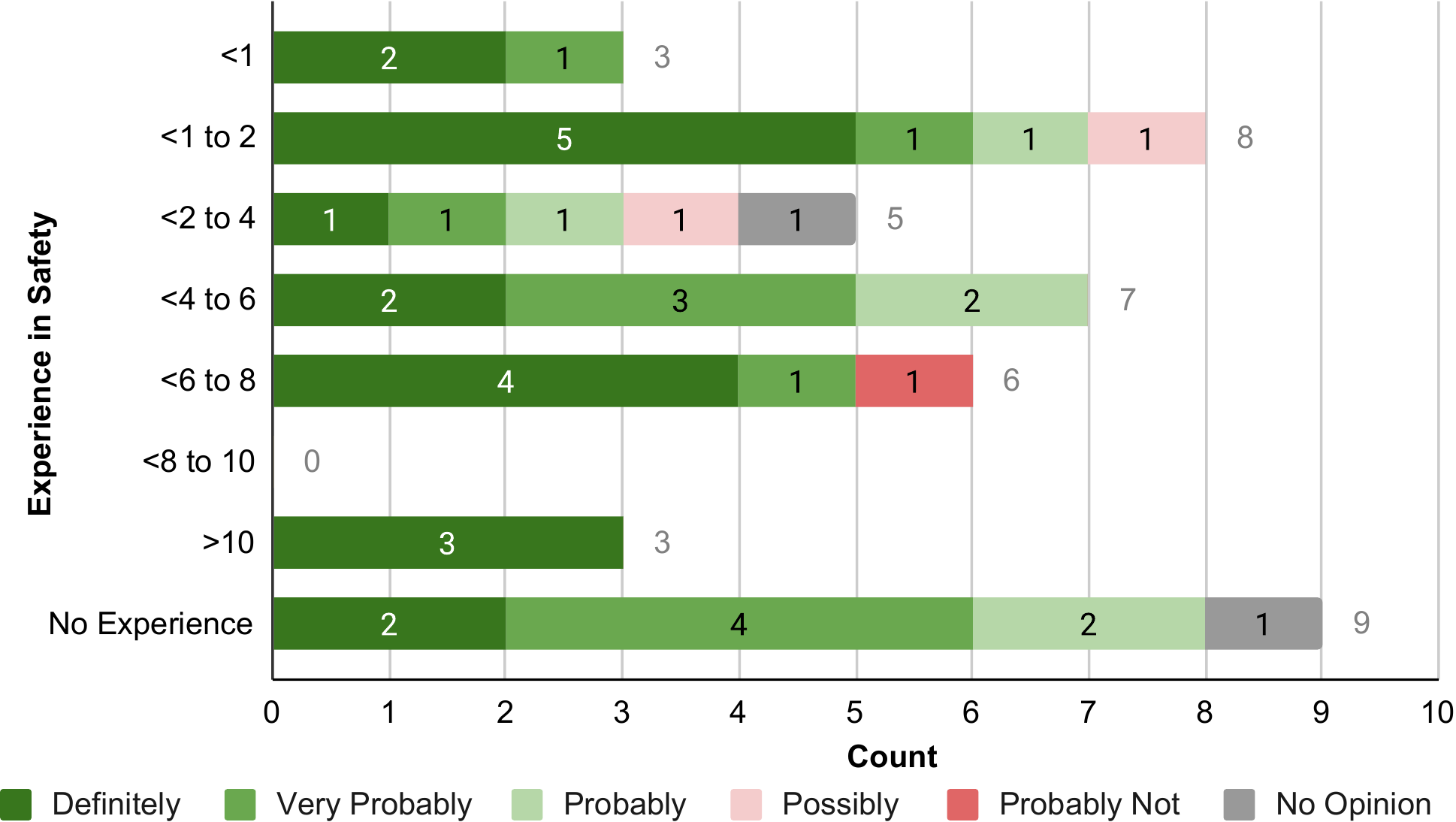}
	\caption{Results for whether using formal verification could be a meaningful add-on to the functional safety methods grouped by the participants' experience in safety.}
	\label{fig:fva_experience}
\end{figure}
\paragraph{Relation to participants' safety experience.}

\fig{fva_experience} depicts the results grouped by the participants' experience in safety (\cf \sect{phase1:participants:experience}). The predominant number of participants having an experience of  6 to $<$8 years and $>$ 10 years answer \textit{definitely}. A \textit{safety manager/engineer} with  6 to $<$8 years of experience states: \textit{\enquote{To me, formal verification is already a method to achieve functional safety. It’s just not applied widely. Formal verification itself cannot ensure safety, it can just contribute.}} Furthermore, a \textit{system/software architect} with $>$ 10 years of experience mentions the challenge that mostly formal verification is not scalable, and very costly. 
Notably, among all participants, only one participant who answers with \textit{probably not} and having between six and eight years of experience states the following:\textit{\enquote{We have two terms: functional safety addresses E/E failures in the systems and SOTIF (safety of intended functionality) addresses the safety issues in absence of E/E failures. Therefore, to ensure safety, we need to reduce the hazards which are related to E/E failures + Performance limitations.}}

\paragraph{Summary.}
19\,participants (49\% of the responding participants) deem formal verification \textit{definitely} to be an add-on to classical functional safety methods. From the free-text responses, most participants state that formal verification together with functional safety methods could address more system safety issues.

\subsubsection{Benefit of Identifying Inconsistent Specifications}
\label{sec:phase1:fm_inconsistent_benificial}

With question \qseighteen, we collect the participants' opinion whether identifying inconsistent formal specifications is beneficial for a safety analysis (possible answers are defined by scale \lsfive in \tab{likert_scale} and allow free-text comments). Forty participants  provided answers and one participant had \textit{no opinion} (\tab{inconsistent_safety}). The majority of 21\,participants (53\%) rates identifying inconsistent formal specifications \textit{definitely} beneficial for a safety analysis. The second most-called answer with 15 votes (38\%) is \textit{very probably}. A \textit{system/software architect} answering \textit{definitely} states that \textit{\enquote{it is beneficial because: (1)~it saves time and effort trying to resolve the inconsistencies in the future, and (2)~avoids possible security threat due to the vagueness in the specification.}}  A research engineer highlights that \textit{\enquote{naturally, safety critical application which rely on inconsistent formal specifications can never be guaranteed to function reliably.}}

\paragraph{Relation to participants' experience in safety.}

\begin{figure}[t]
	\centering
	\includegraphics[width=0.8\textwidth]{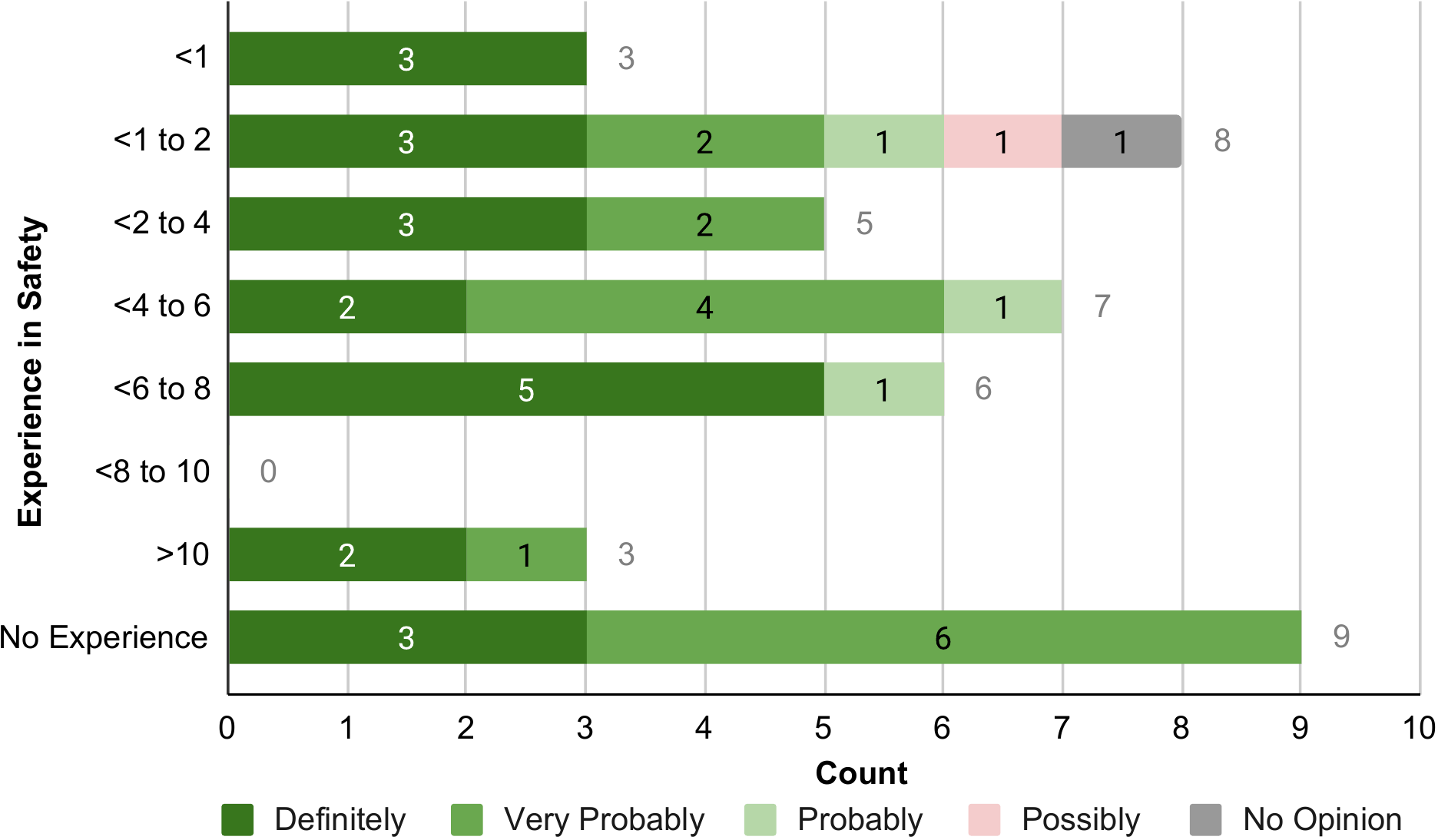}
	\caption{Results for whether identifying inconsistent specifications is beneficial for a safety analysis, grouped by the participants'  experience in safety.}
	\label{fig:inconsistent_safety_experience}
\end{figure}

\fig{inconsistent_safety_experience} depicts the results in relation to the participants' experience in safety. A majority of participants between 0 to 8 years and $>$10 years vote for \textit{definitely}. A \textit{research engineer} answering \textit{definitely} and with 2 to $<$4 years of experience states that \textit{\enquote{if formal methods were to be used extensively in the production domain, it would be very important because it is the starting point of the analysis and will have great influence on the results.}} Further, a \textit{safety manager/engineer} with 6 to $<$8 years experience mentions that \textit{\enquote{[identifying] inconsistent specification is definitely beneficial, not so much for the safety analysis but for the safety of the product itself. Also, it is not limited to the safety but general performance of the product.}} Looking at all benefits, a \textit{safety manager/engineer} answering \textit{definitely} and with 4 to $<$6 years experience mentions that \textit{\enquote{it is indeed very helpful. However, the harder part is to write correct (partial) specifications in the first place. If you fail to do so, there is no benefit from making your (flawed) specification consistent.}}

\begin{table}[t]
	\centering
	\caption{Results for whether identifying inconsistent specifications is beneficial for a safety analysis.}
	\label{tab:inconsistent_safety}
	\resizebox{0.8\textwidth}{!}{%
		\begin{tabular}{|l|l|l|l|l|l|l|l|l|}
			\hline
			Likelihood & Definitely & \begin{tabular}[c]{@{}l@{}}Very\\  Probably\end{tabular} & Probably & \begin{tabular}[c]{@{}l@{}}Neither Probably\\  nor Possibly\end{tabular} & Possibly & \begin{tabular}[c]{@{}l@{}}Probably\\  Not\end{tabular} & \begin{tabular}[c]{@{}l@{}}Definitely\\  Not\end{tabular} & \begin{tabular}[c]{@{}l@{}}No \\ Opinion\end{tabular} \\ \hline
			Count      & 21   (51\%)      & 15   (37\%)                                                    & 3  (8\%)      & 0                                                                       & 1   (2\%)     & 0                                                       & 0                                                         & 1  (2\%)                                                   \\ \hline
		\end{tabular}%
	}
\end{table}

\vspace{-1em}
\paragraph{Summary.}

53\% of the participants who provided responses see that identifying inconsistent formal specifications could \textit{definitely} be beneficial for a safety analysis. Most of the free-text comments highlight that identifying inconsistencies could save time and effort very early and improve system safety.

\vspace{-3em}
\subsection{Using Formal Verification}
\label{sec:phase1:opinion:using}

The questions \qseight, \qsnineteen, and \qstwenty (Table~\ref{tab:survey_questions}) collect the participants' opinions on using formal verification in general, imagining using formal methods if understanding formal notations is eased, and using formal methods in real-world development. The responses to these questions are discussed in the following.

\subsubsection{Opinions on Using Formal Verification in General}
\label{sec:phase1:opinion:using_fm}

Question \qseight collects opinions of the participants on using formal verification with a free-text field. The responses are categorized based on the participants' designations.

\paragraph{System/Software Architects and Safety/Security Experts.}

A \textit{system/software architect} with their knowledge rated as \textit{expert} advises to \textit{\enquote{not try to model your whole system with formal methods. Model relevant aspects and check them.}} In addition, a \textit{system/software architect} as an \textit{advanced beginner} states, that \textit{\enquote{[formal methods] are helpful and can bring great help if used properly. However, there could be some effort initially for translating existing specifications into the representation or notation used by formal verification tools.}} Further two \textit{system/software architects} mention that with the required knowledge it is easy to use formal tools but still the question remains whether it scales to industrial systems.

Responses from \textit{safety/security experts} show that they are interested in different perspectives. For example, a \textit{safety/security expert} who is an \textit{advanced beginner} states the expectation that \textit{\enquote{one very promising solution is needed to compose/decompose safety requirements for component based development.}}
Further, an \textit{expert} highlights that they only \textit{\enquote{see rare use cases, since the benefit most often does not outweigh the costs (yet). This is, however, different when the formal verification is practically hidden and specification is intuitive. See for example type systems for programming languages checked by compilers without support from the user.}}

\paragraph{Verification Experts and Verification Engineers.}

The responses received by verification experts and engineers indicate that they are mainly interested in verifying system architectures and requirements. A \textit{verification expert} highlights that \textit{\enquote{if hard statements about correctness are required, [formal methods] are inevitable if they can be used. In addition, they are very good at finding corner cases that are hard to find (\eg transient errors resulting from concurrency).}}
Further, a \textit{verification engineer} states that formal methods \textit{\enquote{should be used for highly critical parts (only). I’m not sure if it is useful to apply it on higher levels (e.g. system architecture). That might help to get a consistent picture on that level, but will probably not help to ultimately build a safe system, as the properties can usually not be checked on implementation level.}}

\paragraph{Safety Managers/Engineers, Systems Engineers, and Research Engineers.}

Most of the \textit{safety managers/engineer} and \textit{systems engineer} are interested to use formal methods, but the complexity of using and understanding formal tools stands as a barrier. For example, a \textit{safety engineer} states that formal methods \textit{\enquote{can bring a huge benefit to safety engineering, but some fundamental challenges remain (competency of engineers applying formal methods and tools).}} Furthermore, according to the statement by a \textit{systems engineer}, \textit{\enquote{formal verification is powerful, but to get the acceptance we have to keep the formal stuff away from the users. E.g., Astree, Polyspace, QA-C all find defects, that are nearly impossible to detect by hand written tests. Astree and Polyspace find more than QA-C, still, many projects use QA-C, as it is extremely easy to use even for people without knowledge in formal methods.}} In addition, a research engineer highlights a crucial challenge: \textit{\enquote{I do think it makes sense to apply formal verification especially on highly complex and safety-critical system, but the hurdle might be very high for its wide application. We will require not only safety engineers to understand the method, notation, syntax, \etc but also \eg system designers, software developers. If external certification authority is involved in the certification process, it might be an additional challenge to present the safety case (unless they are experts in formal verification).}}

\paragraph{Summary.}
The majority of collected responses are very positive in using formal verification to improve the system safety and design. But, on other hand understanding and scalability of verification (tools) remain as obstacles.

\subsubsection{Opinions on Using Formal Verification if Understanding of Formal Notations is Eased}
\label{sec:phase1:opinion:notation}

Question \qsnineteen collects the opinion of participants on whether making formal notations more understandable could improve the usage of formal methods. The possible answers follow scale \lsfive in \tab{likert_scale} and can be extended with free-text comments. All of the 41\,participants provided responses, which are shown in \tab{fmu}. The majority of participants (23, \ie 56\%) answer that making formal notations more understandable could \textit{definitely} improve the usage of formal methods. Eleven participants (27\%) estimate that it will \textit{very probably} provide an improvement. A verification expert answering \textit{definitely} highlights that \textit{\enquote{increasing understandability of formal specifications is key to bringing them into a more wide-spread use. See for example \cite{Gladisch0HOVP19} lessons learned on SBT and the STL specifications.}} 
Finally, four participants expect that easing understanding of notations will \textit{probably} improve the usage of formal verification while and one participant each votes for \textit{neither probably nor possibly}, \textit{possibly}, and \textit{probably not}.

\begin{table}[b]
	\centering
	\caption{Results of using formal methods if understanding of formal notations is made easier.}
	\label{tab:fmu}
	\resizebox{0.8\textwidth}{!}{%
		\begin{tabular}{|l|l|l|l|l|l|l|l|l|}
			\hline
			Likelihood & Definitely & \begin{tabular}[c]{@{}l@{}}Very\\  Probably\end{tabular} & Probably & \begin{tabular}[c]{@{}l@{}}Neither Probably\\  nor Possibly\end{tabular} & Possibly & \begin{tabular}[c]{@{}l@{}}Probably\\  Not\end{tabular} & \begin{tabular}[c]{@{}l@{}}Definitely\\  Not\end{tabular} & \begin{tabular}[c]{@{}l@{}}No \\ Opinion\end{tabular} \\ \hline
			Count      & 23  (57\%)       & 11   (27\%)                                                     & 4    (10\%)     & 1   (2\%)                                                                    & 1   (2\%)     & 1   (2\%)                                                    & 0                                                         & 0                                                     \\ \hline
		\end{tabular}%
	}
\end{table}

\begin{figure}[t]
	\centering
	\includegraphics[width=0.8\textwidth]{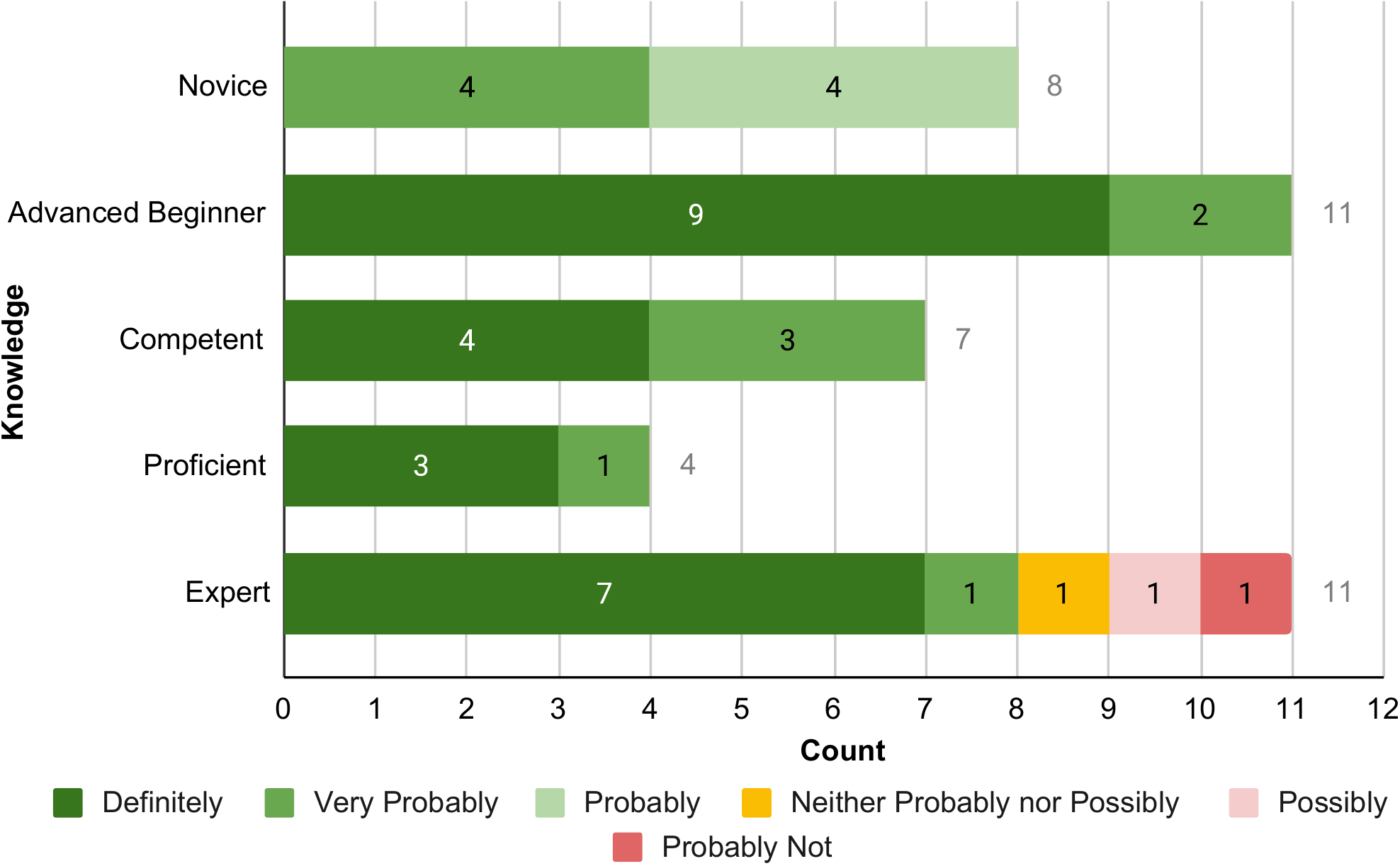}
	\caption{Results for using formal methods if understanding of formal notations is made easier, grouped by the participants knowledge in formal methods.}
	\label{fig:fmu_knowledge}
\end{figure}

\paragraph{Understanding of formal notations is made easier based on formal methods knowledge.}

\fig{fmu_knowledge} depicts the results of whether making formal notations more understandable could improve the usage of formal methods considering the participants' knowledge in formal methods (\cf  \sect{phase1:participants:knowledge}).
Except of \textit{novices}, the majority of participants rated from \textit{advanced beginners} to \textit{experts} expect a \textit{definite} improvement. A highlighting statement from a \textit{safety engineer} with knowledge rated as an \textit{advanced beginner} states that, however, \textit{\enquote{it is not just the decision of the safety team, also the System, SW and maybe the HW experts and also testers needs to understand it to make the investment in formal methods reasonable.}} In addition, a \textit{safety engineer} states the following: \textit{\enquote{To me, understandability is one of the main reasons that it [formal verification] is not applied, besides the huge initial effort (modeling and specifying a system formally).}}

From participants with \textit{expert} knowledge in formal methods, one participant each answers with \textit{neither probably nor possibly}, \textit{possibly}, and \textit{probably not}. From the collected free-text comments, most participants consider the formalization of system models and specifications from informal systems descriptions and requirements as challenging.
For example, a participant answering with \textit{neither probably nor possibly} highlights that \textit{\enquote{the problem is not related to method itself but it is related to how you use the method and what are the inputs, and how you understand the system architecture and artifacts. Also, how you apply the method and so on.}} A further participant answering with \textit{possibly} mentions that \textit{\enquote{the formal notation has to be very simple, it has to [be] comprehensible including small details, and it has to be sufficiently flexible so it can be used in unforeseeable use cases.}} A participant answering with \textit{probably not} states that \textit{\enquote{specifying a formal model is hard. If you can manage that, you can also manage the notation.}}

\paragraph{Summary.}

56\% of all participants think that making formal notations more understandable could \textit{definitely} improve the usage of formal methods. However, understanding and formalizing systems and requirements in the first place remains still a barrier to use formal methods. 

\subsubsection{Opinions on Using Formal Methods in Real-World Development}
\label{sec:phase1:opinion:real_world}

With question \qstwenty, we collect the opinions of participants on whether formal methods are usable in real-world development processes. Possible response follow scale \lsfive in \tab{likert_scale} and could be extended by free-text comments. While one participant has \textit{no opinion}, the remaining 40\,participants provide responses shown in \tab{rwd}. Predominantly, the answers are positive with most participants (13, 33\%) answering with \textit{probably}, 11\,participants (28\%) with \textit{definitely}, and ten participants (25\%) with \textit{very probably}. Only one participant each answers with \textit{neither probably nor possibly} and \textit{definitely not}.

\begin{table}[b]
	\centering
	\caption{Results of using formal methods in real-world development.}
	\label{tab:rwd}
	\resizebox{0.8\textwidth}{!}{%
		\begin{tabular}{|l|l|l|l|l|l|l|l|l|}
			\hline
			Likelihood & Definitely & \begin{tabular}[c]{@{}l@{}}Very\\  Probably\end{tabular} & Probably & \begin{tabular}[c]{@{}l@{}}Neither Probably\\  nor Possibly\end{tabular} & Possibly & \begin{tabular}[c]{@{}l@{}}Probably\\  Not\end{tabular} & \begin{tabular}[c]{@{}l@{}}Definitely\\  Not\end{tabular} & \begin{tabular}[c]{@{}l@{}}No \\ Opinion\end{tabular} \\ \hline
			Count      & 11 (27\%)        & 10     (25\%)                                                 & 13 (32\%)       & 1  (2\%)                                                                     & 4   (10\%)     & 0                                                      & 1  (2\%)                                                        & 1    (2\%)                                                  \\ \hline
		\end{tabular}%
	}
\end{table}

\begin{figure}[b]
	\centering
	\includegraphics[width=0.8\textwidth]{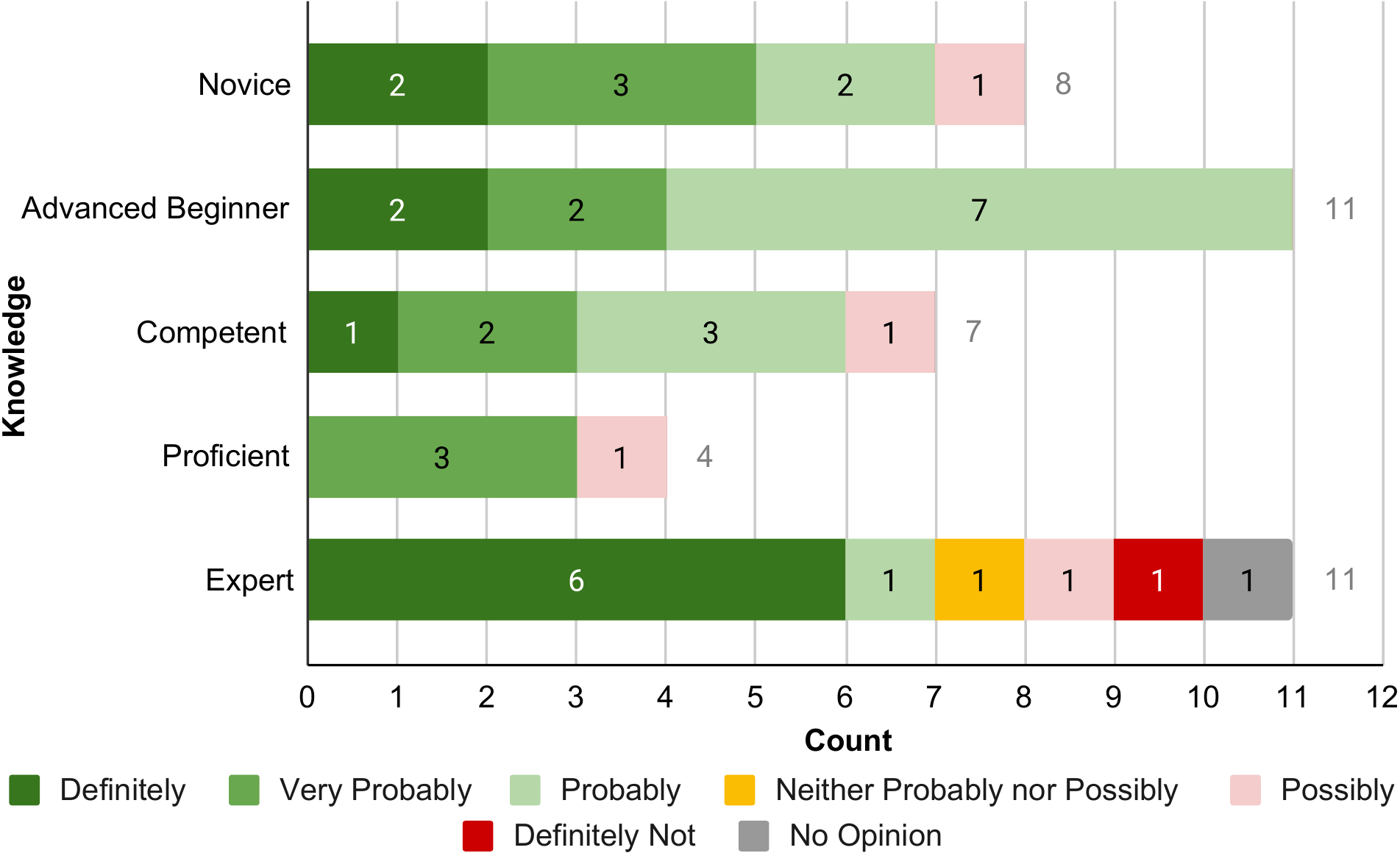}
	\caption{Results of whether formal methods are usable in real-world development processes grouped by the participants' knowledge in formal methods.}
	\label{fig:rwd_knowledge}
\end{figure}

\paragraph{Usability in real-world development processes based on formal methods knowledge.}

\fig{rwd_knowledge} depicts the results of whether formal methods are usable in real-world development processes considering the participants' knowledge in formal methods (\cf  \sect{phase1:participants:knowledge}).
Most answers received as free-text comments highlight the interest to use formal methods, however, understanding and familiarizing with the notations and tools still need to be improved. From the majority of 13 answers selecting \textit{probably}, seven participants rated their knowledge as an \textit{advanced beginner}. Such a participant highlights that \textit{\enquote{the huge WHY NOT in my opinion is the frontloading/enabling. Application of formal methods usually requires more time in the beginning to become familiar with the formalization method (and maybe tools). Often, representation is not intuitive on the first glance and requires a few people who actually ‘dig themselves into the problem’. This makes it unattractive for fast applications/scouting. As soon as you started with another type of analysis in the beginning, the technical debt still changing to a more formal method becomes higher and higher. Thus, in my opinion, formal methods can be very helpful but require a good visualization and explanation as – in the beginning – these methods are often an investment of time for a later quality return.}} 
Furthermore, notably, six of the 11 participants answering with \textit{definitely} rated their knowledge as an \textit{expert}. For instance, such participants see the benefits but also the prerequisites of using formal methods: \textit{\enquote{[Formal methods] are usable and can provide a huge benefit. However, the necessary competences have to be created, the management has to support this.}}

\paragraph{Usability in real-world development processes based on designations.}

\begin{figure}[b]
	\centering
	\includegraphics[width=0.8\textwidth]{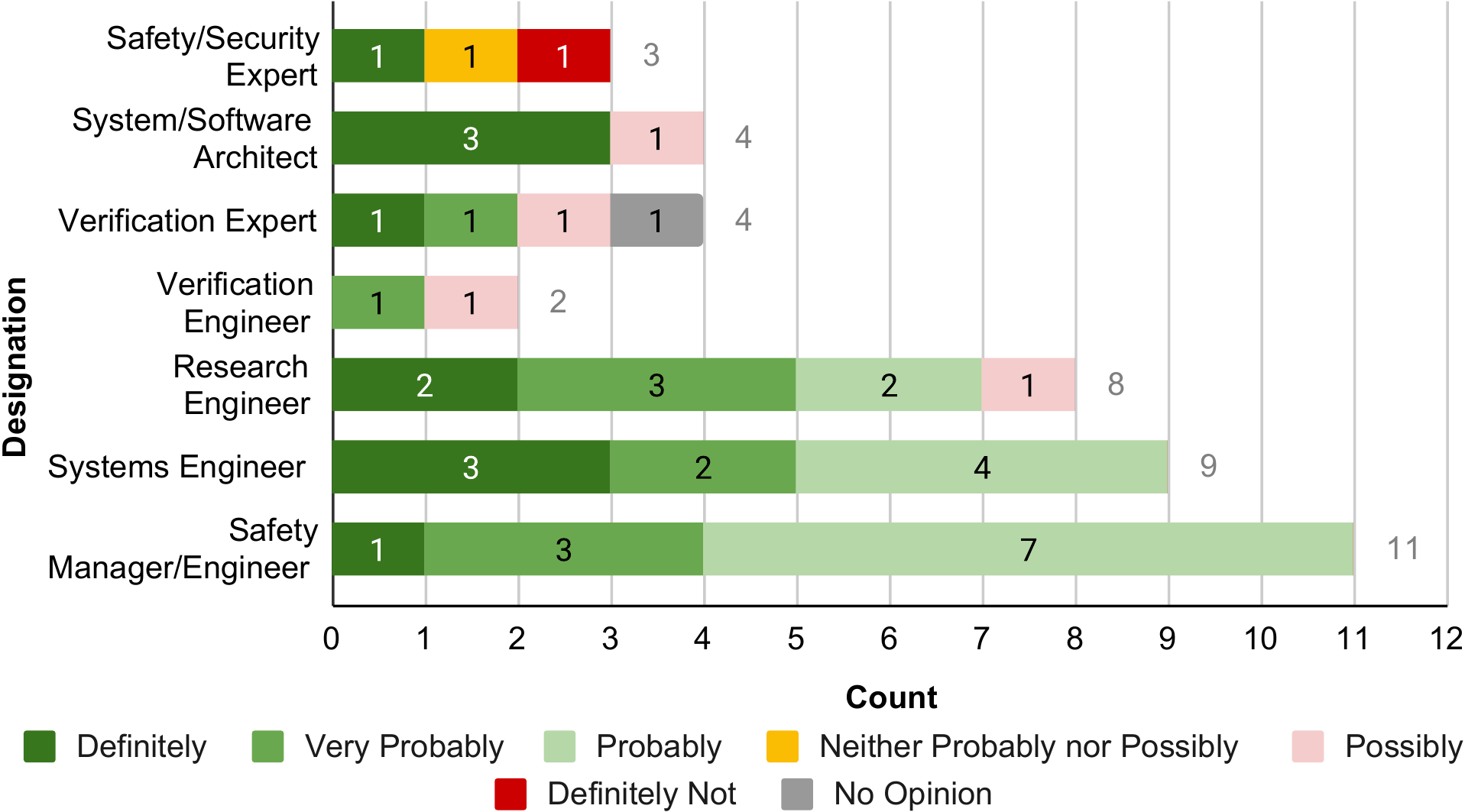}
	\caption{Results of whether formal methods are usable in real-world development processes grouped by the participants' designations.}
	\label{fig:rwd_designation}
\end{figure}

\fig{rwd_designation} depicts the results of whether formal methods are usable in real-world development processes grouped by participants' designations (\sect{phase1:participants:designation}).
Among 11\,participants answering with \textit{definitely}, eight participants are \textit{system/software architects}, \textit{systems engineers}, and \textit{research engineers}. An example statement from a \textit{systems engineer} is that \textit{\enquote{formal methods are already in use, \eg in tools like Astree, Polyspace, \etc This only works, if we try to make it as easy for the user as possible. I don’t see people formalizing requirements but, I think if we lower the hurdle this might also change on the left side of the V.}} Similarly, eight of ten participants answering with \textit{very probably} are \textit{systems engineers}, \textit{safety manager/engineers}, and \textit{research engineers}. A \textit{verification engineer} highlights the necessity of easing the semantics: \textit{\enquote{An engineer can make use of such tools only if it serves a user base. Formal methods offer great metrics to assert real-world processes, however it is limited today by its very semantic nature. One needs something on top to make this human understandable.}} Further, 11 of 13\,participants answering with \textit{probably} are \textit{systems engineers} and \textit{safety manager/engineers}. A \textit{system engineer} states that \textit{\enquote{it is necessary to integrate currently used/recommended methods. Changes from well known methods at software engineering team must be as easy as possible.}}

Apart from the positive opinions, a \textit{safety/security expert} highlights that it \textit{\enquote{depends on what you consider formal methods. Type systems and similar methods are used heavily by compilers in the background already. Formal specifications in the form of LTL formulas for complex systems seem extremely challenging from a cost/benefit point of view}}. 
All of the participants answering with \textit{possibly} consider scalability as the major barrier. For example, a \textit{verification expert} states that \textit{\enquote{scalability, time cost and, required technical knowledge constitute a  barrier to usability.}} Finally, the only participant answering with \textit{definitely not} is a \textit{safety/security expert} who states that \textit{\enquote{only a small amount of people will accept it, as too many people are not able to understand notations used.}}

\paragraph{Summary.}

To sum up, 34 out of all 41\,participants (83\%) have positive opinions regarding the use of formal methods in real-world development processes (formal methods are \textit{definitely}, \textit{very probably}, and \textit{probably} usable in such a context). According to the free-text comments, the main barriers for the application of formal methods in the real world are that they are not scalable, not suitable for all kinds of engineers, not easily understandable, and not in the state of ``plug and play'', yet.

\section{One-Group Pretest-Posttest Experiment (Part\,2): Results and Analysis}
\label{sec:phase2}

The questionnaire listed in \tab{onegroup}, gathered from 13\,participants, corresponds to \textit{Part\,2} of our study, the one-group pretest-posttest experiment. The results of this experiment are summarized in the following.

\subsection{Participants}
\label{sec:phase2:participants}

We collected demographic information about the participants using the questions \dqone and \dqtwo. 

\subsubsection{Participants' Experience in Formal Methods}
\label{sec:phase2:participants:experience}

\begin{figure}[h]
	\centering
	\includegraphics[width=0.8\textwidth]{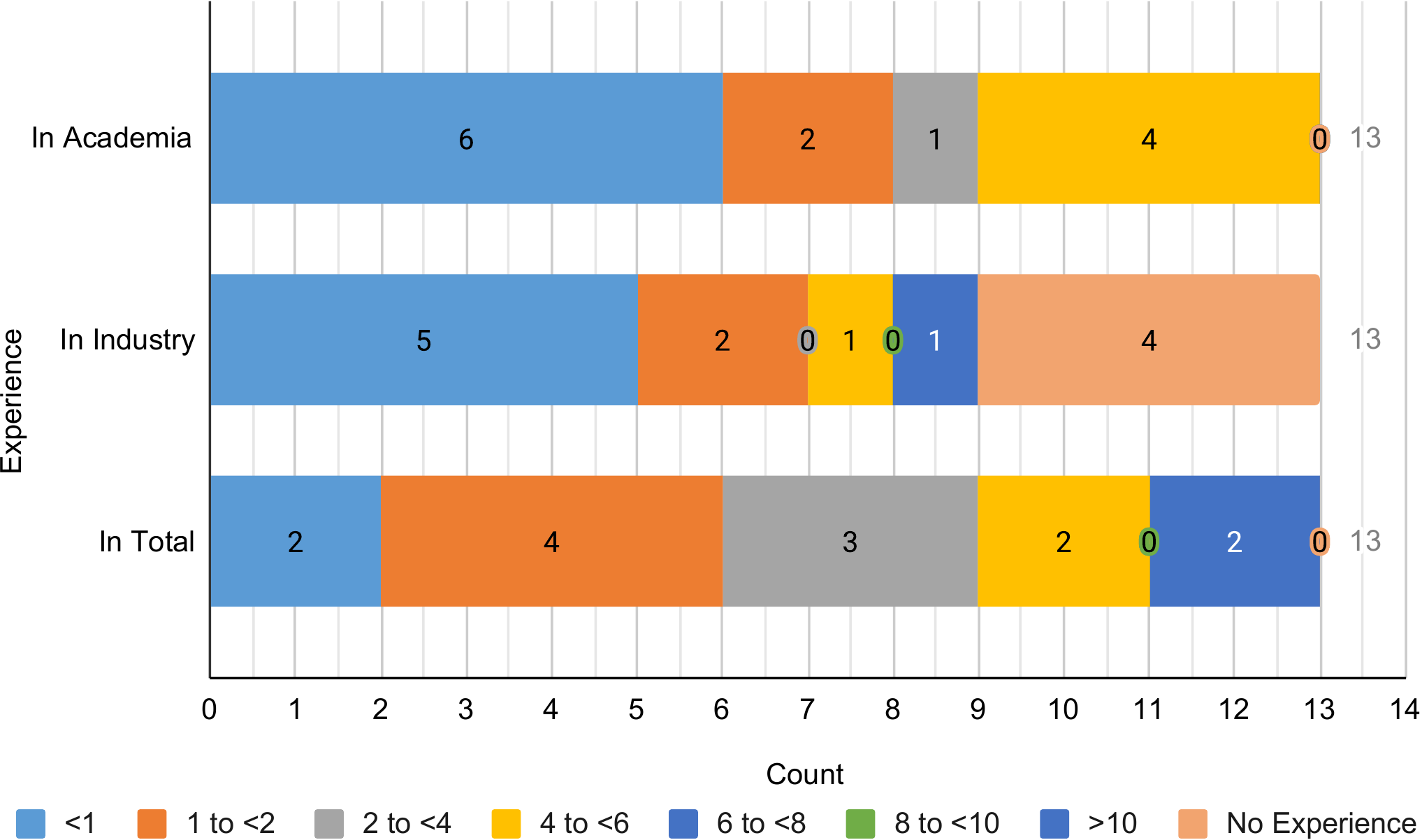}
	\caption{Experience of the participants in formal methods gained in academia, industry, and overall/in total.}
	\label{fig:experience2}
\end{figure}

The experience of participants in using formal methods is collected through question \dqtwo (\tab{onegroup}). Possible answers correspond to scale \lstwo (\tab{likert_scale}). Likewise to \qsthree in \tab{survey_questions} (\cf \sect{phase1:participants:experience}), the participants are asked to fill-in their experience in formal methods gained individually in academia and industry as well as their overall experience. \fig{experience2} shows the responses.

All of the 13\,participants have gained experience in formal methods in academia, industry, or both. Particularly, all participants have academic experience. Six participants have $<$1 year and four participants have 4 to $<$6 years of experience gained in academia. 
Concerning industrial experience, four participants have \textit{no experience} and five participants have $<$1 year of experience. 
Finally, focusing on the overall experience, four participants have 1 to $<$2 years of experience, and three participants have 2 to $<$4 years of experience. Furthermore, two participants each have $<$1, 4 to $<$6, and $>$10 years of experience.

\subsubsection{Participants' Knowledge in Formal Methods}
\label{sec:phase2:participants:knowledge}

\begin{figure}[h]
	\centering
	\includegraphics[width=0.8\textwidth]{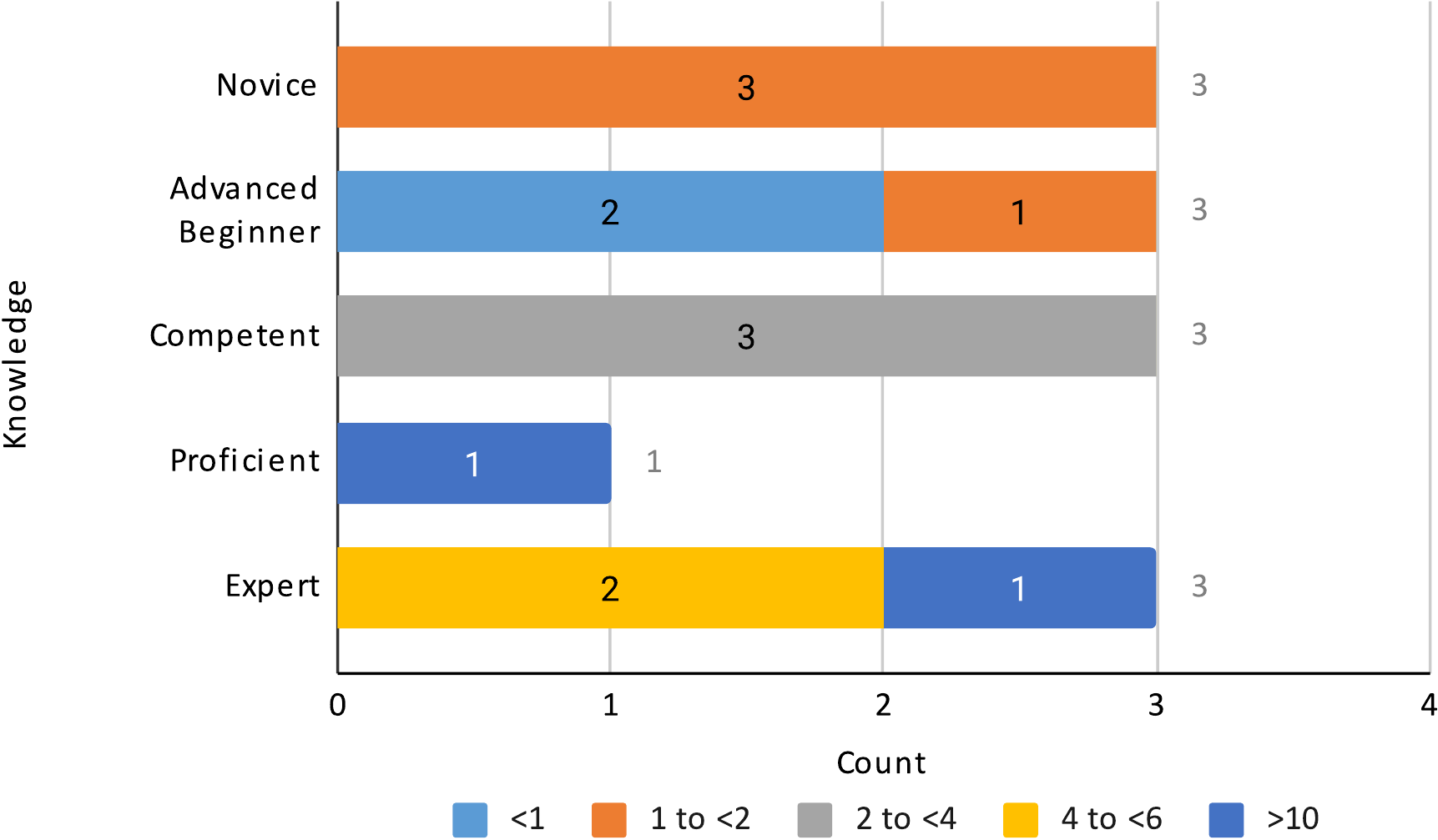}
	\caption{Participants' knowledge in formal methods, along with their total experience in formal methods in terms of years.}
	\label{fig:expertise2}
\end{figure}

Question \dqone (\tab{onegroup}) asks the 13\,participants to rate their knowledge in formal methods (answer according to scale \lsone in \tab{likert_scale}). The responses of the participants are shown in \fig{expertise2}. Similar to the result of \qstwo in \textit{Part\,1} (\cf \sect{phase1:participants:knowledge}), all participants have rated their knowledge within the scale \textit{novice} to \textit{expert} while no answers are received for the scale \textit{mastery} and \textit{practical wisdom}. Thus, both \textit{mastery} and \textit{practical wisdom} scales will not be considered for the rest of the discussion. Among the 13\,participants, three participants each rated themselves as \textit{novices}, \textit{advanced beginners}, \textit{competent}, and \textit{experts} while only one participant is rated as \textit{proficient}. 
\fig{expertise2} depicts the participants' knowledge together with their total experience in formal methods. All participants having $<$2 years of experience are rated as \textit{novice} and \textit{advanced beginners}, 2 to $<$4 years as \textit{competent}, and 4 to $<$6 as well as $>$ 10 years as \textit{proficient} and \textit{experts}.

\subsection{Use Case for the Pretest}
\label{sec:phase2:use_case:pretest}

In this section, we discuss the use case for the pretest, a formally specified airbag system, that has been presented to the participants, as well as the collected responses for questions \tqone and \tqtwo (\cf \tab{onegroup}).

\subsubsection{Airbag System}
\label{sec:phase2:use_case:pretest:as}

\begin{figure}[t]
	\centering
	\includegraphics[width=\textwidth]{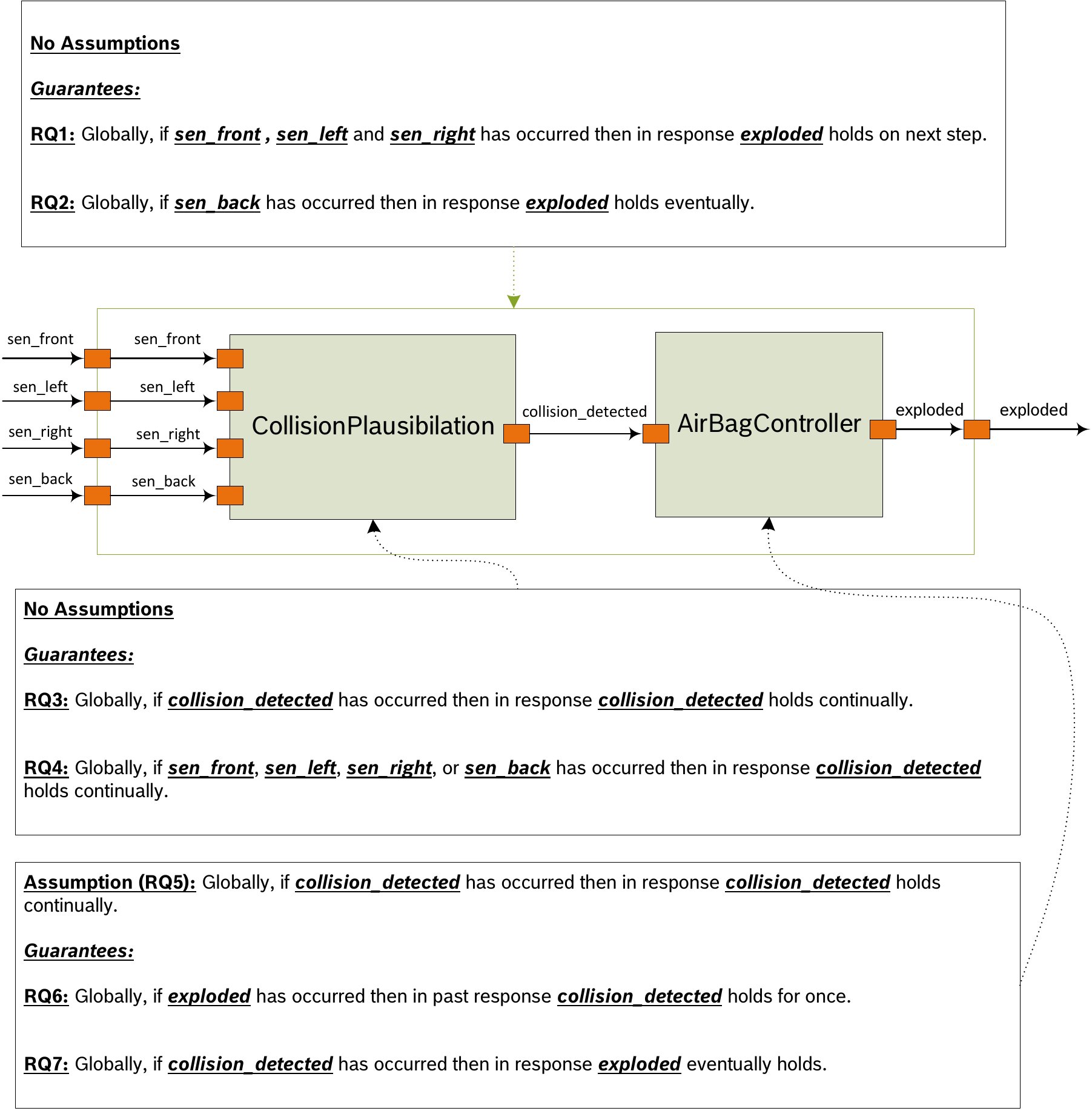}
	\caption{The component model and specifications of the airbag system.}
	\label{fig:airbagSystem}
\end{figure}

The component model and specifications of the airbag system used in the pretest are shown in \fig{airbagSystem}. An explanation of the component model and their corresponding specifications were provided in-detail in a video to the participants. The airbag system consists of one parent component and two sub-components, \textit{CollisionPlausibilation} and \textit{AirbagController}. The behavior of the parent component is to activate the airbag system via the \textit{exploded} signal whenever any of the sensor signals (\textit{sen\_front}, \textit{sen\_right}, \textit{sen\_left}, or \textit{sen\_back}) holds. To achieve this, the sub-component \textit{CollisionPlausibilation} processes the sensor signals and provides the detection signal \textit{collision\_detected} as output. Finally, the sub-component \textit{AirbagController} takes the signal \textit{collision\_detected} as input and provides the \textit{exploded} signal to activate the actuator of the airbag system.

\subsubsection{Difficulty of the Use Case and Understanding it}
\label{sec:phase2:use_case:pretest:difficulty}

\begin{figure}[t]
	\centering
	\includegraphics[width=0.8\textwidth]{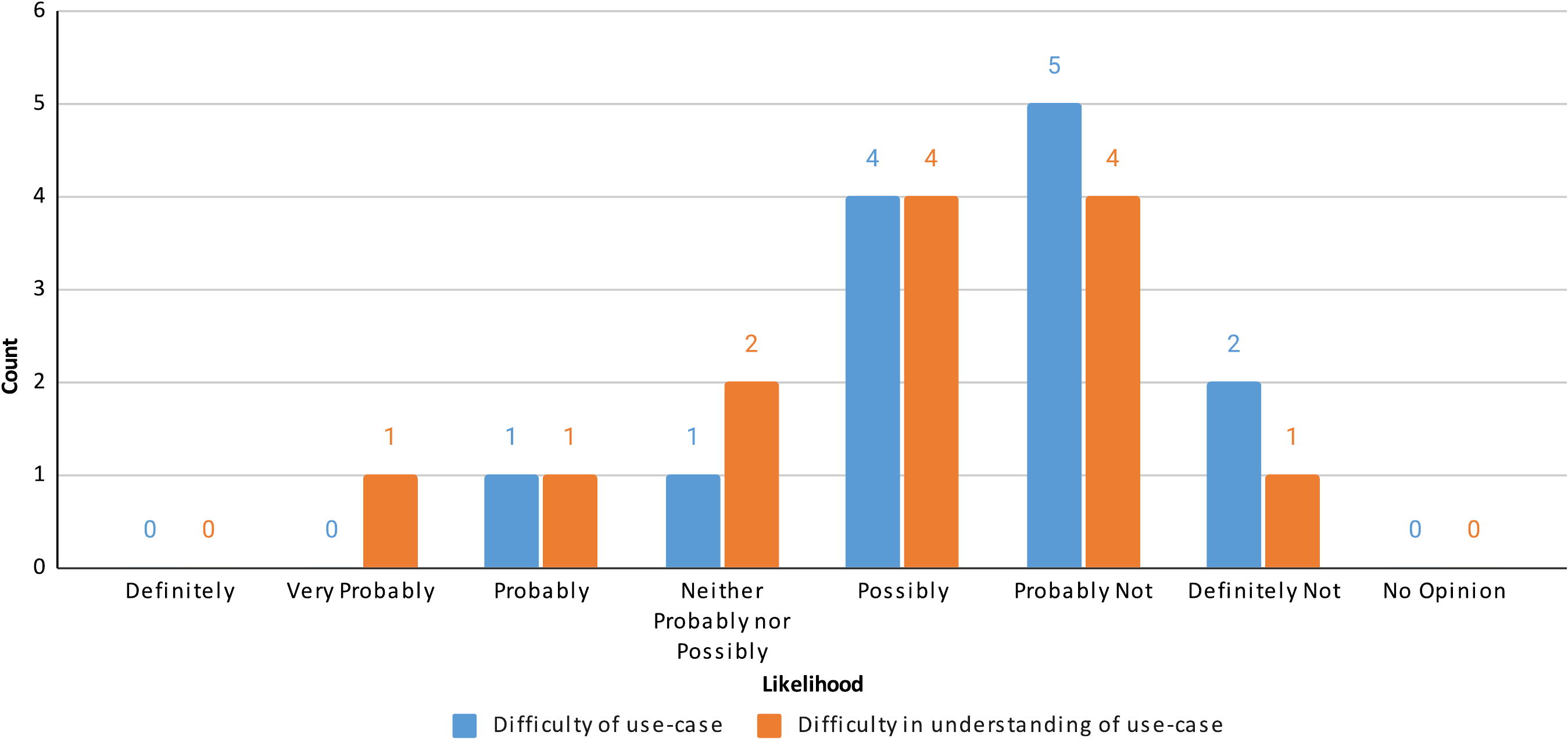}
	\caption{Difficulty of the airbag system use case and of understanding it.}
	\label{fig:usecase_pretest}
\end{figure}

Questions \tqone and \tqtwo (\tab{onegroup}) assess the participants' difficulty  and understanding of the airbag system use case. The participants' responses are shown in \fig{usecase_pretest} (responses follow scale \lsfive in \tab{likert_scale}).

\begin{figure}[t]
	\begin{subfigure}{0.48\textwidth}
		\includegraphics[width=\textwidth]{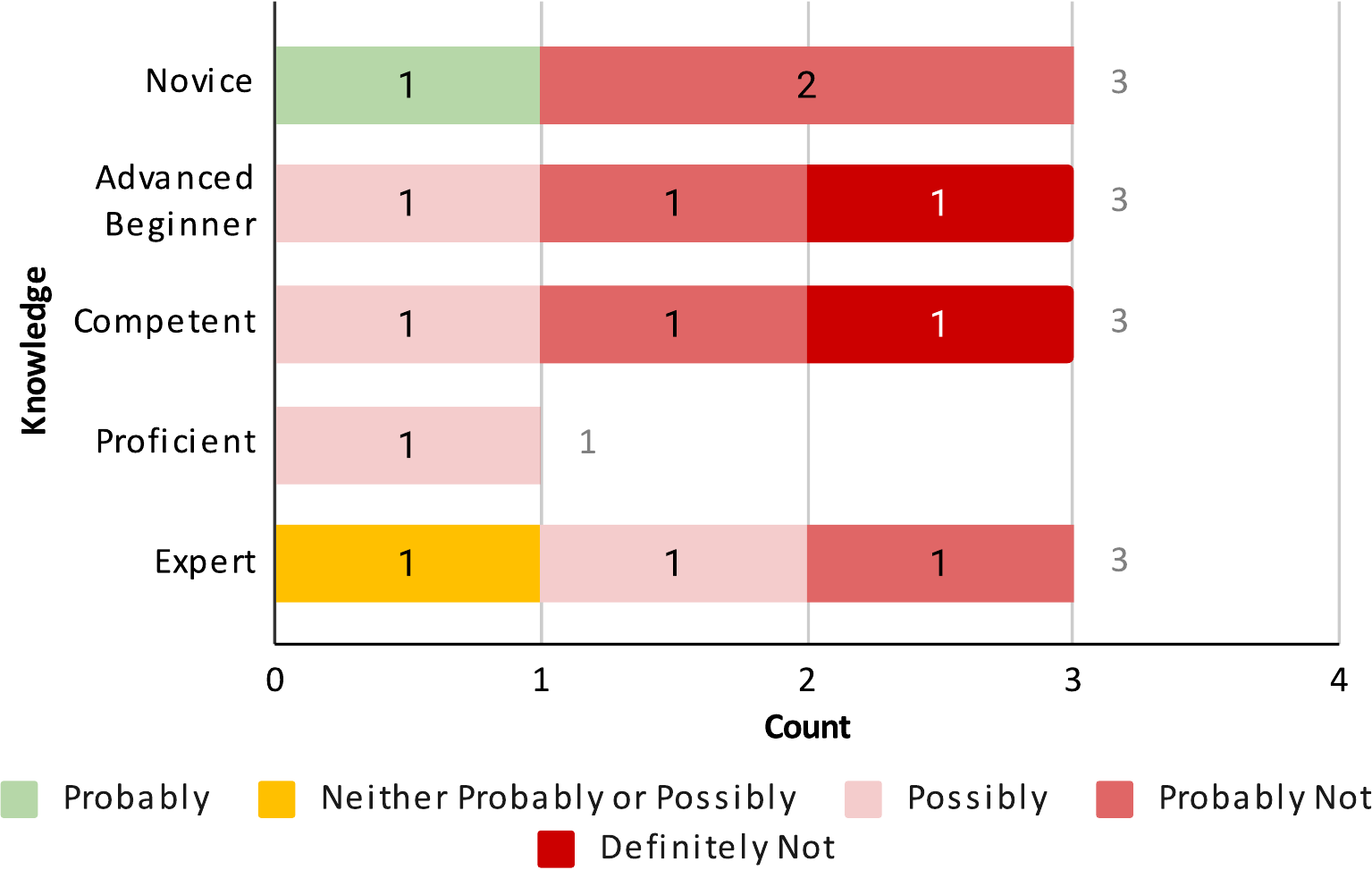}
		\caption{} 
		\label{fig:usecase_difficulty_pretest_knowledge}
	\end{subfigure}
	\hspace{0.3cm}
	\begin{subfigure}{0.48\textwidth}
		\includegraphics[width=\textwidth]{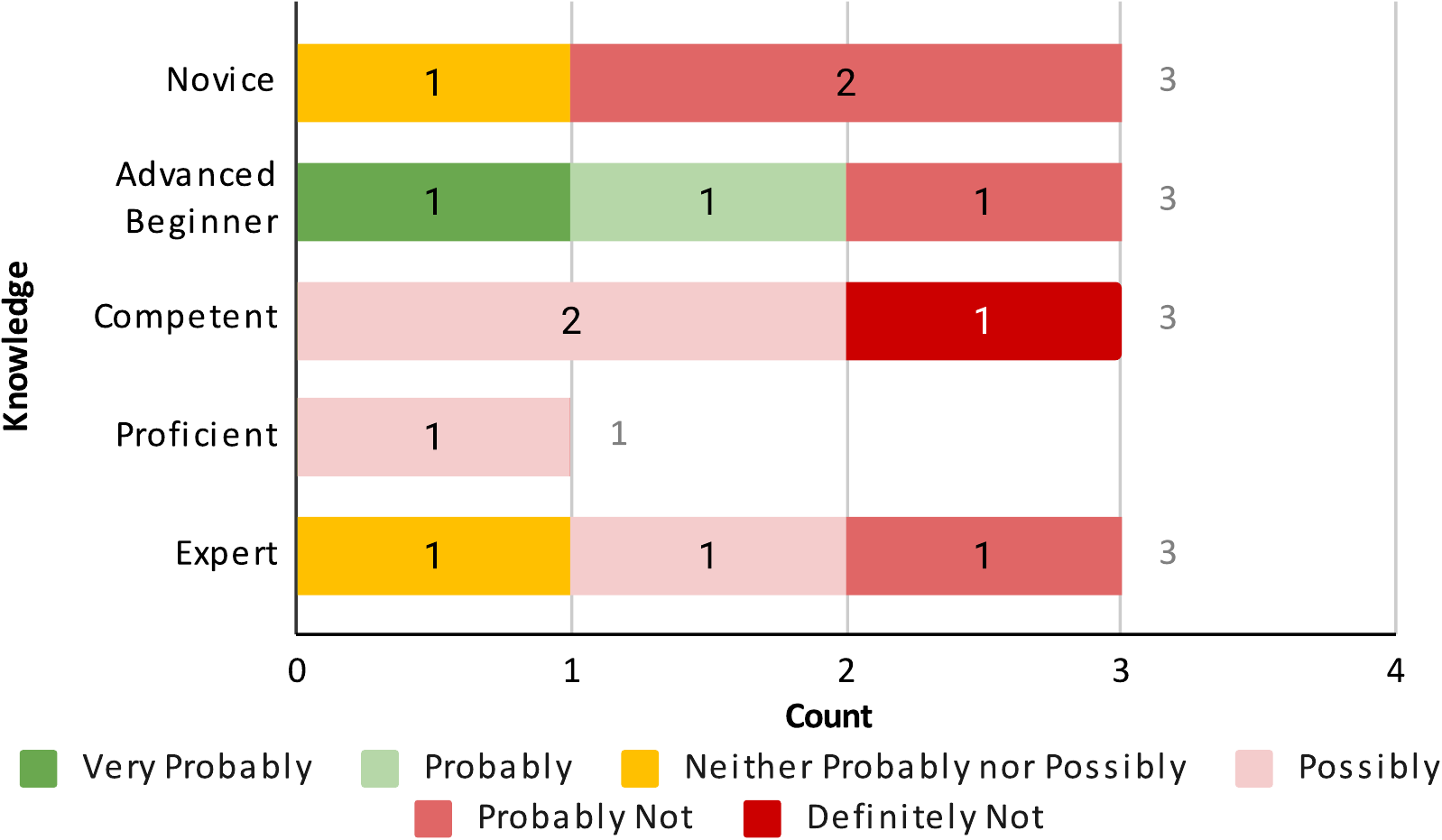}
		\caption{} 
		\label{fig:usecase_understanding_pretest_knowledge}
	\end{subfigure}
	\caption{Difficulty of (a) the use case and (b) of understanding it grouped by the participants' knowledge in formal methods.}
\end{figure}
\paragraph{Difficulty of the use case.}

Among 13\,participants, 11\,participants perceive the airbag system use case as not difficult by answering within the scale \textit{possibly}, \textit{probably not}, or \textit{definitely not}.
\fig{usecase_difficulty_pretest_knowledge} shows responses for the difficulty of the airbag system use case based on the participants' knowledge in formal methods (\cf \sect{phase2:participants:knowledge}).

\paragraph{Difficulty in understanding the use case.}

The responses to question \tqtwo is almost similar to those for \tqone (\cf \fig{usecase_pretest}). A majority of participants (9 of 13) rate their understanding of the airbag system use case as not difficult by answering within the scale \textit{possibly}, \textit{probably not}, or \textit{definitely not}. \fig{usecase_understanding_pretest_knowledge} shows the responses for the difficulty of understanding the use case based on the participants' knowledge in formal methods. Among three participants as \textit{advanced beginners}, one participant each answers \textit{very probably} and \textit{probably}. One participant each answering \textit{very probably} and \textit{definitely not} (two extreme ends) are an \textit{advanced beginner} and a \textit{competent} person in formal methods.

\paragraph{Summary.}

The answers show that most of the participants (11 of 13) perceive the airbag system as not being a difficult use case. Similarly, understanding the airbag system use case is also not rated as difficult (9 of 13).

\subsection{Use Case for the Posttest}
\label{sec:phase2:use_case:posttest}

For the posttest, we use a more complex use case, an electronic power steering systems. As for the use case in the pretest, we ask the participants questions \tqone and \tqtwo (\cf \tab{onegroup}).

\subsubsection{Electronic Power Steering System}
\label{sec:phase2:use_case:posttest:eps}

\begin{figure}[t]
	\centering
	\includegraphics[width=0.8\textwidth]{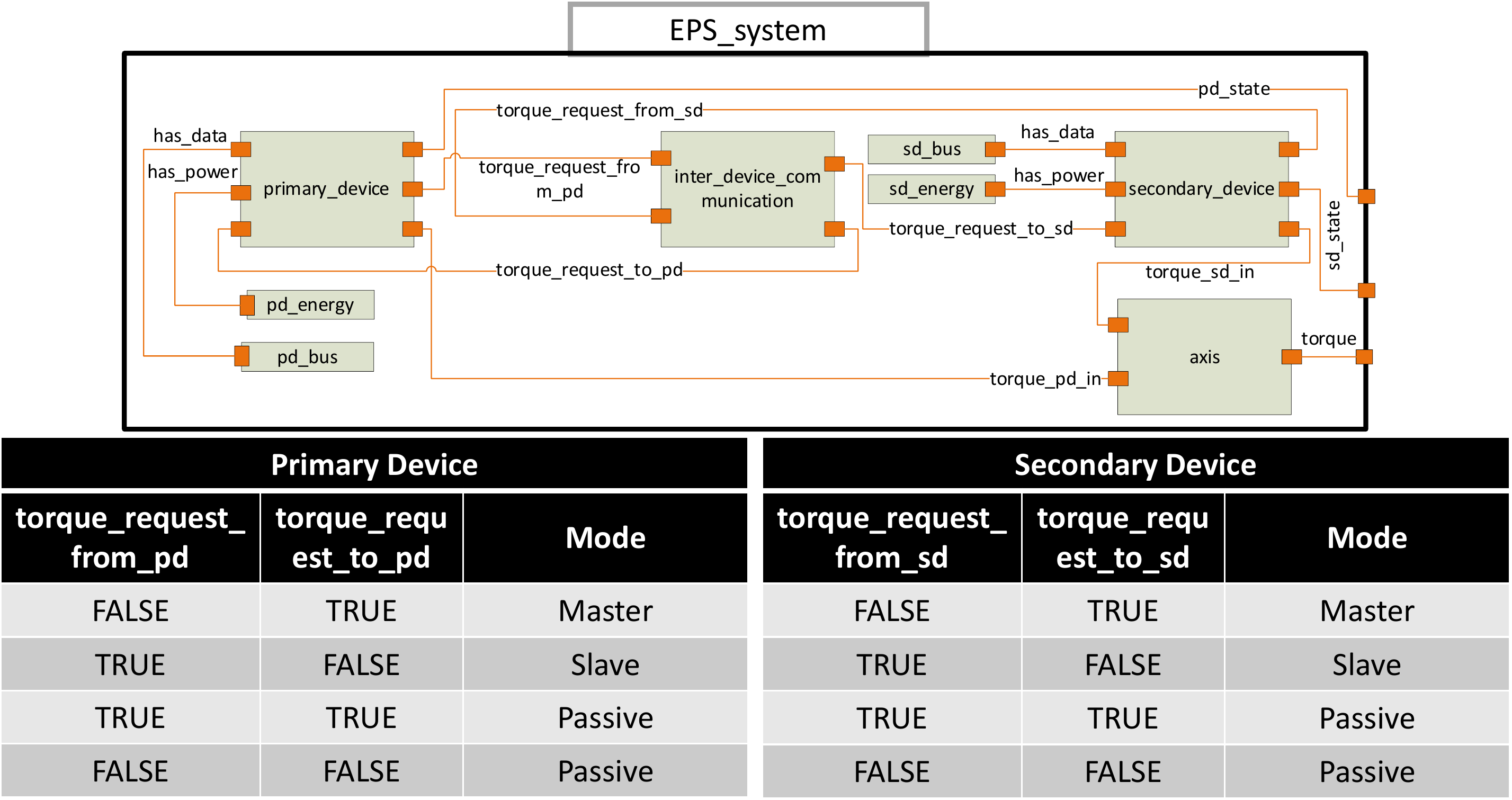}
	\caption{The component model and specifications of the electronic power steering (EPS) system.}
	\label{fig:EPS}
\end{figure}

The Electronic Power Steering (EPS)~\citep{BozzanoMSTV20} system is a Bosch product designed for highly-automated driving vehicles. It steers either based on input from the driver or commands from the vehicle bus. The ECU component of the EPS system, shown in \fig{EPS} and used for the posttest, has two redundant channels: a primary and a secondary channel. Each channel consists of three modes: master, slave, and passive. The nominal behavior is that one channel is master and the other one is slave. This synchronization is taken care of by the sub-component \textit{interComDevice} in between the primary and secondary devices. As shown in \fig{EPS}, if \textit{torque\_request\_from\_pd} does not hold and \textit{torque\_request\_to\_pd} holds, then the sub-component \textit{primary\_device} is in master mode. Similarly, if \textit{torque\_request\_from\_sd} holds and \textit{torque\_request\_to\_pd} does not hold, then the sub-component \textit{secondary\_device} is in slave mode. In these two cases, \textit{torque} is given to the system actuator.

\subsubsection{Difficulty of the Use Case and Understanding it}
\label{sec:phase2:use_case:posttest:difficulty}

\begin{figure}[t]
	\centering
	\includegraphics[width=0.8\textwidth]{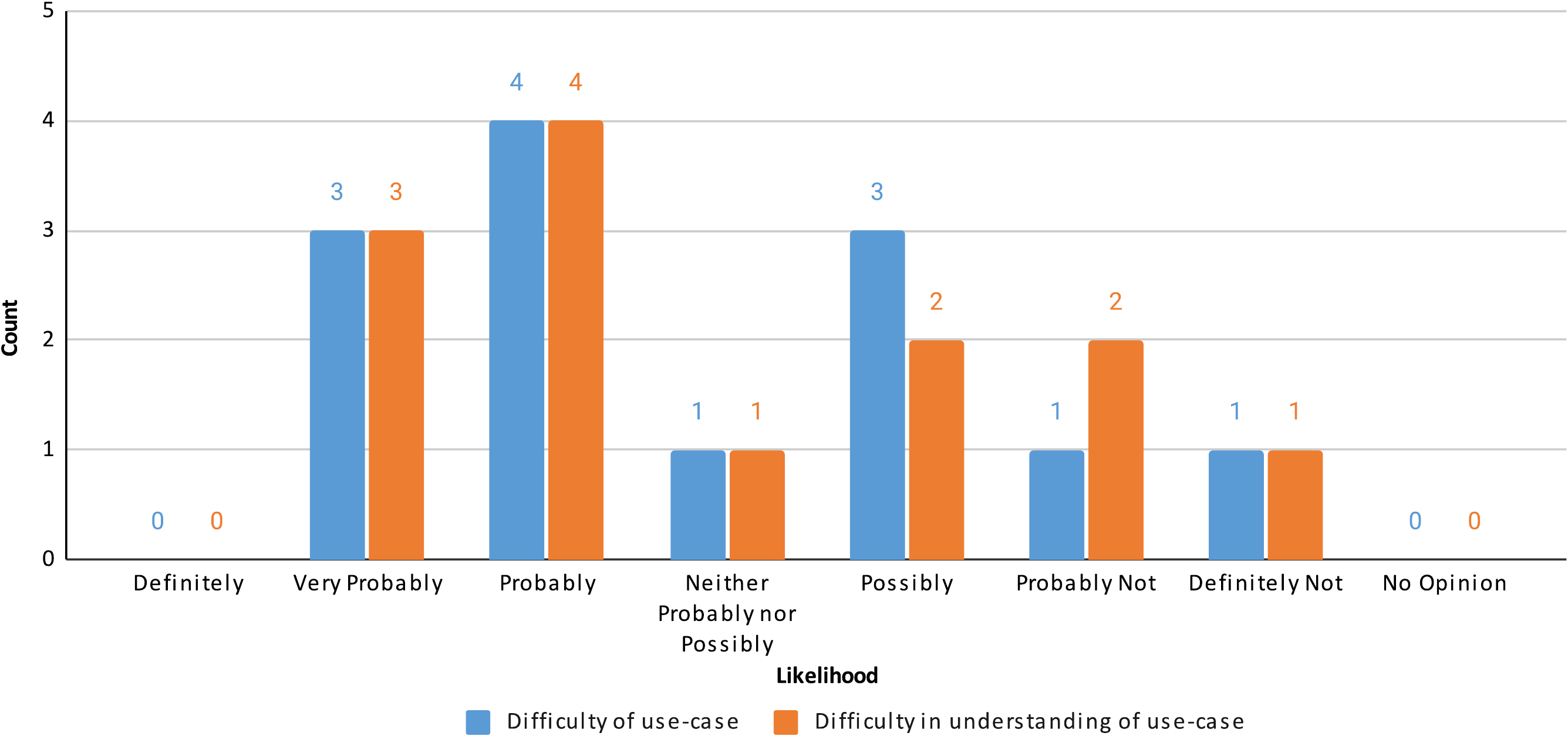}
	\caption{Difficulty of the EPS use case and of understanding it.}
	\label{fig:usecase_posttest}
\end{figure}

As mentioned previously, the same questions \tqone and \tqtwo in \tab{onegroup} used in the pretest are again used in the posttest to collect the participants' opinions in assessing the difficulty of the electronic power steering system use case and of understanding it. The participants' responses are shown in \fig{usecase_posttest} (answers follow scale \lsfive in \tab{likert_scale}). To avoid any bias, we have used a less complex system for the pretest and more complex real-world project for the posttest. Thus, our early hypothesis is that a majority of participants perceives the use case and its understanding of the posttest as more complex. This is confirmed by the results shown in \fig{usecase_posttest}. Overall, the motive of using the EPS use case for the posttest is to identify whether the proposed counterexample explanation approach is suitable for real-world systems with the use of formal methods.

\begin{figure}[t]
	\begin{subfigure}{0.48\textwidth}
		\includegraphics[width=\textwidth]{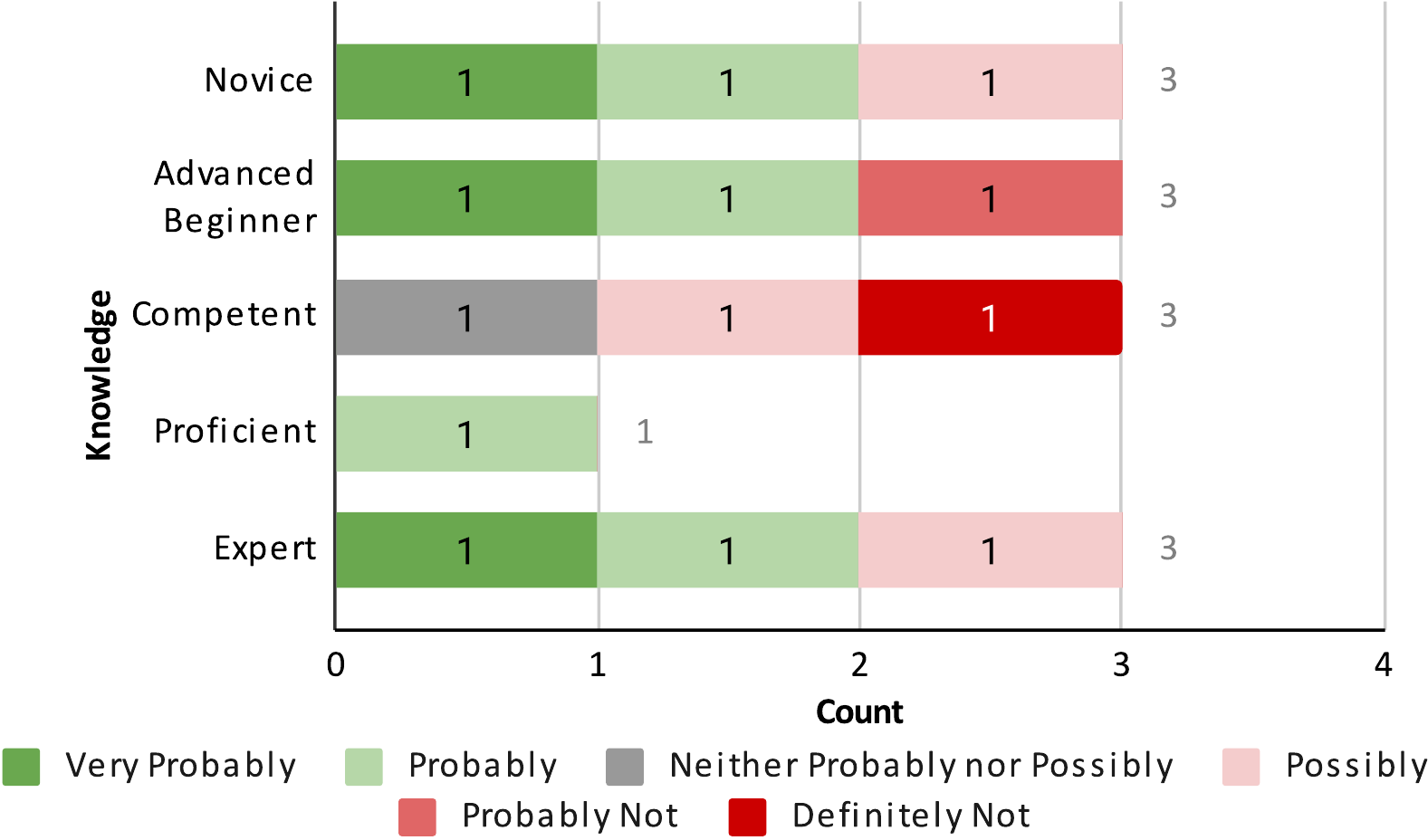}
		\caption{} 
		\label{fig:usecase_difficulty_posttest_knowledge}
	\end{subfigure}
	\hspace{0.3cm}
	\begin{subfigure}{0.48\textwidth}
		\includegraphics[width=\textwidth]{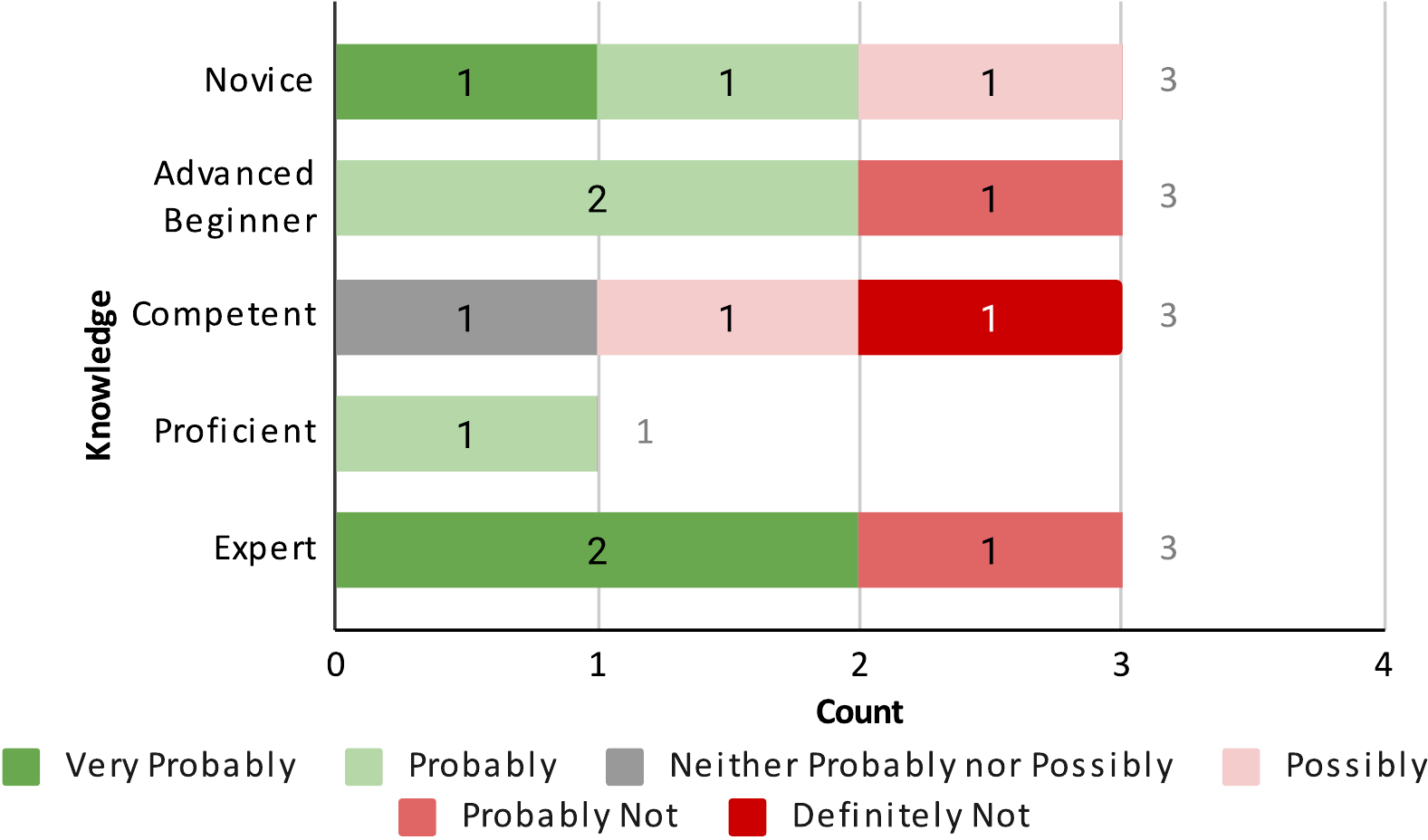}
		\caption{} 
		\label{fig:usecase_understanding_posttest_knowledge}
	\end{subfigure}
	\caption{Difficulty of (a) the EPS use case and of (b) understanding it grouped by the participants' knowledge in formal methods.}
\end{figure}

\paragraph{Difficulty of the use case.}

\fig{usecase_posttest} shows that seven out of 13\,participants perceive the EPS use case as difficult by answering with \textit{very probably} or \textit{probably}. One further participant answers with \textit{neither probably nor possibly} and the remaining five participants perceive it as not difficult (\textit{possibly}, \textit{probably not}, and \textit{definitely not}). \fig{usecase_difficulty_posttest_knowledge} groups the responses by the participants' knowledge in formal methods. Except for the knowledge level \textit{competent}, all other knowledge levels predominantly perceive the use case as difficult.

\paragraph{Difficulty in understanding the use case.}

The results for \tqtwo on the difficulty of understanding the use case are  similar to \tqone. Seven of 13\,participants perceive understanding the EPS system as difficult, one participant answers with \textit{neither probably nor possibly}, and the remaining five participants perceive it as not difficult. The results are also similar to \tqone taking the participants' knowledge levels into account (\fig{usecase_understanding_posttest_knowledge}).

\paragraph{Summary.}

The results for both the difficulty of the EPS use case and of understanding it are quite similar. The majority of participants (7 of 13) perceive both the EPS use case and understanding it as difficult.

\begin{figure}[t]
	\centering
	\includegraphics[width=0.9\textwidth]{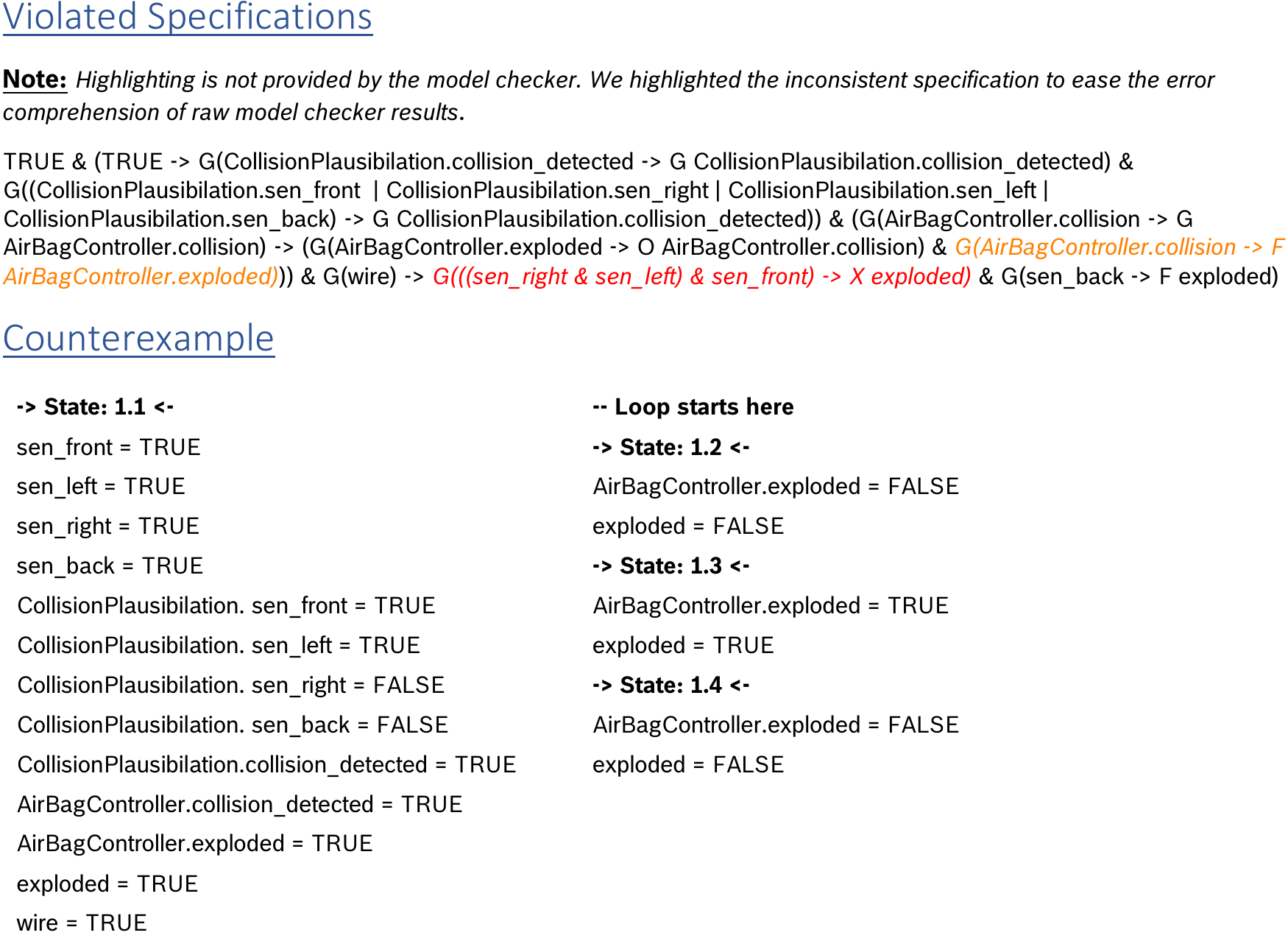}
	\caption{Result of the model checker when verifying a refinement of the airbag system.}
	\label{fig:AS_violated_specification}
\end{figure}

\subsection{Results of the Model Checker and Counterexample Explanation, and Understanding it}
\label{sec:phase2:postest:result}

We use questions \tqthree and \tqfour (\tab{onegroup}) to collect the participants' responses on understanding the model checker and counterexample explanation results shown during the pretest and protest to the participants. These results and responses are discussed in the following.

\paragraph{Model checker output.} 

\fig{AS_violated_specification} is the result generated by the model checker for the airbag system use case described in \sect{phase2:use_case:pretest:as}. This result is used during the pretest. If an inconsistency is identified by the model checker, the whole violated refinement specification and the counterexample to illustrate the erroneous behavior are shown to the participant. We have highlighted the inconsistent specifications in the whole violated refinement specification to avoid any bias. For example, the model checker output could be hard to interpret on first glance, which is not necessarily the case for our counterexample explanation approach since the relevant information is highlighted explicitly. This could mislead the participant at first glance in deciding the model checker output to be too difficult. Thus, to avoid this bias, highlighting the inconsistent specification in the model checker output would trigger the participant to understand and identify the inconsistency.

\begin{figure}[t]
	\centering
	\includegraphics[width=0.9\textwidth]{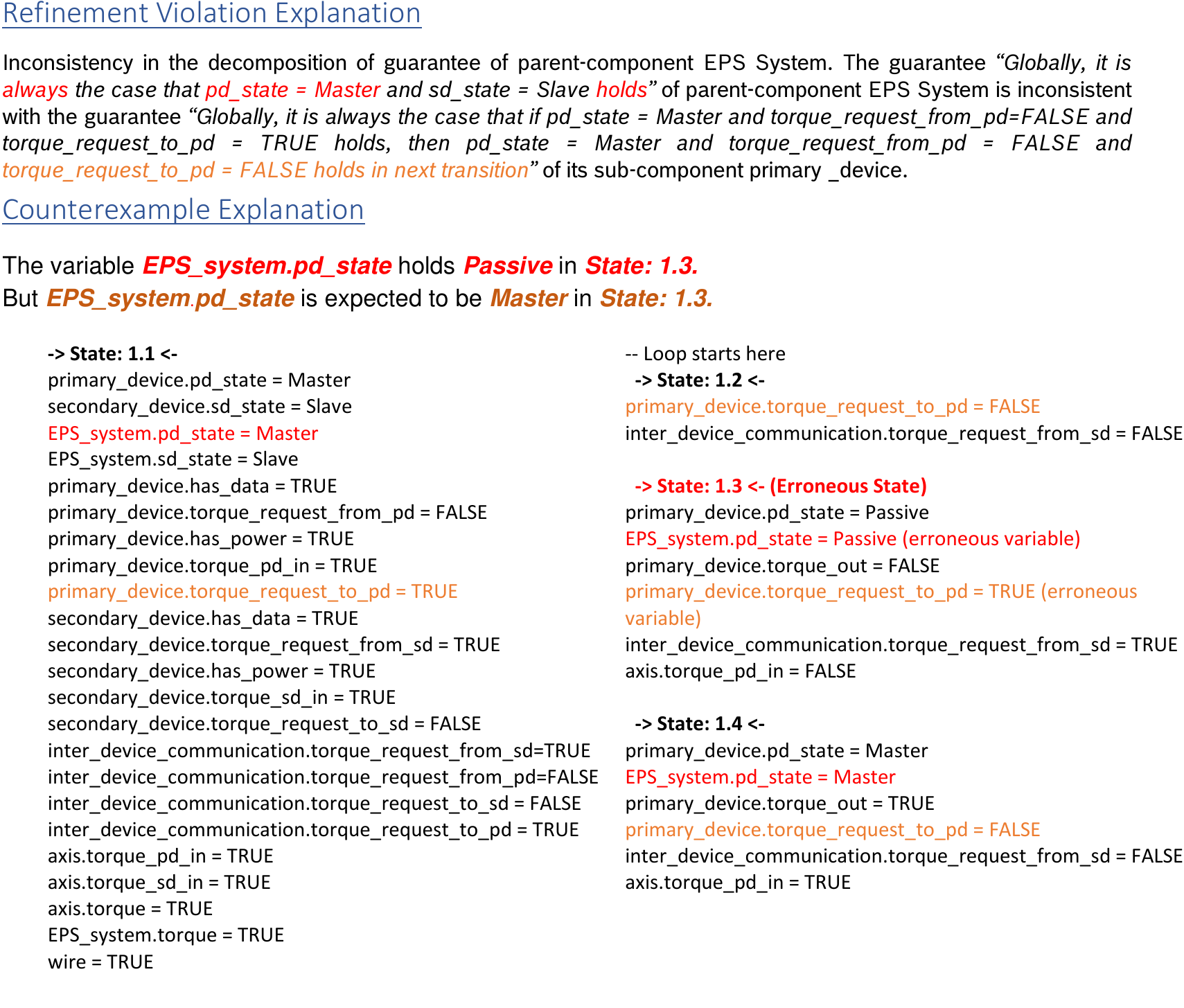}
	\caption{Result of the counterexample explanation approach when verifying a refinement of the EPS system.}
	\label{fig:EPS_violated_specification}
\end{figure}

\paragraph{Counterexample explanation.} 

The counterexample explanation shown in \fig{EPS_violated_specification} is used for the posttest. This explanation is the result of verifying a refinement and explaining a refinement inconsistency of the EPS use case (\cf \sect{phase2:use_case:posttest:eps}). Instead of showing the complete violated refinement specification, the explanation presents the type of violation, list of inconsistent specifications, and their corresponding components. Further, instead of a concrete counterexample, highlighting of erroneous states and variables in the counterexample as well as explanations of the erroneous and expected nominal behavior are shown to ease the error comprehension for the participants.

\paragraph{Understanding the results.}

\begin{figure}[t]
	\centering
	\includegraphics[width=0.8\textwidth]{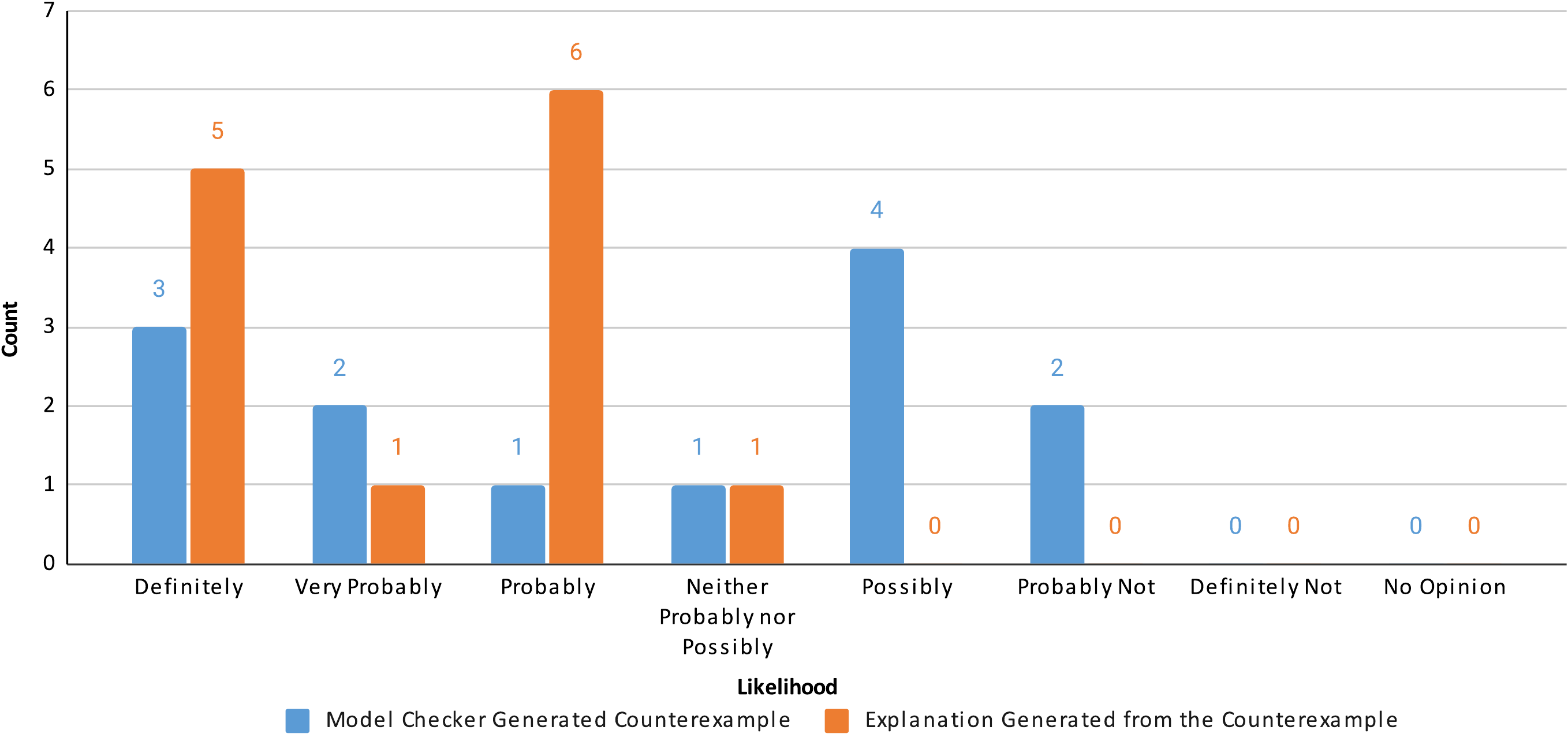}
	\caption{Results for understanding the results generated by the model checker and counterexample explanation approach.}
	\label{fig:understanding_result}
\end{figure}

With questions \tqthree and \tqfour, we assess the understanding of the model checker output and counterexample explanation approach. Only a minority (6 of 13) of participants perceives the model checker output as easy to understand (these participants answer with \textit{definitely}, \textit{very probably}, and \textit{probably}), while a clear majority participants (12 of 13) perceives understanding of the counterexample explanation approach as easy (\fig{understanding_result}). Notably, no participant answers that understanding the results of the counterexample explanation approach is hard, while six of 13 participants still find the model checker output to be hard to understand (corresponding responses are \textit{possibly} and \textit{probably not}), even with the less complex use case (airbag system) and additional highlighting of the inconsistent specifications in the model checker output.

\paragraph{Understanding the results grouped by the participants' knowledge in formal methods.}

\begin{figure}[t]
	\begin{subfigure}{0.48\textwidth}
		\includegraphics[width=\textwidth]{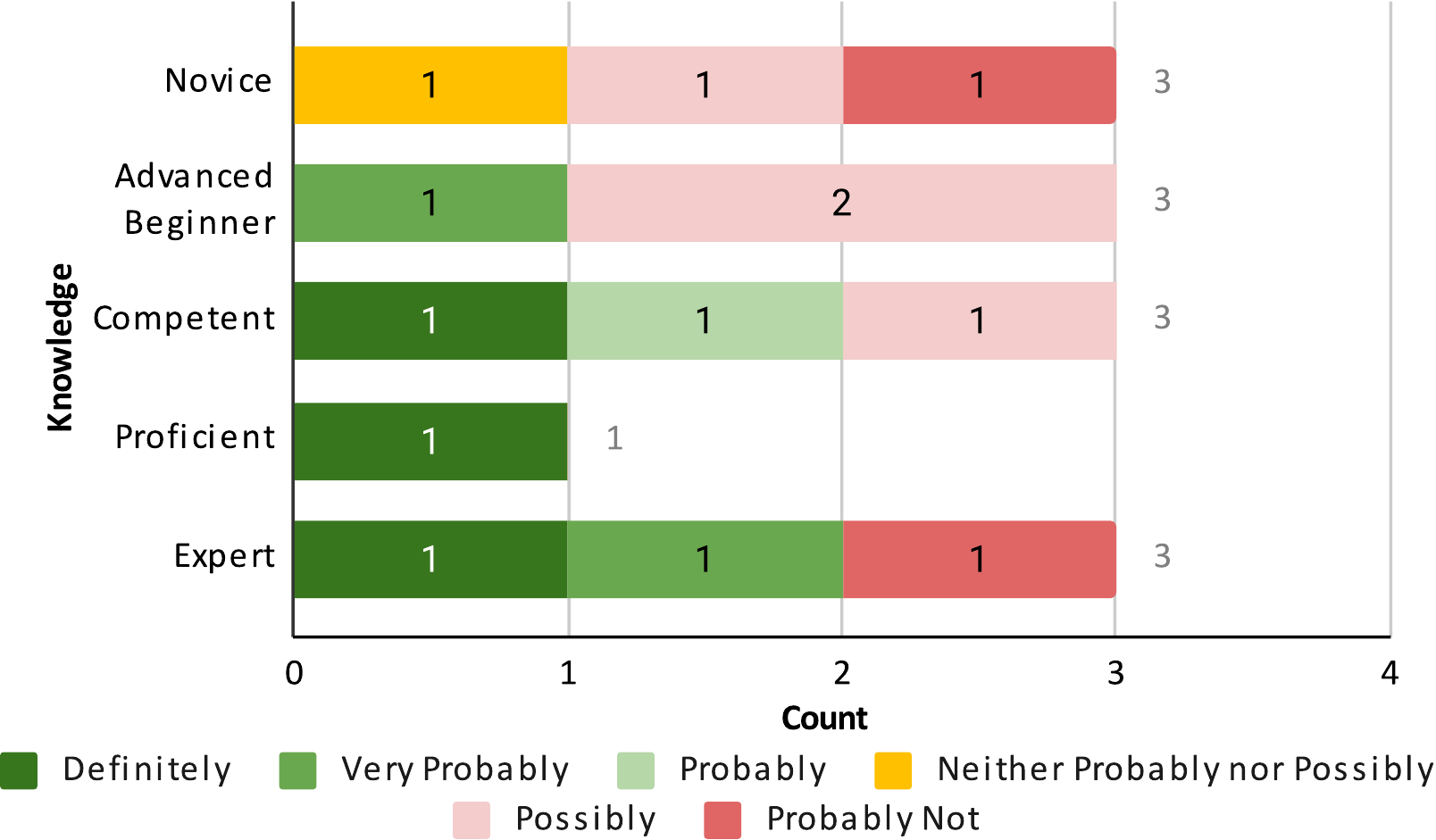}
		\caption{} 
		\label{fig:understanding_result_mc}
	\end{subfigure}
	\hspace{0.3cm}
	\begin{subfigure}{0.48\textwidth}
		\includegraphics[width=\textwidth]{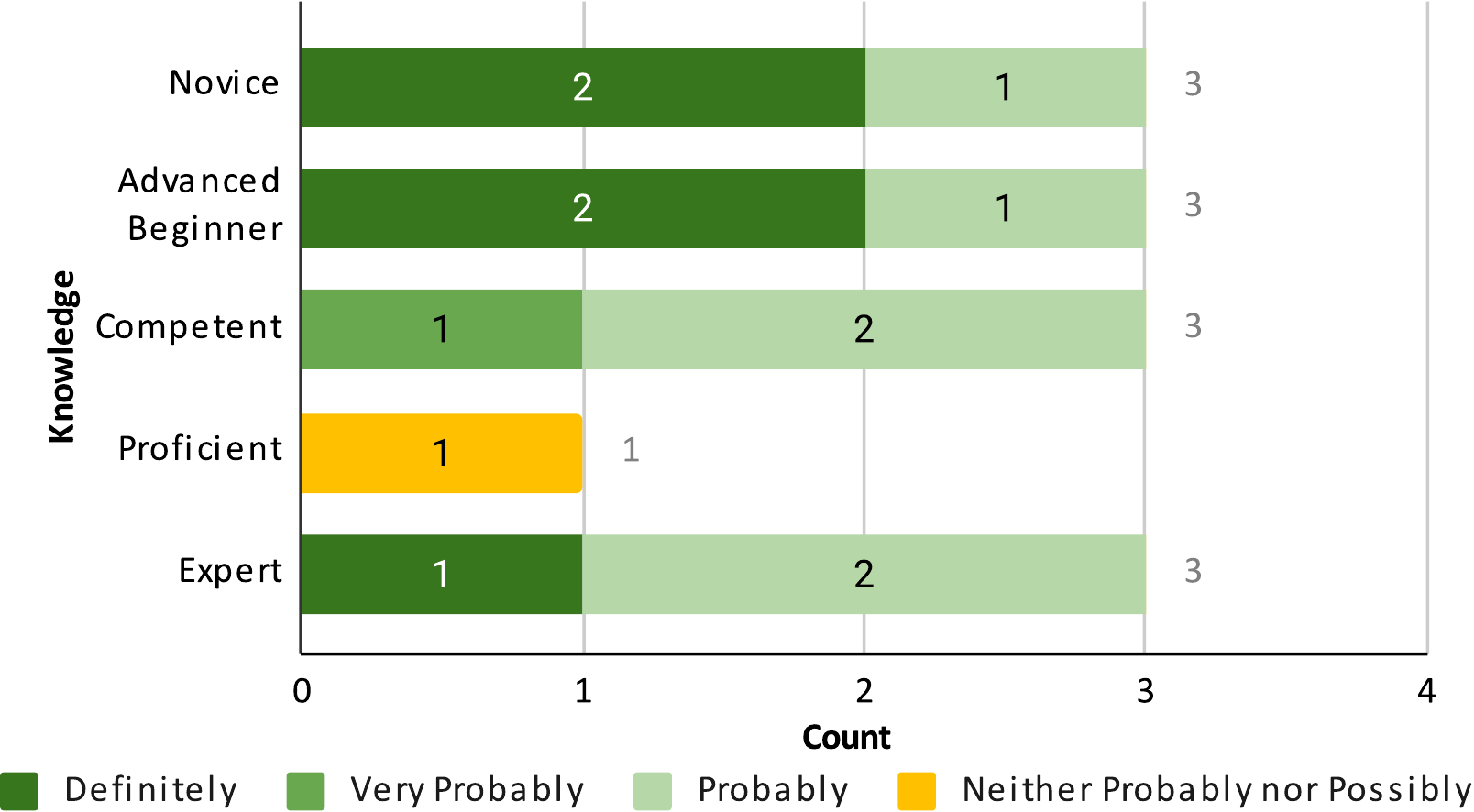}
		\caption{} 
		\label{fig:understanding_result_exp}
	\end{subfigure}
	\caption{Responses for understanding the results generated (a) by the model checker and (b) by the counterexample explanation approach grouped by the participants' knowledge in formal methods.}
\end{figure}

\fig{understanding_result_mc} and \fig{understanding_result_exp} show how the participants understand the model checker output and counterexample explanation based on their knowledge in formal methods. The understanding of the model checker output (\fig{understanding_result_mc}) aligns with the participants' knowledge. For example, a majority of the participants answering that understanding is hard (\textit{possibly} and \textit{probably not}) are either \textit{novice} or \textit{advanced beginners}. On other hand, a majority of the participants answering that understanding is easy (\textit{definitely}, \textit{very probably}, and \textit{probably}) belongs to the \textit{competent}, \textit{proficient}, or \textit{expert} groups. 
However, as shown in \fig{understanding_result_exp}, a majority of the participants belonging to the \textit{novice} and \textit{advanced beginner} groups perceive that understanding the counterexample explanation is \textit{definitely} easy.

\paragraph{Summary.}

The raw output generated by the model checker contains the whole violated refinement specification and the counterexample to illustrate the erroneous behavior. The counterexample explanation approach, however, generates an explanation containing the type of violation, the inconsistent specification, and their corresponding components. Additionally, erroneous states and erroneous variables are highlighted in the counterexample. From the collected responses, 12 out of 13 participants perceive the understanding of the counterexample explanation result as easy, which contrasts the six out of the 13 participants who perceive the model checker output to be easy to understand.

\subsection{Participants Responses for Task-Related Questions}
\label{sec:phase2:posttest:tq}

Questions from \tqfive to \tqnine are task-related, in which participants answer based on their understanding of the model checker and counterexample explanation results. For these questions during the pretest and posttest, the participants should identify the inconsistent components (\tqfive) and specifications (\tqsix), the reason for the inconsistency (\tqseven), a solution to fix the inconsistency (\tqeight), and a nominal behavior of the system in terms of a correct state transition in place of the erroneous state transition (\tqnine). 

\paragraph{Identifying inconsistent components.}

\begin{figure}[b]
	\begin{subfigure}{0.48\textwidth}
		\includegraphics[width=\textwidth]{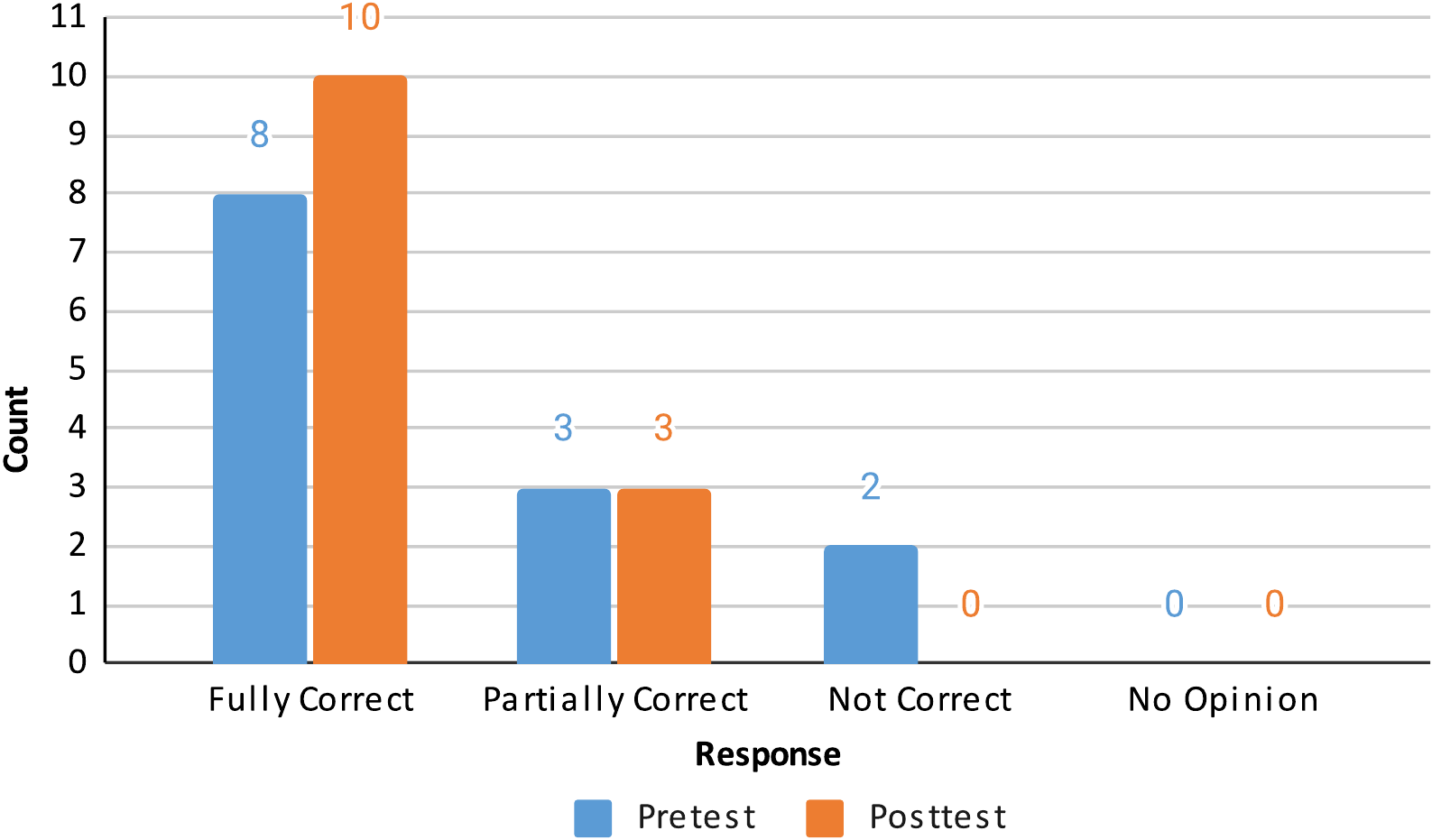}
		\caption{} 
		\label{fig:response_component}
	\end{subfigure}
	\hspace{0.3cm}
	\begin{subfigure}{0.48\textwidth}
		\includegraphics[width=\textwidth]{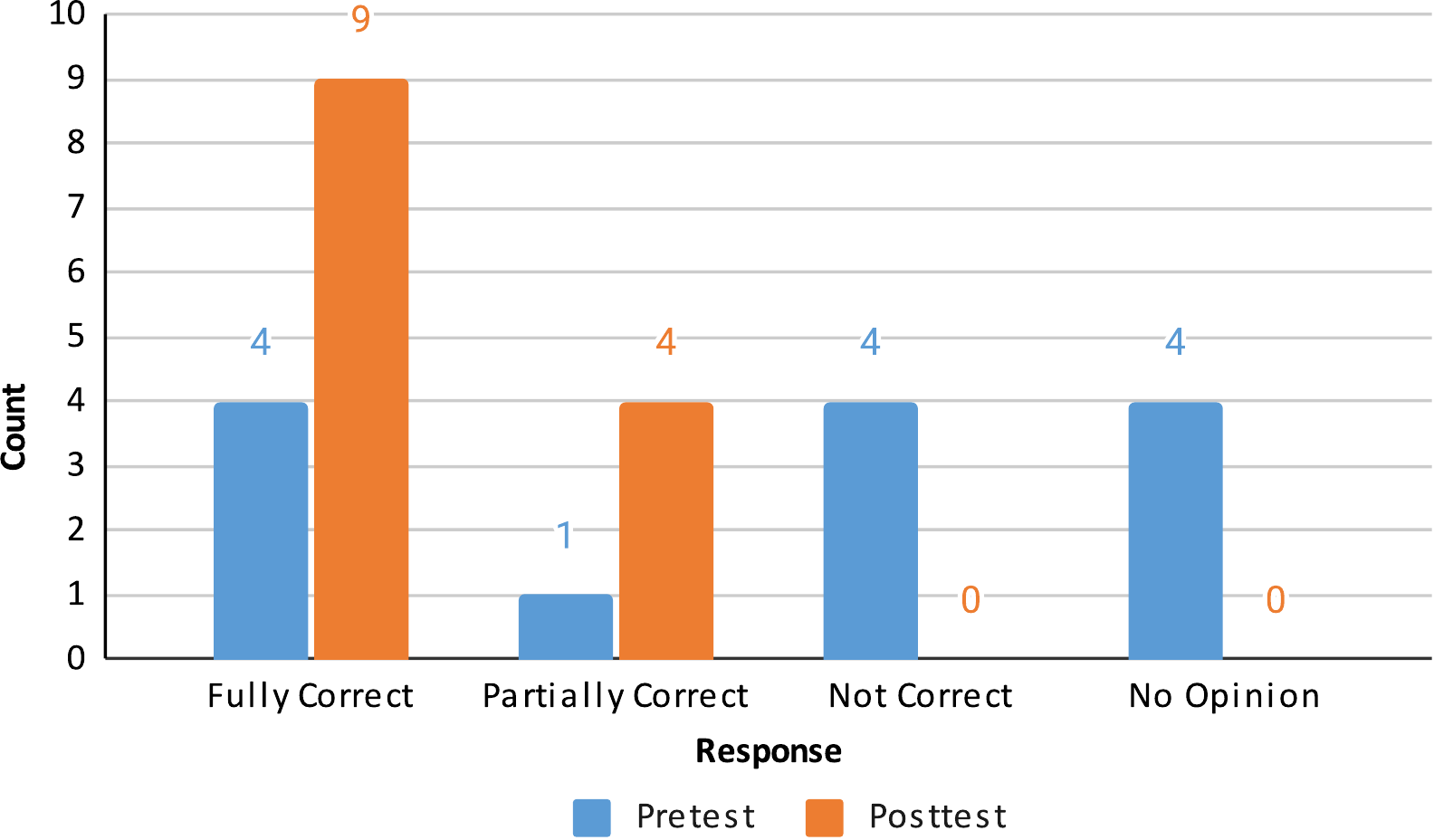}
		\caption{} 
		\label{fig:response_req}
	\end{subfigure}
	\caption{Results of identifying (a) inconsistent components and (b) inconsistent specifications from the model checker output (pretest) and counterexample explanation (posttest).}
\end{figure}

\fig{response_component} shows the responses to \tqfive, which requires from the participants to identify the inconsistent components based on the model checker result during the pretest and counterexample explanation during the posttest. Eight out of 13\,participants identify both the inconsistent parent and sub-components during the pretest and ten participants during the posttest. Further three participants each identify \emph{either} the sub-component or parent component correctly during both the pretest and posttest. Notably, no participant completely identifies the incorrect components based on the counterexample explanation. 

\paragraph{Identifying inconsistent specifications.}

\fig{response_req} shows responses to \tqsix, for which participants should identify the inconsistent specification from the  model checker output and based on the counterexample explanation. Among the 13\,participants, four participants have \textit{no opinion} from the model checker output. From the remaining nine participants, four participants each identify the inconsistent specifications fully correct and completely incorrect, and one participant identifies parts of the inconsistent specification correctly. Concerning the responses of identifying the inconsistent specifications using the counterexample explanation, nine of 13\,participants identify all inconsistent specifications correctly. The remaining four participants identify parts of specifications correctly.

Among the five participants who identify the inconsistent specifications fully or partially based on the model checker output, the responses by four participants are fully correct in terms of the reason for the inconsistency (\tqseven) and the appropriate fix (\tqeight). On other hand, among 13\,participants who identify the inconsistent specifications fully or partially using the counterexample explanation, the responses by 11\,participants correctly identify the reason (\tqseven) and by nine participants correctly identify the fix (\tqeight).

\paragraph{Nominal system behavior in the counterexample.}

Among the 13 participants, six participants attempt to answer \tqnine in the pretest and nine participants in the posttest to identify the expected system behavior in the counterexample, that is, to name a correct state transition in place of the erroneous state transition. 
Among the six participants in the pretest, four participants correctly name  the expected behavior based on the model checker output, while seven of the nine participants in the posttest correctly name  the expected behavior based on the counterexample explanation.

\paragraph{Summary.}

A majority of participants identifies both the inconsistent components and specifications correctly based on the counterexample explanation approach in the posttest, significantly more than based on the model checker output in the pretest. This shows that the counterexample explanation is well suited to identify inconsistent components and specifications, and that participants are able to explain the reason for and to fix the inconsistencies.

\subsection{Participants' Opinion on Understanding the Model Checker Output and Counterexample Explanation}
\label{sec:phase2:posttest:opinion}

\begin{figure}[t!]
	\includegraphics[width=\textwidth]{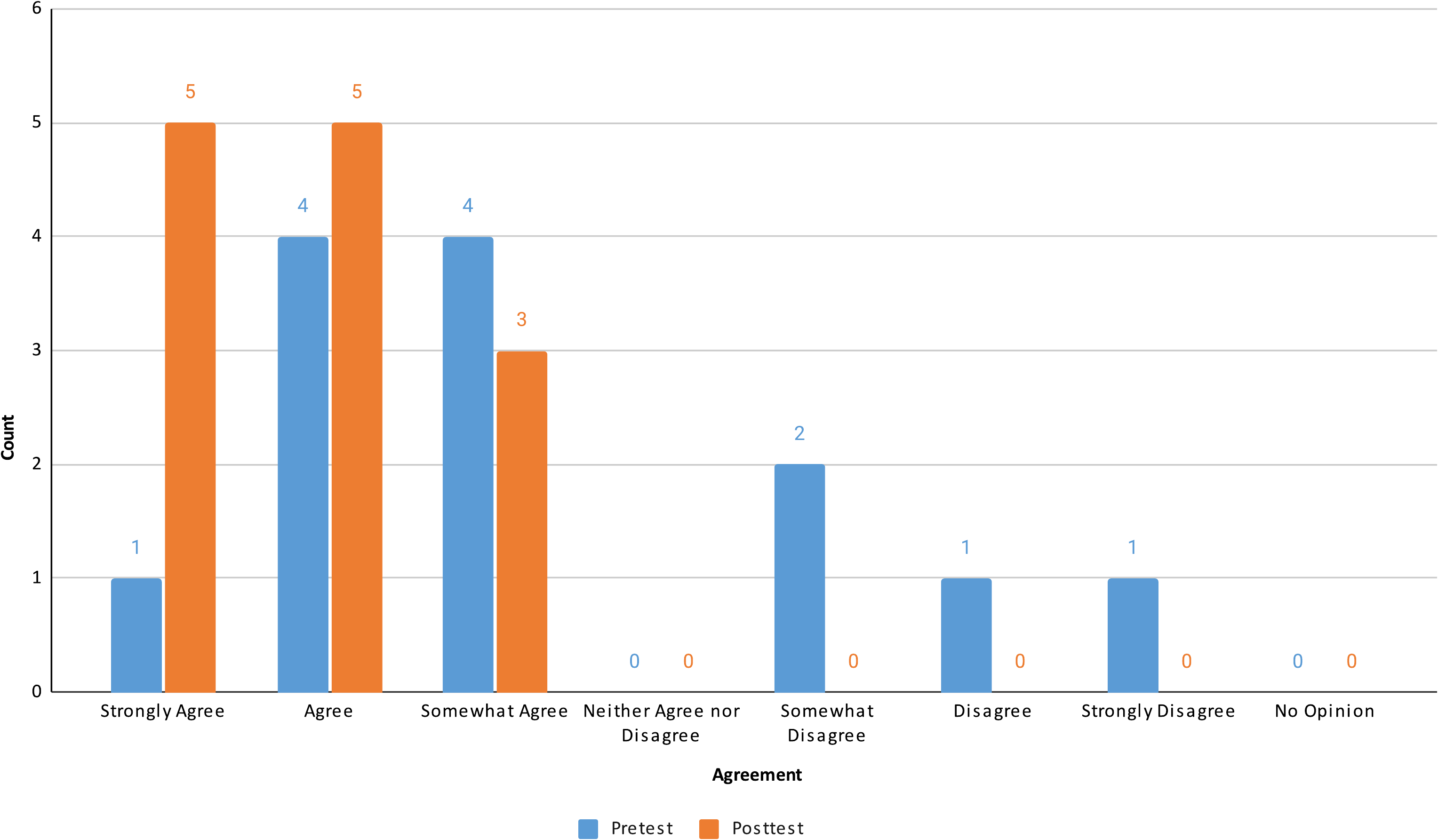}
	\caption{Better understanding.}
	\label{fig:better_understanding}
\end{figure}

\begin{figure}[b!]
	\includegraphics[width=\textwidth]{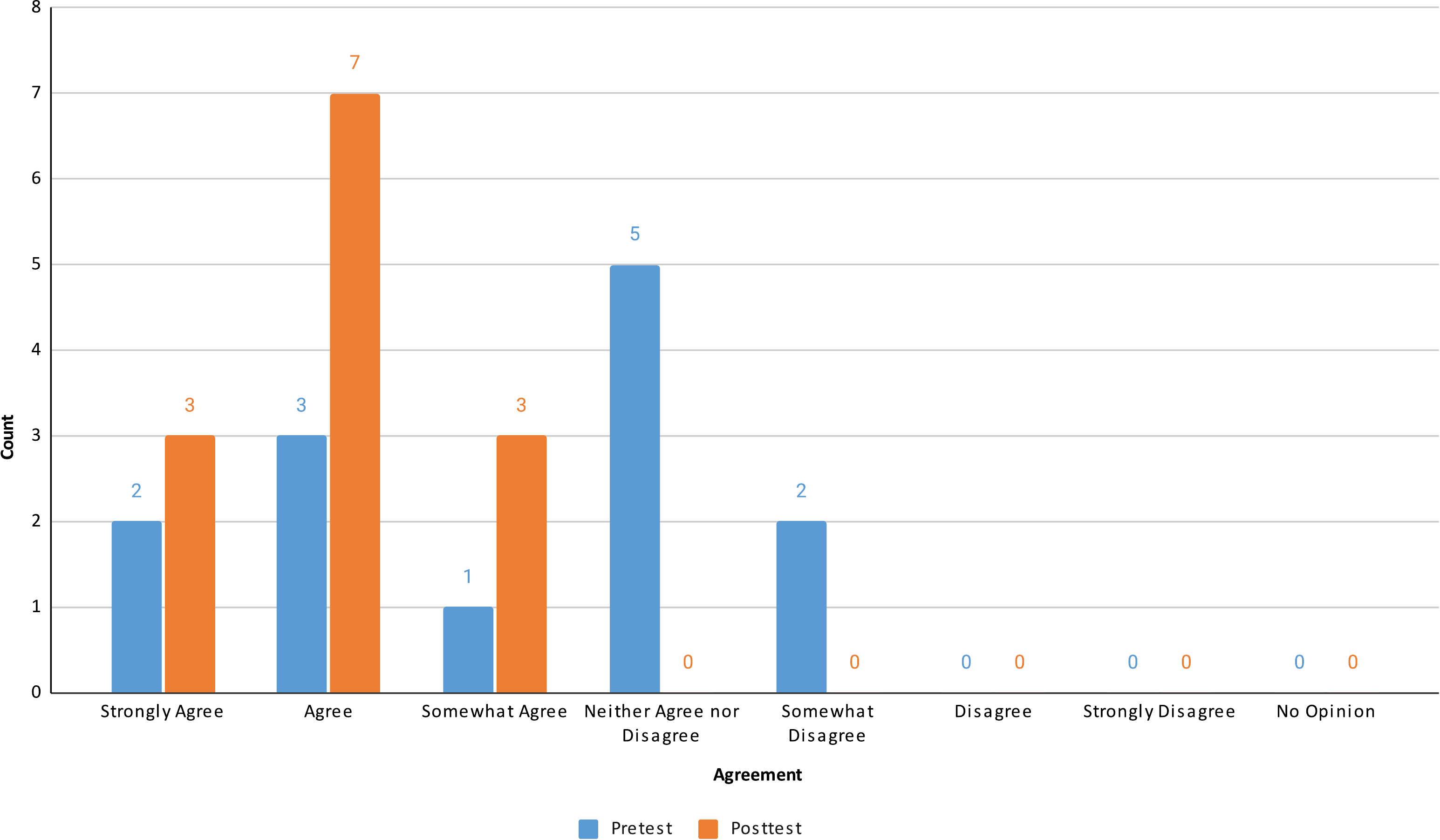}
	\caption{Quicker understanding.}
	\label{fig:quicker_understanding}
\end{figure}

\begin{figure}[t!]
	\includegraphics[width=\textwidth]{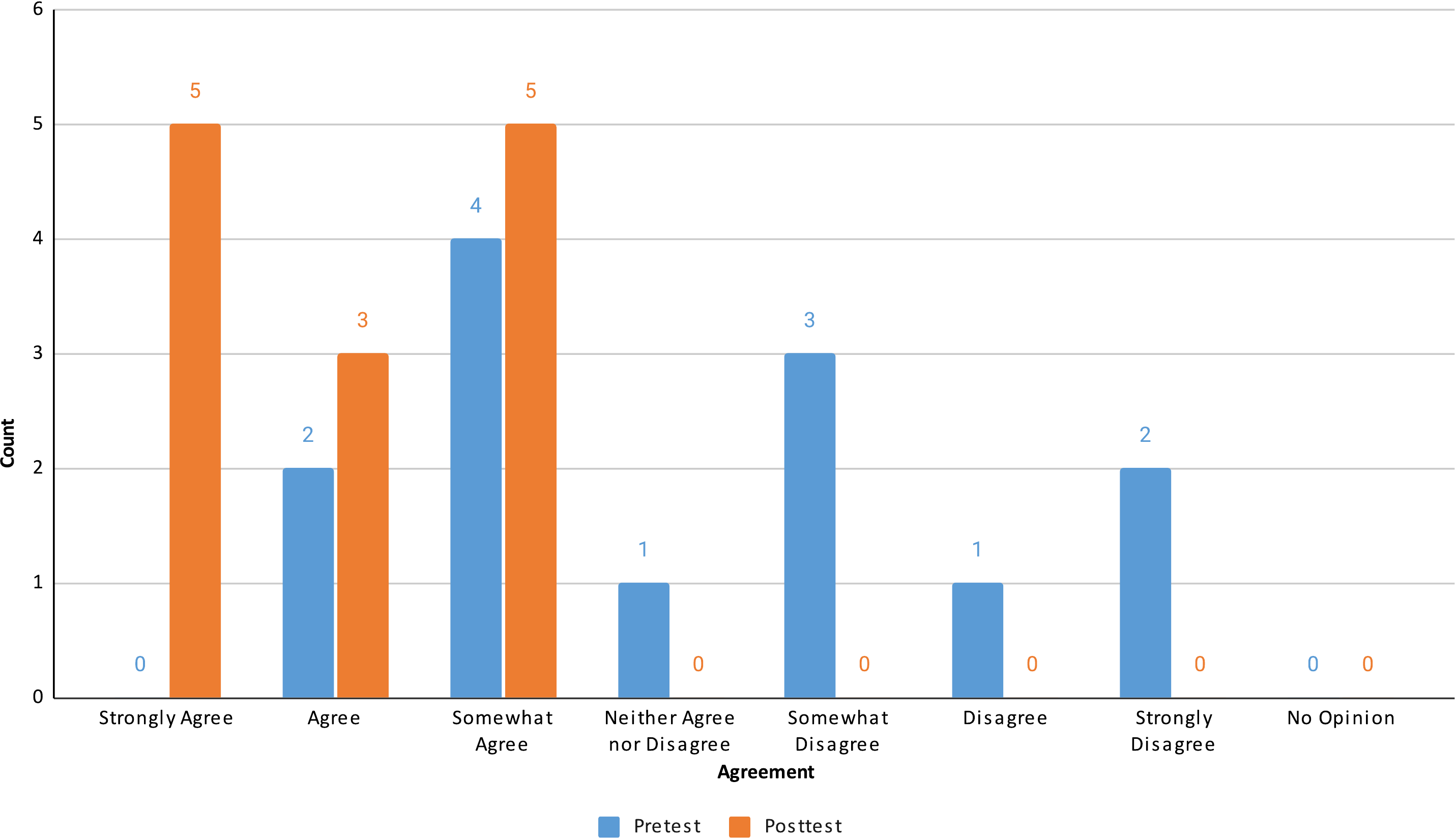}
	\caption{Confidence of understanding.}
	\label{fig:more_confident}
\end{figure}

\begin{figure}[t!]
	\includegraphics[width=\textwidth]{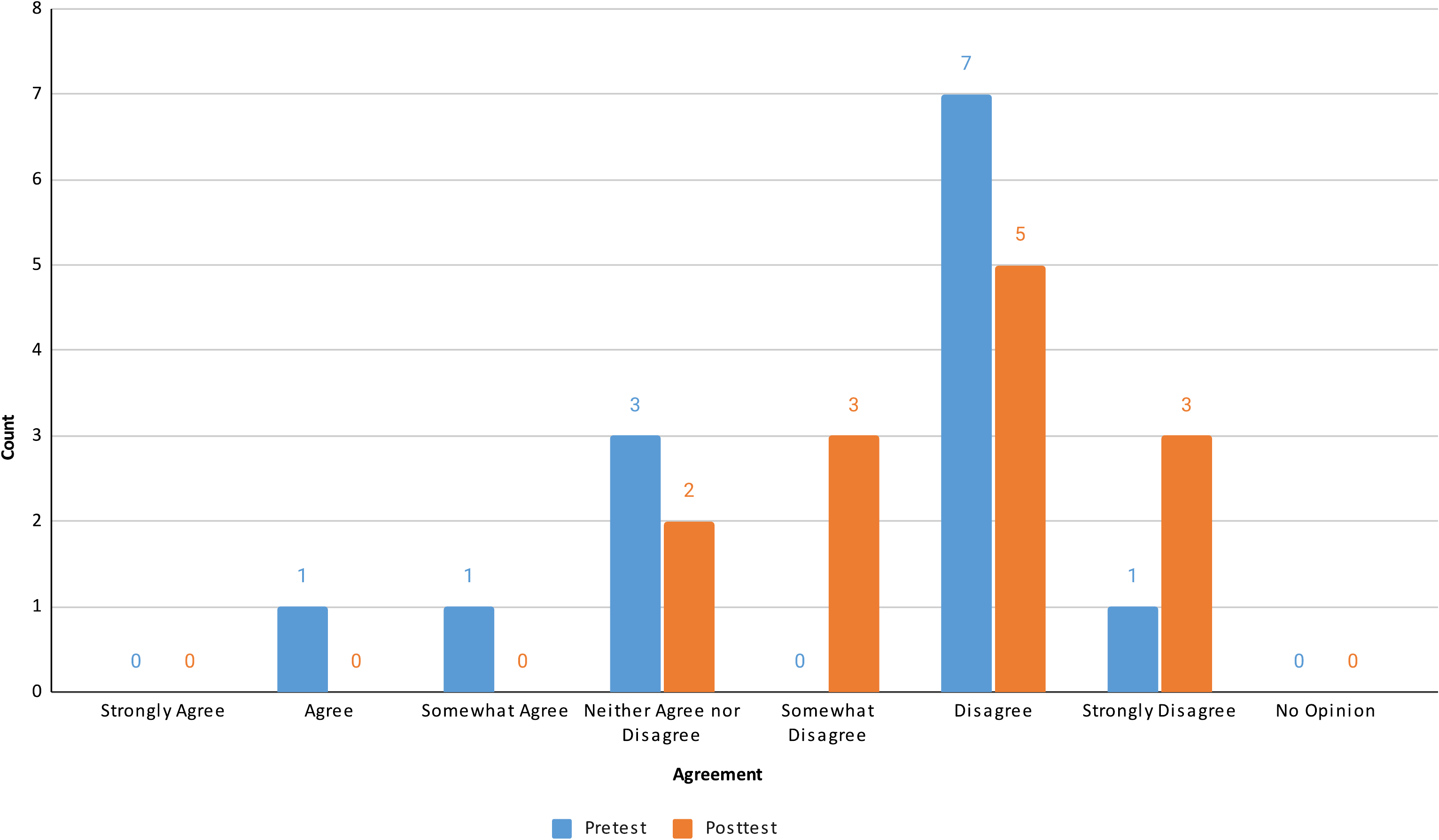}
	\caption{Minimal added value.}
	\label{fig:minimal}
\end{figure}

Questions \prqone to \prqfour are used to collect the participants' opinions on understanding the model checker output during the pretest (\lssix scale). Similarly, questions \poqone to \poqfour are used to collect opinions on understanding the counterexample explanation during the posttest. Each group of four questions (\prqone to \prqfour and \poqone to \poqfour) mainly focus on four aspects: (1)~better understanding, (2)~quicker understanding, (3)~confidence, and (4)~added value. The responses for these four aspects are shown in Figures\,\ref{fig:better_understanding}--\ref{fig:minimal}.

\paragraph{Better understanding.}

\fig{better_understanding} shows the results whether the participants think that the model checker output resp. the counterexample explanation helps understanding refinement inconsistencies. Nine out of 13\,participants answer this question positive (\textit{strongly agree}, \textit{agree}, and \textit{somewhat agree}) with respect to the model checker output in the pretest.
On other hand, all of the 13\,participants answer the question positive with respect to the counterexample explanation in the posttest. Out of these 13\,participants, five participants each \textit{strongly agree} and \textit{agree}, proving that the counterexample explanation allows participants to better understand inconsistencies of refinements. 

\paragraph{Quicker understanding.}

\fig{quicker_understanding} shows the participants' responses regarding time saved based on the model checker output in the pretest resp. the counterexample explanation in the posttest. 
Regarding the model checker output (pretest), most participants \textit{neither agree nor disagree} (5\,participants). Among the remaining eight participants, six participants answer positively (\textit{strongly agree}, \textit{agree}, and \textit{somewhat agree}) and two participants answer negatively (\textit{somewhat disagree}). 

Based on the counterexample explanation (posttest), a majority of the participants \textit{agree} that this can save time (7\,participants). From the remaining participants, three participants each answer with \textit{strongly agree} and \textit{somewhat agree}. No participants answers negatively, which strongly indicates that the provided counterexample explanation does indeed support a quicker understanding of inconsistencies.

\paragraph{Confidence of understanding.}

\fig{more_confident} shows the participants' responses and depicts whether the output from the model checker or the counterexample explanation makes participants confident in their understanding of inconsistencies.
An equal number of participants (6\,participants each) answer either positively (\textit{agree} and \textit{somewhat disagree}) or negatively (\textit{somewhat disagree}, \textit{disagree}, and \textit{strongly disagree}) for the pretest.
For the posttest, however, all participants agree that the counterexample explanation helps to gain confidence with five participants responding \textit{strongly agree}, five participants answering \textit{somewhat agree}, and the remaining three participants responding \textit{agree}.

\paragraph{Minimal added value.}

\fig{minimal} shows whether participants believe in an added value of the model checker output resp. the counterexample explanation by asking whether they think that the added value is minimal. Seven out of 13\,participants answer with \textit{disagree} for the statement that the model checker output provides only the minimal added value. Further, only one participant each answers \textit{agree} and \textit{somewhat agree}.
For the counterexample explanation, the largest share of participants (5\,participants) \textit{disagree}s that the explanation only adds minimal value to real-world projects. Further three participants each \textit{somewhat disagree} and \textit{strongly disagree}, two participants \textit{neither agree nor disagree}. No participant agrees that the counterexample explanation does only provide a minimal added value.

\begin{figure}[b!]
	\centering
	\includegraphics[width=\textwidth]{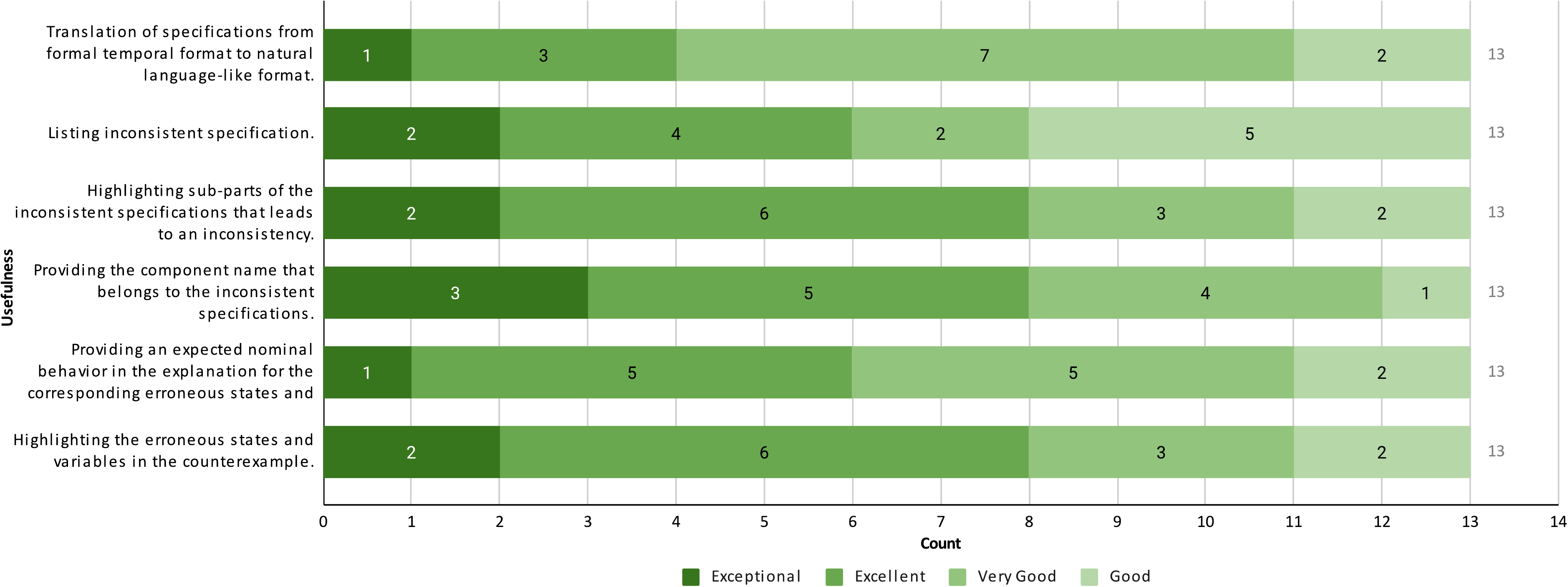}
	\caption{Participants' rating of counterexample explanation features.}
	\label{fig:features}
\end{figure}

\paragraph{Summary.}

Figures\,\ref{fig:better_understanding}--\ref{fig:minimal} show that the participants think that the counterexample approach helps in (1)~better and (2)~quicker understanding of inconsistencies, that it (3)~raises their confidence in the analysis, and that (4)~the provided value is not minimal. For all four aspects, answers were positive for the counterexample explanation in the  posttest than for the model checker output in the pretest.

\subsection{Rating of Counterexample Explanation Features}
\label{sec:phase2:posttest:rating}

Questions \fqone to \fqsix are designed to collect the participants' opinions on the different features of the counterexample explanation approach (possible responses follow scale \lsseven). The responses are shown in \fig{features}. Notably, all provided features are rated positively ranging from \textit{good} to \textit{exceptional}. No participant rated any feature negatively, that is, to be \textit{fair}, \textit{poor}, or \textit{very poor}.
In \fig{features}, the first four features correspond to explaining the violated specification. The last two features correspond to explaining the counterexample.

The three features regarding highlighting the erroneous sub-parts in the inconsistent specification (feature \#3 in \fig{features}), listing the component name of the corresponding inconsistent specifications (feature \#4), and highlighting the erroneous state in the counterexample (feature \#6) were rated particularly helpful, each with eight participants answering either \textit{exceptional} or \textit{excellent}.

\subsection{Feedback}
\label{sec:phase2:posttest:feedback}

Questions \feone to \feeight are used to collect feedback from the participants on the counterexample explanation approach and using formal methods. Suggestions and responses are discussed in the following.

\begin{table}[b]
	\centering
	\caption{Participants' opinions on the ease of understanding inconsistencies with a counterexample explanation in contrast to a model checker output.}
	\label{tab:comp_understanding}
	\resizebox{0.8\textwidth}{!}{%
		\begin{tabular}{|l|l|l|l|l|l|l|l|l|}
			\hline
			\textbf{Likelihood} & Definitely & \begin{tabular}[c]{@{}l@{}}Very \\ Probably\end{tabular} & Probably & \begin{tabular}[c]{@{}l@{}}Neither Probably \\ nor Possibly\end{tabular} & Possibly & \begin{tabular}[c]{@{}l@{}}Probably \\ Not\end{tabular} & \begin{tabular}[c]{@{}l@{}}Definitely \\ Not\end{tabular} & \begin{tabular}[c]{@{}l@{}}No \\ Opinion\end{tabular} \\ \hline
			\textbf{Count}      & 6 (46\%)         & 1    (8\%)                                                    & 5   (38\%)     & 0                                                                       & 1   (8\%)     & 0                                                       & 0                                                         & 0                                                     \\ \hline
		\end{tabular}%
	}
\end{table}

\begin{figure}[b]
	\centering
	\includegraphics[width=0.6\textwidth]{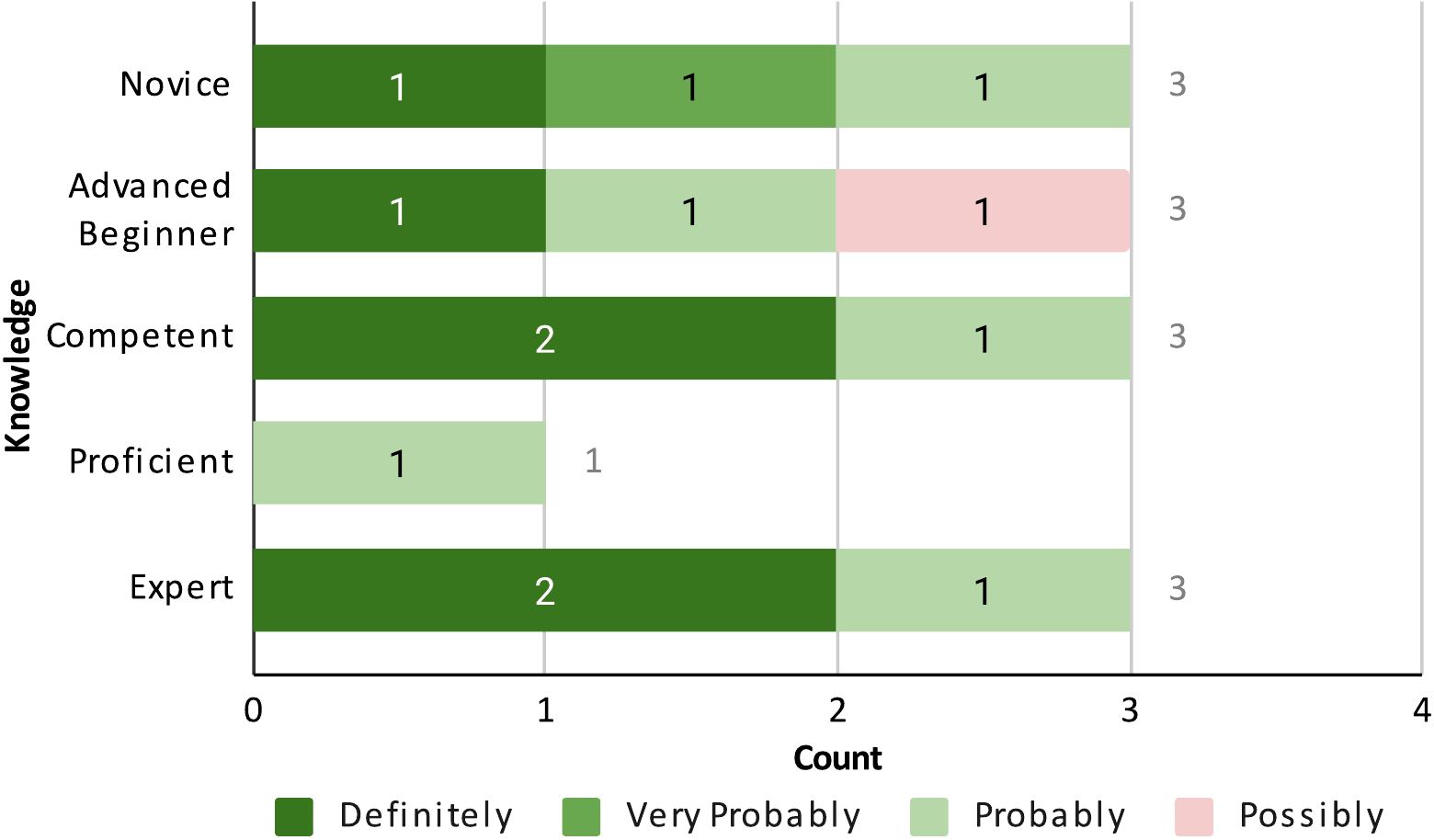}
	\caption{Participants' opinions on the ease of understanding inconsistencies with a counterexample explanation in contrast to a model checker output, grouped by the participants' knowledge in formal methods.}
	\label{fig:inconsistencies_easier_knowledge}
\end{figure}

\subsubsection{Comparison of Understanding Inconsistencies}
\label{sec:phase2:posttest:feedback:comparision}

With question \feone, we asked the participants for feedback according to scale \lsfive whether it is easier to understand the inconsistencies with the counterexample explanation approach than based on the original model checker output. From the 13\,participants, the largest share with six participants answer with \textit{definitely}, further five participants with \textit{probably} (\tab{comp_understanding}).
\fig{inconsistencies_easier_knowledge} shows the results grouped by the participants' knowledge in formal methods.

A majority of participants answer that understanding inconsistencies is easier with the counterexample explanation than with the original model checker output. Only one participant answers slightly hesitant (\textit{possibly}).

\subsubsection{Challenges in Analyzing Inconsistencies}
\label{sec:phase2:posttest:feedback:challange_inconsistency}

With question \fetwo, we want to identify challenges that participants see in analyzing inconsistencies based on the counterexample explanation approach (answers as free-text comments). Nine out of 13\,participants provided responses. Among them, four participants state that the main challenge in understanding stems from the complexity of the EPS system specifications and components. Further three participants state that the main challenge stems from not using formal methods frequently in their daily work. An example statement from a participant is the following: \textit{\enquote{I think, the second example is much more complicated. As I did not work with formal specifications within the last 5 years, it was hard to get into the topic.}}. The remaining two participants highlight that it is still hard to fix the issues despite the counterexample explanation: \textit{\enquote{Highlights and explanation do not provide a solution to the issue so the challenge to identify the root cause in the specification and removing it in a way to expresses the intended behavior still remains a challenge. But the approach helps to identify the root cause.}}

\begin{table}[b]
	\centering
	\caption{Participants' opinions on the ease of maintaining refinement consistency during a refinement step with the counterexample explanation approach.}
	\label{tab:maintaining_consistency}
	\resizebox{0.8\textwidth}{!}{%
		\begin{tabular}{|l|l|l|l|l|l|l|l|l|}
			\hline
			\textbf{Agreement} & \begin{tabular}[c]{@{}l@{}}Extremely \\ Hard\end{tabular} & Hard & \begin{tabular}[c]{@{}l@{}}Slightly \\ Hard\end{tabular} & \begin{tabular}[c]{@{}l@{}}Neither Hard \\ nor Easy\end{tabular} & \begin{tabular}[c]{@{}l@{}}Slightly \\ Easy\end{tabular} & Easy & \begin{tabular}[c]{@{}l@{}}Extremely \\ Easy\end{tabular} & \begin{tabular}[c]{@{}l@{}}No \\ Opinion\end{tabular} \\ \hline
			\textbf{Count}     & 0                                                         & 0    & 0                                                        & 1  (8\%)                                                              & 4  (31\%)                                                      & 6  (46\%)  & 2 (15\%)                                                        & 0                                                     \\ \hline
		\end{tabular}%
	}
\end{table}

\begin{figure}[b]
	\centering
		\includegraphics[width=0.6\textwidth]{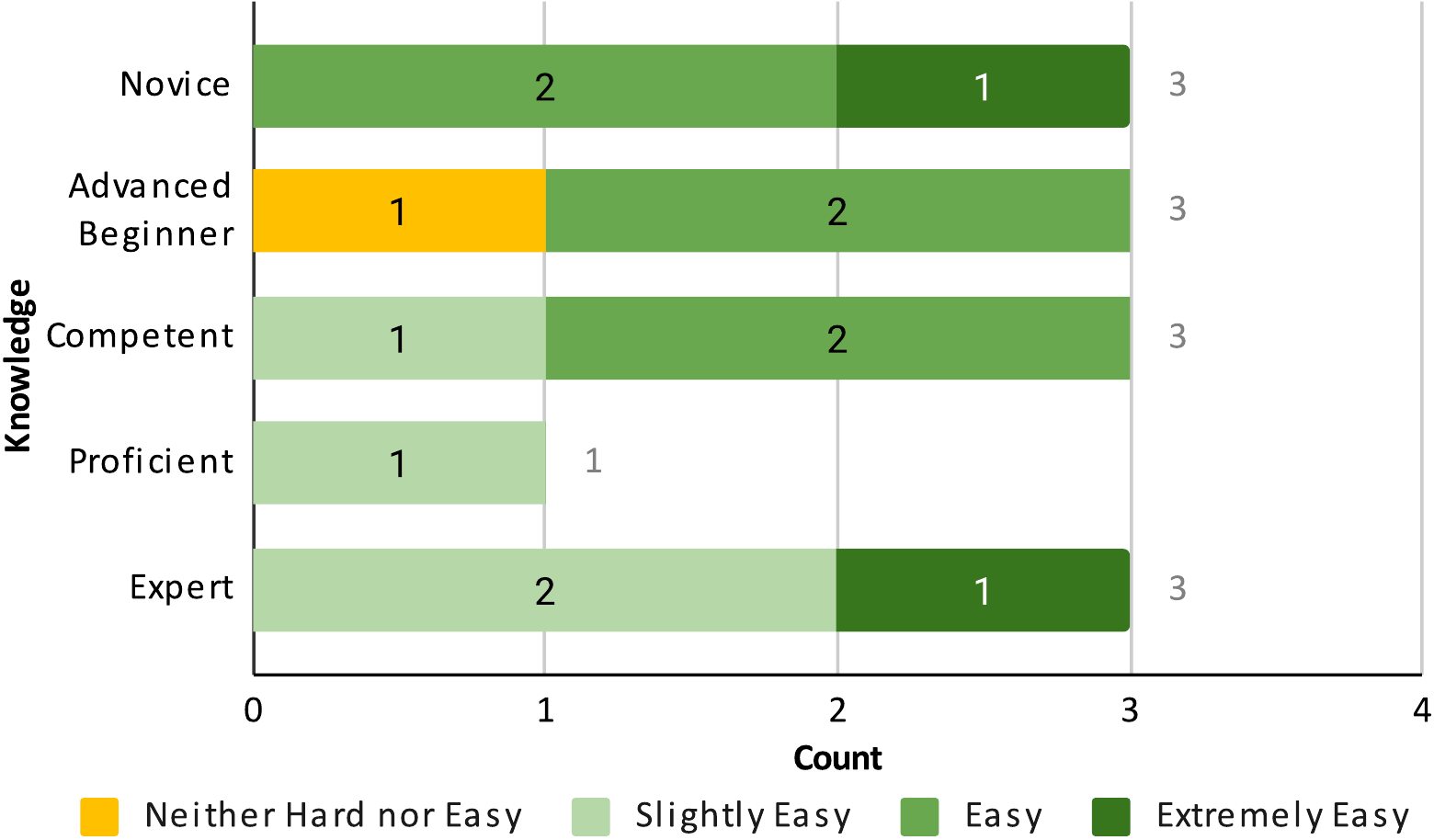}
		\caption{Participants' opinions on the ease of maintaining refinement consistency during a refinement step with the counterexample explanation approach, grouped by the participants' knowledge in formal methods.} 
		\label{fig:maintain_consistency_knowledge}
\end{figure}

\subsubsection{Maintaining Refinement Consistency}
\label{sec:phase2:posttest:feedback:maintaining}

With question \fethree, we collect feedback according to scale \lsthree on whether it is easier to maintain the refinement consistency during a refinement step with the counterexample explanation approach. A majority of the participants answers positively (\tab{maintaining_consistency}). Two participants answer with \textit{extremely easy}, six with \textit{easy}, four with \textit{slightly easy}, and one with \textit{neither hard nor easy}.
\fig{maintain_consistency_knowledge} shows the results based on participants' knowledge in formal methods.

\subsubsection{Counterexample Explanation in Real-world Development}
\label{sec:phase2:posttest:feedback:realworld_ce}

\begin{figure}[b]
	\centering
		\includegraphics[width=0.6\textwidth]{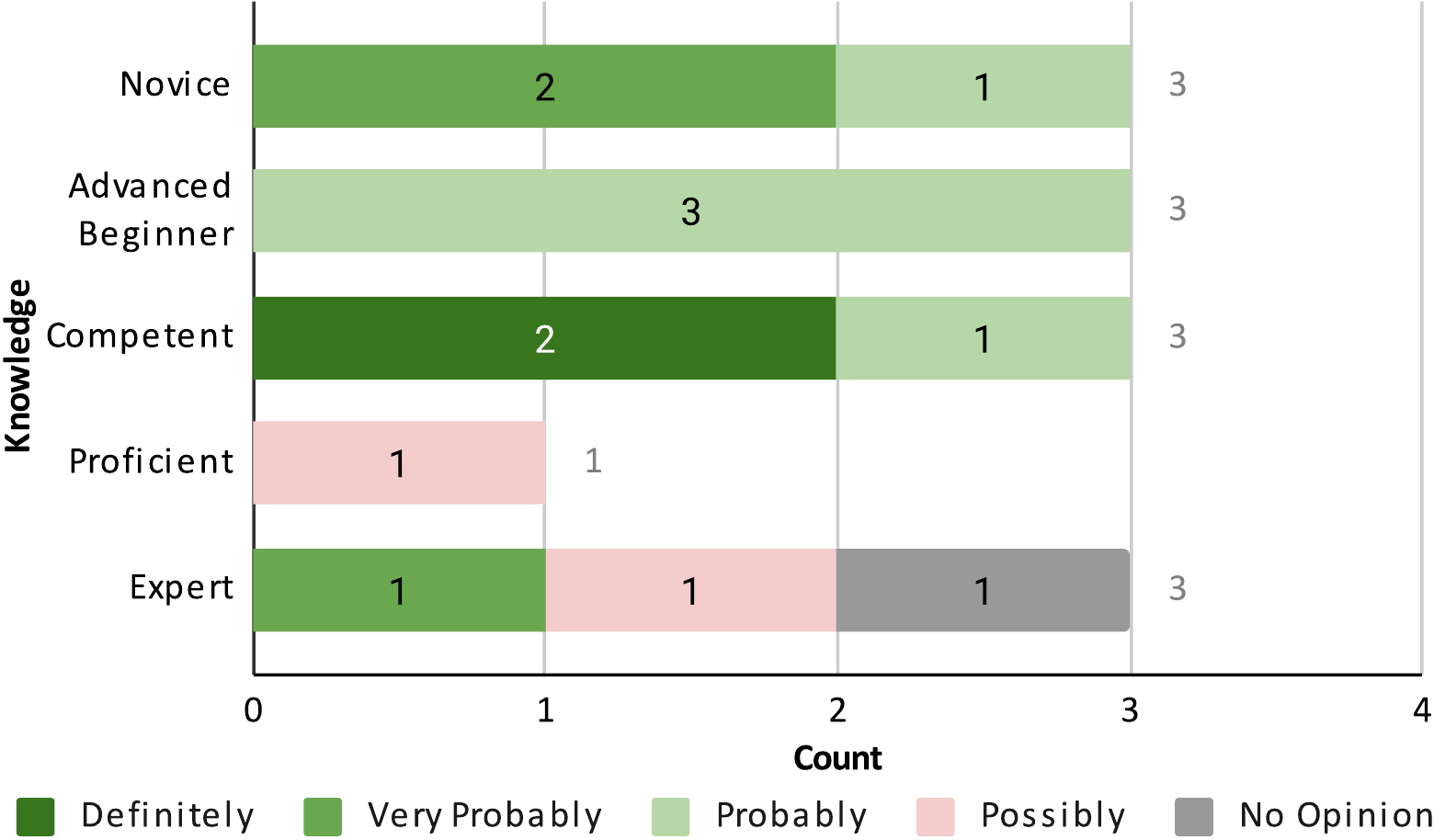}
		\caption{Participants' opinions on using the counterexample explanation approach in real-world development, grouped by the participants' knowledge in formal methods.}
		\label{fig:proposed_counterexample_knowledge}
\end{figure}

With question \fefour, we collect feedback according to scale \lsfive from the participants on whether the proposed counterexample explanation approach is usable in the real-world development processes. Twelve participants provided responses. According to \tab{real_world_dev}, a majority of participants (5) vote for \textit{probably}, with ten participants in total answering positively (\textit{definitely}, \textit{very probably}, and \textit{probably}). 
\fig{proposed_counterexample_knowledge} shows the participants' opinions on using the counterexample explanation approach in real-world development based on their knowledge in formal methods.

\begin{table}[b]
	\centering
	\caption{Participants' opinions on using the counterexample explanation approach in real-world development.}
	\label{tab:real_world_dev}
	\resizebox{0.8\textwidth}{!}{%
		\begin{tabular}{|l|l|l|l|l|l|l|l|l|}
			\hline
			\textbf{Likelihood} & Definitely & \begin{tabular}[c]{@{}l@{}}Very \\ Probably\end{tabular} & Probably & \begin{tabular}[c]{@{}l@{}}Neither Probably \\ nor Possibly\end{tabular} & Possibly & \begin{tabular}[c]{@{}l@{}}Probably \\ Not\end{tabular} & \begin{tabular}[c]{@{}l@{}}Definitely \\ Not\end{tabular} & \begin{tabular}[c]{@{}l@{}}No \\ Opinion\end{tabular} \\ \hline
			\textbf{Count}      & 2   (15\%)       & 3    (23\%)                                                    & 5 (38\%)& 0       & 2   (15\%)                                                                   & 0      & 0                                                       & 1    (8\%)                                               \\ \hline
		\end{tabular}%
	}
\end{table}

\subsubsection{Counterexample Explanation with Formal Methods in Real-World Development}
\label{sec:phase2:posttest:feedback:realworld_fm}

With question \fefive, we collect feedback from participants according to answer scale \lsfive on whether engineers prefer to use the proposed counterexample explanation while using formal methods. Among 13\,participants, 12 have responded (\tab{using_formal_methods}). Among these 12\,participants, the largest share of seven participants answer with \textit{probably}. One further participant each answers with \textit{definitely} and \textit{very probably}, the remaining three participants answer with \textit{possibly}.

\begin{table}[t]
	\centering
	\caption{Participants' opinions on using the counterexample explanation approach while using formal methods in real-world development.}
	\label{tab:using_formal_methods}
	\resizebox{0.8\textwidth}{!}{%
		\begin{tabular}{|l|l|l|l|l|l|l|l|l|}
			\hline
			\textbf{Likelihood} & Definitely & \begin{tabular}[c]{@{}l@{}}Very \\ Probably\end{tabular} & Probably & \begin{tabular}[c]{@{}l@{}}Neither Probably \\ nor Possibly\end{tabular} & Possibly & \begin{tabular}[c]{@{}l@{}}Probably \\ Not\end{tabular} & \begin{tabular}[c]{@{}l@{}}Definitely \\ Not\end{tabular} & \begin{tabular}[c]{@{}l@{}}No \\ Opinion\end{tabular} \\ \hline
			\textbf{Count}      & 1 (8\%)         & 1  (8\%)                                                      & 7 (54\%) & 0       & 3  (23\%)                                                                    & 0      & 0                                                       & 1  (8\%)                                                 \\ \hline
		\end{tabular}%
	}
\end{table}

\subsubsection{Counterexample Explanation in Your Project}
\label{sec:phase2:posttest:feedback:project_ce}

With question \fesix, we collect feedback from the participants according to answer scale \lsfive on whether engineers prefer to use the proposed counterexample explanation approach in their real-world projects. Four participants have \textit{no opinion} and the remaining nine participants provided feedback (\tab{presented_approach}). Among these nine participants, five participants answer negatively (\textit{possibly}, \textit{probably not}, and \textit{definitely not}), and four participants answer positively (\textit{definitely}, \textit{very probably}, and \textit{probably}). These answers contrast the previously discussed feedback questions \feone to \fefive where responses are more positive while for \fesix, the majority of parti\-cipants answer negatively. The reason for this negative rating as mentioned by the majority of participants are the currently used development processes and tools that do not fit well with formal methods or the counterexample explanation approach.

\begin{table}[h!]
	\centering
	\caption{Participants' opinions on using the counterexample explanation approach in the participants' projects.}
	\label{tab:presented_approach}
	\resizebox{0.8\textwidth}{!}{%
		\begin{tabular}{|l|l|l|l|l|l|l|l|l|}
			\hline
			\textbf{Likelihood} & Definitely & \begin{tabular}[c]{@{}l@{}}Very \\ Probably\end{tabular} & Probably & \begin{tabular}[c]{@{}l@{}}Neither Probably \\ nor Possibly\end{tabular} & Possibly & \begin{tabular}[c]{@{}l@{}}Probably \\ Not\end{tabular} & \begin{tabular}[c]{@{}l@{}}Definitely \\ Not\end{tabular} & \begin{tabular}[c]{@{}l@{}}No \\ Opinion\end{tabular} \\ \hline
			\textbf{Count}      & 2 (15\%)         & 0                                                        & 2 (15\%)& 0       & 1  (8\%)                                                                    & 3   (23\%)   & 1   (8\%)                                                   & 4   (31\%)                                                \\ \hline
		\end{tabular}%
	}
\end{table}

\subsubsection{Further Improvements}
\label{sec:phase2:posttest:feedback:improvements}

The motive of questions \feseven and \feeight are to collect suggestions for further improvement of the counterexample explanation approach.
Question \feseven mainly focuses on collecting participants' feedback on whether providing a list of possible solutions/fixes would be helpful for engineers. This feedback is collected according to answer scale \lsfive. \feeight is an open question that allows participants to suggest further improvements as free-text comments. The responses to \feseven are shown in \tab{possible_suggestions}. Most of the answers are positive, meaning that providing a list of possible solutions/fixes would be helpful to participants.

With question \feeight, we collected general suggestions from the participants. We received eight responses that suggest improvement for using the counterexample explanation approach. A majority suggests to provide training to use formal methods, and improve or develop the visualization of the explanation by associating and integrating it with the tools used within Bosch, \eg IBM Rhapsody and DOORS Next Generation (DNG).

\begin{table}[t!]
	\centering
	\caption{Participants' opinions on the usefulness of providing a list of possible solutions/fixes to a refinement inconsistency.}
	\label{tab:possible_suggestions}
	\resizebox{0.8\textwidth}{!}{%
		\begin{tabular}{|l|l|l|l|l|l|l|l|l|}
			\hline
			\textbf{Likelihood} & Definitely & \begin{tabular}[c]{@{}l@{}}Very \\ Probably\end{tabular} & Probably & \begin{tabular}[c]{@{}l@{}}Neither Probably \\ nor Possibly\end{tabular} & Possibly & \begin{tabular}[c]{@{}l@{}}Probably \\ Not\end{tabular} & \begin{tabular}[c]{@{}l@{}}Definitely \\ Not\end{tabular} & \begin{tabular}[c]{@{}l@{}}No \\ Opinion\end{tabular} \\ \hline
			\textbf{Count}      & 5  (38\%)        & 4       (31\%)                                                 & 2 (15\%)& 1    (8\%)   & 1    (8\%)                                                                    & 0      & 0                                                      & 0                                                   \\ \hline
		\end{tabular}%
	}
\end{table}

\section{Discussion}
\label{sec:discussion}

In this section, we discuss the findings of the two user study phases following the research questions and points to be investigated (\cf \sect{rq}).

\subsection{RQ1 -- Challenges in Identifying Inconsistent Specifications}
\label{sec:discussion:rq1}

Research question \textit{RQ1} gathers challenges in identifying inconsistent formal specifications that are introduced during the refinement of a system.

\paragraph{Understanding formal notations is difficult for engineers.}
Formal methods may play a crucial role on the left-hand side of the V-model~\citep{Weber09}, where \textit{systems engineers} and \textit{safety manager/engineers} are mainly involved, to avoid major flaws in early design decisions. In our \textit{user survey} Phase 1, 20 of 41\,participants are \textit{system engineers} and \textit{safety managers/engineers} (\sect{phase1:participants:designation}). The majority of them perceives understanding of formal notations as hard to some degree (\cf \sect{phase1:notation}). Additionally, we noticed that the complexity of understanding depends on years of experience in using formal methods. Results further show that introducing formal methods to engineering teams with little experience in formal methods is challenging.

\paragraph{\change{Identifying inconsistent specifications is difficult for engineers.}}
Results in \sect{phas1:inconsistent_spec} clearly show that a majority of Phase 1 participants perceive the identification and understanding of inconsistent specifications as \textit{hard} to some degree and that it consumes significant time. Results are similar for maintaining and verifying the refinement consistency (\sect{phase1:refinement}).

Qualitative statements by the participants note that identifying and understanding of inconsistent specifications highly depends on the size of the system and the number of its requirements. 
These statements emphasize the question whether the usage of formal methods for industrial system and at industrial scale is possible. For example, the automobile sector is now developing systems that are quickly expanding in size and complexity as a result of highly automated driving.
System and requirements are not only complex but also frequently changing, for instance, due to security demands.

Our initial expectation was that more of our participants perform \textit{manual inspections/reviews} rather than using automated tools like \textit{model checker}, \textit{simulators}, and \textit{reasoners} in their development projects. Thus, our hypothesis is that a majority of participants will answer that identifying inconsistent specifications and understanding is very complex. However, on the contrary, a majority of participants use \textit{model checkers} (\sect{phase1:methods}). Even though a majority uses automated tools like \textit{model checkers}, a majority still answers that the identification and understanding of inconsistent specifications are \textit{hard}.
A reason for this might be the complexity of verification tools and their generated results, raising the need to make them more user-friendly to be used by engineers without having in-depth knowledge of formal methods.

\subsection{RQ2 -- Benefit of Formal Methods to Development Processes}
\label{sec:discussion:rq2}

\textit{RQ2} gathers insights on whether the identification of inconsistent specifications and usage of formal methods are beneficial to real-world development processes.

\paragraph{\change{Using formal methods can make the system safer.}}

Functionalities of automotive systems increase expeditiously, resulting in more (safety) requirements to avoid any unintended behavior.
Thus, performing safety analysis early on the left-hand side of the V-model~\citep{Weber09} is crucial to help reducing the number of errors identified later during the validation. A majority of participants agrees that formal verification can make the system safer and be a benefit to the functional safety (\cf \sect{phase1:fm_safety} and \sect{phase1:opinion:using}). Although a majority of the participants have a positive opinion on using formal verification, based on the qualitative answers from the participants, there is a discussion whether formal verification is usable and scalable to real-world systems. Specifically, a majority participants indicate that the usage of formal methods could be improved by making formal notations easier to understand (\cf \sect{phase1:opinion:notation}). 

\paragraph{\change{Identifying inconsistent specifications is beneficial in real-world development processes.}}

Eliciting requirements, refining requirements, and developing system architectures are the initial steps in the V-model~\citep{Weber09}. Thus, requirement elicitation and refinement of those requirements are crucial as they serve as a basis for further system development and safety analysis. Errors and inconsistencies introduced in these early phases, identified only in later development stages, become costly and may lead to catastrophic events. There is a high possibility that most safety-critical errors are identified late during the validation phase in industry~\citep{PohlR11}.
Therefore, to identify errors in the requirements during the initial stages performing manual reviews (\eg inspections) does not seem to be sufficient or an efficient approach. This motivates the usage of automated methods like formal verification and simulation to help identifying errors in requirements and overcoming challenges of manual reviews. Although a majority of participants answers that using and understanding formal verification are complex, the majority agrees that identifying inconsistent formal specifications is beneficial for safety analysis and makes a system safer (\cf \sect{phase1:fm_inconsistent_benificial}). 

\subsection{RQ3 -- Easing the Use of Formal Methods}
\label{sec:discussion:rq3}

Insights for \textit{RQ3} are drawn from Phase\,2, the one-group pretest-posttest experiment. \textit{RQ3} gathers insights whether engineers prefer to use formal methods (model checkers particularly) if the difficulty for understanding verification results to identify inconsistent specifications is reduced, in particular with the counterexample explanation approach.

\paragraph{The counterexample explanation approach eases the comprehension when compared to the interpretation of the raw model checker output.} 

Six of 13\,participants of the one-group pretest-posttest experiment answer that they understand the verification result generated by the model checker for the airbag system (\sect{phase2:postest:result}). Our initial hypothesis was that by proving an additional user-friendly counterexample explanation, it would be easier for engineers to understand the error as well as that it can ease the usage of formal methods among engineers. Indications for this are already drawn from \textit{Part\,1}, where a majority answers that making formal notations easier to understand can improve the usage of formal methods (\sect{phase1:opinion:notation}). Finally, the results discussed in \sect{phase2:postest:result} support this hypothesis, as a stark majority of 12 of 13\,participants prefer the counterexample explanation compared to understanding the model checker output. In summary, improving the usability and understandability aspects of formal notations can promote the use of formal methods.

Additionally, the results collected for the four aspects of a better understanding, quicker understanding, confidence of understanding, and added value (\cf \sect{phase2:posttest:opinion}) validate this hypothesis, as the counterexample explanation improves all four aspects compared to the raw model checker output presented otherwise to engineers.
	
\paragraph{It is possible for engineers to identify and fix inconsistent specifications based on the counterexample explanation approach.}

Apart from collecting the participants' \emph{opinions}, we can also rely on the experiment, which let participants perform tasks for the provided use cases. The results presented in \sect{phase2:posttest:feedback:comparision} evaluate the difference of correct and incorrect answers in finding inconsistent components between pretest and posttest and shows only a minor difference between working with raw results and the counterexample explanation.

However, this is not the case for the identification of inconsistent specifications. With the raw model checker output, only five of 13\,participants identify either fully or partially correct the inconsistent specifications. While with the counterexample explanation, nine of 13\,participants were able to identify the complete set of inconsistent specifications and also correctly explained the reason of the inconsistency by understanding the explanation. This strongly shows that a counterexample explanation can indeed improve the error comprehension and providing such an explanation can promote the use of formal methods among engineers.   

\paragraph{The counterexample explanation approach can promote formal verification and usage of model checking in real-world development processes.}

From the collected responses, a majority of participants have a positive opinion as a counterexample explanation could support maintaining refinement consistency, could be usable in real-world development process, and could be used while using formal methods (\cf \sect{phase2:posttest:feedback}). These results clearly indicate that a counterexample explanation approach could be one possible way along with other possible options like property specification patters to improve the usability aspects of formal methods. However, for the question whether the participants are interested to use the counterexample explanation approach in their project, the response is contradictory where only a minimal number of participants are interested to use it. A major challenge mentioned by the participants is that tools currently used in their projects do not support integrating the proposed counterexample explanation approach. This shows that integrating the existing verification tools with industrial tools needs to be one of the prime focus to improve the usage of formal methods. Nevertheless, such an integration cannot be achieved easily since larger organizations such as Bosch typically use different tools for different projects. Thus, focusing on adaption and integration of each tool individually is not a trivial task.

\section{Threats to Validity}
\label{sec:ttv}

In this section, we discuss threats that might jeopardize the validity of our study results as well as on measures we take to reduce these threats.
We consider threats to validity as discussed by \cite{WohlinRHO12}, \cite{KitchenhamP08}, and \cite{CampbellS63}. In the following, we structure them according to construct, internal, and external validity.

\paragraph{Construct validity.}

The prime threats to construct validity are related to the completeness of the questionnaire and in phrasing questions in a way that is understood by all participants in the same way. To mitigate these threats, we have considered the following steps in our research method: (i)~we incorporated feedback from two senior engineers having background in formal methods and model checking, (ii)~we incorporated feedback regarding unbiased questions from a psychologist, and (iii)~we performed a pilot test with five research engineers to check for completeness and understandability.

\paragraph{Internal validity.}

The critical internal threat to be considered for the \textit{user survey} is the selection of participants. Since we followed snowball sampling for the participant selection, there could be a possibility of several participants working in the same project, which could bias the final result. Therefore, we considered at most first four participants from each project and neglected further project members.

We consider threats to internal validity listed by \cite{CampbellS63} for the pretest-posttest experiment. To mitigate the \emph{history} and \emph{maturation} threats, we performed the posttest experiment within fifteen days following the pretest experiment. The most severe threats to be considered in this experimental design are \emph{testing} and \emph{instrumentation}. Those threats arise because participants get overwhelmed with the intervention \change{including the fact that we have developed the counterexample explanation approach. Consequently, participants could answer more positively in the posttest experiment than the actual value due to the intervention. To mitigate these threats, participants conduct the study anonymously and we explicitly emphasized to the participants that the obtained study results would serve as a reference in the future to use our counterexample explanation approach for real-world projects at Bosch.}
Additionally, to avoid overwhelmed responses and accept only valid responses, we have added the task questions \tqone--\tqnine (\tab{onegroup}); And the response is accepted as valid only if the participant attempted to answer at least some part of these questions. Further, to reduce biasing between the pretest and posttest experiment, the use case of an airbag system (a toy example) used in the pretest is significantly less complex than the use case of the Bosch EPS system used in the posttest. However, to adjust the difficulty level of the systems used for the experiment, we used feedback from the pilot study with five research engineers. Basically, adjustment of difficulty is done by increasing or decreasing the number of components and size of the specifications which have to be understood by the participants.

Finally, another internal threat is to present the model checker’s raw output with the inconsistent specification highlighted by us to the participants in the pretest. This could bias the participants’ opinions that the model checker's output included the highlighted parts is easier to interpret than it actually is in practice where the highlighted parts are not available. As such, we can rather expect larger benefits of our counterexample explanation approach in practice than we observed it in the one-group pretest-posttest experiment.

\paragraph{External validity.}

To avoid polluting results, we do not force the participants to select an option from an answer scale for every question. For example, the participants could choose the option \emph{No Opinion}, which supports in achieving actual results. However, the participants have the choice to enter a reason as a qualitative statement if they do not want to select any option.
One of the severe drawbacks of the one-group pretest-posttest experiment is its generalization. However, the benefit of our study is that we used a real-world EPS system for the posttest experiment, and the participants are professional engineers who work on real-world automotive projects at Bosch.

\section{Conclusion}
\label{sec:conclusion}

Our user study was designed to i) identify the motivation, challenges, and applicability of formal methods in industry and ii) evaluate if the proposed counterexample explanation approach matches the identified challenges.

To identify the motivation, challenges, and applicability of formal methods in industry, we conducted an extensive survey with 41\,participants of various business units and disciplines within Bosch as a first phase of the user study. Responses show that the majority of the participants is positive regarding the use of formal methods in real-world development processes. However, participants identify that
incomplete formal models in industry,
understanding formal notations,
as well as
understanding verification results, e.g., produced by a model checker,
still remain a challenge for adoption of formal methods in practice.
Identifying refinement inconsistencies gets more complex with the system getting more complex and the number of requirements increasing. Apart from understanding and scalability challenges, one of the major challenge in using verification tools is a lack of training for engineers.

As a second part of the user study, we performed a one group pretest-posttest study with 13\,participants of various Bosch business units to evaluate if the proposed counterexample explanation approach is capable of supporting the use of formal methods in industry.
Results from the experiments as well as collected opinions from participants prove that the approach helps in
i)~better understanding and
ii)~quicker understanding inconsistencies, that it
iii)~raises their confidence in the analysis, and that
iv)~it provide value for the development of safety-critical projects in industry. 

\begin{center}
	\textit{As researchers and educators in formal methods, we should strive to make our notations and tools accessible to non-experts.} -- \cite{ClarkeW96}
\end{center}

\paragraph{Future directions.}
\label{sec:future_work}

To leverage formal methods in real-world development processes, one of the most suitable means is to provide education and training in formal methods. On the one hand, universities can teach the foundations of formal methods to students (\eg temporal logics). On the other hand, companies can teach the skills required in the specific industrial context (\eg considering the domain and tooling) to people entering industry as well as upskill existing employees to understand the foundations of formal methods. 
By providing education structured training in formal methods either in universities or companies, hesitancy in using formal methods could be reduced and the benefits of formal methods could be reaped. 

Looking at the results of evaluating our counterexample explanation approach, it is clear that understanding of verification results is easier with a counterexample explanation than with the direct output of model checker. In future, similar explanations need to be generated for different model checkers, domain-specific system models and requirements, as well as integrated with project-specific tool chains. Furthermore, instead of only providing explanations that illustrate the error, providing suggestion to fix those errors could help to improve the agility to perform verification iteratively and thus, to support round-trip engineering.

\begin{acknowledgements}
We would like to express our gratitude to all engineers who participated in our study for their time spent and their valuable responses. Furthermore, we would like to thank Amalinda Post, Igor Menzel, and Kevin Heiner for their support in designing the study and improving the questionnaire.
\end{acknowledgements}

%
\section*{Declarations}

\textbf{Conflict of Interests} The authors declare that they have no conflict of interest.
\\~\\
\noindent\textbf{Data Availability} The data collected during the studies is not openly available due to reasons of sensitivity to Bosch and the privacy of the engineers. The data can be made available from the corresponding author upon reasonable request.

\bibliography{1_main}  
\end{document}